\documentclass[preprint, 12pt]{elsarticle}

\usepackage{graphicx}
\usepackage{dcolumn}
\usepackage{bm}
\usepackage{psfrag}
\usepackage{subcaption}
\usepackage{siunitx}
\usepackage{amsmath}
\usepackage{ragged2e}
\usepackage{float}
\usepackage{tikz}
\usetikzlibrary{plotmarks}
\usepackage{rotating}
\usepackage{multirow}
\usepackage{comment}
\usepackage{soul}
\newcommand\marksymbol[2]{\tikz[#2,scale=1.5]\pgfuseplotmark{#1};}

\definecolor{magenta2}{RGB}{255,0,255}
\definecolor{blue2}{RGB}{0, 127, 255}
\definecolor{grey2}{RGB}{85, 85, 85}
\definecolor{purp2}{RGB}{133, 11, 255}
\tikzstyle{longdashed}=                  [dash pattern=on 6pt off 2pt]
\tikzstyle{dashdotdot}=              [dash pattern=on 4pt off 2pt on \the\pgflinewidth off 1pt on \the\pgflinewidth off 2pt]
\tikzstyle{dotdotted}=              [dash pattern=on 1.5pt off 1.5pt on 1.5pt off 1.5pt on 1.5pt off 1.5pt]
\DeclareRobustCommand{\pupdotdotted}{\raisebox{2pt}{\tikz{\draw[purp2,dotdotted,line width=1.0pt](0,0) -- (5mm,0);}}}
\DeclareRobustCommand{\redsolid}{\raisebox{2pt}{\tikz{\draw[red,solid,line width=1.0pt](0,0) -- (5mm,0);}}}

\DeclareRobustCommand{\bluedashdotted}{\raisebox{2pt}{\tikz{\draw[blue,dashdotted,line width=1.0pt](0,0) -- (5mm,0);}}}
\DeclareRobustCommand{\blackdashed}{\raisebox{2pt}{\tikz{\draw[black,dash pattern=on 5pt off 4pt,line width=1.0pt](0,0) -- (5mm,0);}}}

\DeclareRobustCommand{\greendashed}{\raisebox{2pt}{\tikz{\draw[green,dashed,line width=1.0pt](0,0) -- (5mm,0);}}}
\DeclareRobustCommand{\magentadashed}{\raisebox{2pt}{\tikz{\draw[magenta2,longdashed,line width=1.0pt](0,0) -- (5mm,0);}}}
\DeclareRobustCommand{\blackdashdotdot}{\raisebox{2pt}{\tikz{\draw[black,dashdotdot,line width=1.0pt](0,0) -- (5mm,0);}}}
\DeclareRobustCommand{\blackdashdot}{\raisebox{2pt}{\tikz{\draw[black,dashdotted,line width=0.7pt](0,0) -- (5mm,0);}}}
\DeclareRobustCommand{\orangedashdotdot}{\raisebox{2pt}{\tikz{\draw[orange,dashdotdot,line width=1.0pt](0,0) -- (5mm,0);}}}
\DeclareRobustCommand{\reddashdotdot}{\raisebox{2pt}{\tikz{\draw[red,dashdotdot,line width=1.0pt](0,0) -- (5mm,0);}}}

\journal{Journal of Computational Physics}

\begin{document}

\begin{frontmatter}

\title{Sensitivity analysis of wall-modeled large-eddy simulation for separated turbulent flow}

\author{Di Zhou\corref{cor1}}
\ead{dizhou@caltech.edu}
\author{H. Jane Bae}
\ead{jbae@caltech.edu}
\cortext[cor1]{Corresponding author}
\affiliation{organization={Graduate Aerospace Laboratories},
            addressline={California Institute of Technology}, 
            city={Pasadena},
            postcode={91125}, 
            state={CA},
            country={USA}}



\begin{abstract}
In this study, we conduct a parametric analysis to evaluate the sensitivities of wall-modeled large-eddy simulation (LES) with respect to subgrid-scale (SGS) models, mesh resolution, wall boundary conditions and mesh anisotropy. While such investigations have been conducted for attached/flat-plate flow configurations, systematic studies specifically targeting turbulent flows with separation are notably sparse. To bridge this gap, our study focuses on the flow over a two-dimensional Gaussian-shaped bump at a moderately high Reynolds number, which involves smooth-body separation of a turbulent boundary layer under pressure-gradient and surface-curvature effects. In the simulations, the no-slip condition at the wall is replaced by three different forms of boundary condition based on the thin boundary layer equations and the mean wall-shear stress from high-fidelity numerical simulation to avoid the additional complexity of modeling the wall-shear stress. Various statistics, including the mean separation bubble size, mean velocity profile, and  {dissipation} from SGS model, are compared and analyzed. The results reveal that capturing the separation bubble strongly depends on the choice of SGS model.  {While simulations approach grid convergence with resolutions nearing those of wall-resolved LES meshes}, above this limit, the LES predictions exhibit intricate sensitivities to mesh resolution. Furthermore, both wall boundary conditions and the anisotropy of mesh cells exert discernible impacts on the turbulent flow predictions, yet the magnitudes of these impacts vary based on the specific SGS model chosen for the simulation.

\end{abstract}



\end{frontmatter}



\section{\label{sec:1}Introduction}

In recent years, wall-modeled LES (WMLES) has been treated as an essential technology in the field of computational fluid dynamics (CFD)  {\citep{rumsey2022over}}. It offers enhanced accuracy over Reynolds-averaged Navier-Stokes (RANS) simulations and facilitates intricate simulations of high-Reynolds-number turbulent flows without the prohibitive computational costs of demanding near-wall mesh resolutions. WMLES can be broadly categorized as either hybrid RANS/LES or LES based on a wall-stress model \citep{larsson2016large,bose2018wall}. The present research focuses on the latter, where the LES equations are solved throughout the entire computational domain, and the wall-shear stress, accounting for the effects of the unresolved inner layer of the boundary layer, is provided by the wall-stress model at the wall. Many wall-stress models operate on the assumption of equilibrium turbulence near the wall, either explicitly or implicitly incorporating the law of the wall for attached turbulent flow. To go beyond the equilibrium assumption, the development of wall models capable of addressing broader non-equilibrium flows has become a focal point in the wall-modeling community \citep{wang2002dynamic,kawai2013dynamic,bose2014dynamic, park2014improved, bae2019dynamic}. Given the ongoing advancements in wall-modeling techniques, there is an escalating interest in using WMLES for simulating turbulent flows with separation in both academia and industry \citep{bodart2013large, park2017wall, whitmore2021large, iyer2022wall}. Separated turbulent flows are commonplace in realistic applications across various engineering fields, including aerospace, automotive, and environmental systems. Their accurate numerical prediction, however, remains a significant challenge due to the inherent complexity and multi-scale nature of the flow physics. In addition, there is a noticeable absence of comprehensive guidance for applying WMLES to separated turbulent flows, which highlights the crucial need to assess and understand the sensitivities of WMLES for such flows. 

Recent studies have significantly advanced our understanding of the sensitivities associated with WMLES. For instance, \citet{rezaeiravesh2019systematic} systematically explored the predictive accuracy of WMLES in turbulent channel flow simulations. This research investigated how wall models, grid resolution, grid anisotropy, numerical convective schemes, and SGS modeling influence the simulation outcomes. The findings highlighted that these factors play an interconnected role in shaping the overall simulation accuracy. Taking a slightly different perspective, \citet{lozano2019error} analyzed the LES error scaling for the mean velocity profiles, turbulence intensities, and energy spectra in the outer region of wall-bounded flows. The analysis was carried out via LES of turbulent channel flows, employing the precise mean wall-stress at the wall to isolate the influence from the near-wall region. Their discoveries suggested that LES errors in the mean velocity profile are inversely proportional to the grid resolution while being unaffected by the Reynolds number. Building on a comparable methodology, \citet{whitmore2020requirements} delves into the sensitivities of near-wall WMLES solutions, particularly focusing on SGS modeling, boundary condition types, numerical methods, and mesh topologies. Their observations emphasized that WMLES's sensitivity to the SGS model is closely tied to the detailed implementation of the wall model. While these studies primarily focus on attached turbulent flows, research on separated turbulent flows remains sparse. Noteworthy contributions include the grid convergence study by \citet{park2017wall} regarding the WMLES of flow over the NASA wall-mounted hump, and \citet{iyer2022wall}'s exploration of WMLES for flows over a Gaussian bump, which examined the impacts of grid resolution and the SGS model. To bridge this research gap, our current study aims to provide an in-depth parametric analysis, evaluating the sensitivities of WMLES in simulating separated turbulent flow.

The flow over a Gaussian-shaped bump, proposed by Boeing Research \& Technology \citep{slotnick2019integrated}, is a widely investigated separated turbulent flow. This configuration mimics the smooth junctions between an aircraft wing and fuselage that involves smooth-body separation of a turbulent boundary layer (TBL) subject to pressure-gradient and surface-curvature effects. As a canonical flow configuration with extensive study, a wealth of experimental data exists \cite{williams2020experimental, gray2021new, gray2022experimental, gray2022benchmark, gluzman2022simplified}, establishing it as a benchmark for validating CFD techniques. Diverse computational approaches, including RANS methods \citep{williams2020experimental}, direct numerical simulation (DNS) \citep{balin2021direct,uzun2021simulation}, wall-resolved LES (WRLES) \citep{rizzetta2023wall}, hybrid DNS-WRLES \citep{wright2021unstructured,uzun2022high}, WMLES \citep{whitmore2021large,iyer2022wall,agrawal2022non,zhou2023large,arranz2023wall} and detached-eddy simulation \citep{balin2020wall} have been evaluated. These computational studies have consistently highlighted the challenge of precisely predicting the extent and location of smooth-body separation. In particular, WMLES investigations of \citet{iyer2022wall} and \citet{agrawal2022non} suggest that the performance of the equilibrium wall model can be affected by the SGS model, especially in the prediction of the separation bubble.  {Additionally, \citet{iyer2022wall} also evaluated the effect of variations in wall modeling, including scenarios devoid of any wall model. Their findings indicate a notable influence of wall models on the prediction of flow separation; however, compared to the equilibrium wall model, the enhancements derived from employing non-equilibrium wall model \citep{park2014improved} were found to be marginal.} Given the intricate physics of smooth-body separation and the availability of high-fidelity reference data, the flow over a Gaussian-shaped bump emerges as an apt choice for our examination of the sensitivity of WMLES for separated turbulent flow. To avoid the complexity of spanwise variations, our focus of the current study narrows to the flow over a two-dimensional Gaussian bump with periodic boundary conditions in the spanwise direction. The flow configuration and simulation setup for this investigation are based on the hybrid DNS-WRLES study by \citet{uzun2022high}. In order to preclude the complications of wall modeling, the no-slip condition at the wall is substituted with three unique boundary conditions based on the thin boundary layer equations and the mean wall-shear stress from the hybrid DNS-WRLES \citep{uzun2022high}. In our study, we rigorously examine the effects of SGS models, mesh resolutions, mesh anisotropy, and wall boundary conditions on the simulation results. By delving into these multifaceted sensitivities of WMLES, we hope to set the stage for advancing the practice and understanding of WMLES in simulating complex turbulent flows of real-world applications. 

The remainder of this paper is organized as follows. In Sec.~\ref{sec:2}, the numerical methodology and simulation set-up employed to assess the sensitivities of WMLES are introduced. Section~\ref{sec:3} is dedicated to demonstrating the results of the sensitivity analysis, illustrating the effects of the SGS model, mesh resolution, wall boundary condition, and mesh anisotropy. Finally, we conclude in Sec.~\ref{sec:4}, summarizing the key findings and their implications for future work.

\section{Numerical Approach}\label{sec:2}

\subsection{Flow solver}

Flow simulations in this study were executed using an unstructured-mesh,  {finite-volume}, LES code tailored for incompressible flow \cite{you2008discrete}. This cell-centered numerical scheme is energy-conserving and low-dissipative, providing an accurate capture of a broad spectrum of turbulence scales. It maintains second-order accuracy temporally and spatially. For momentum equations, a fully implicit, fractional-step method is utilized for time advancement. Meanwhile, pressure computations are facilitated using a Poisson solver based on the bi-conjugate gradient stabilized method \citep{van1992bi}. The effect of SGS motions is modeled using an SGS model.  {This study evaluates four SGS models: the Vreman model \citep{vreman2004eddy}, the dynamic Smagorinsky model (DSM) \citep{germano1991dynamic,lilly1992proposed}, the anisotropic minimum dissipation (AMD) model \citep{rozema2015minimum}, and the mixed-similarity model (MSM) \citep{bardina1983improved, sarghini1999scale}.}  {The first three models are based on the eddy-viscosity closure assumption, whereas the last one is a nonlinear model capable of accounting for the anisotropy of SGS stress \citep{meneveau2000scale}.} For the Vreman model, two model constant values, $c=0.025$ and $c=0.07$, are considered. The value of $c=0.07$ is derived theoretically for homogeneous isotropic turbulence, while $c=0.025$ is recommended for more complicated cases. For the AMD model, the model constant, which is determined solely by the choice of numerical method, is equal to 0.3.  {The MSM combines the classical Smagorinsky model \citep{smagorinsky1963general} with a scale-similarity term computed using explicit filtering \citep{meneveau2000scale}, and it is given by
\begin{equation}\label{MSMeq}
\tau_{ij} = -2(C_s\Delta)^2\lvert S\rvert\tilde{S}_{ij}+\left(\widehat{\tilde{u}_i\tilde{u}_j}-\hat{\tilde{u}}_i\hat{\tilde{u}}_j\right),
\end{equation} where the first term on the right-hand side corresponds to the Smagorinsky model, and the second term introduces the scale-similarity term. In the equation, $\tau_{ij}$ signifies the SGS stress, $u_i$ denotes the components of instantaneous velocity, $S_{ij}$ represents the strain-rate tensor, and $\lvert S\rvert=(2\tilde{S}_{ij}\tilde{S}_{ij})^{1/2}$. The operator $\widetilde{(\cdot)}$ signifies grid-filtered quantities, while $\widehat{(\cdot)}$ denotes a secondary filtering operation, chosen here as Gaussian filtering. Furthermore, in the Smagorinsky model, the eddy viscosity ($\nu_t$) is represented by $\nu_t=(C_s\Delta)^2|S|$, where $\Delta$ denotes the grid filter width, assumed to be the geometric mean of the local grid size. The Smagorinsky coefficient, $C_s$, is typically selected to lie between 0.1 and 0.2. For the MSM, we have set $C_s$ to a fixed value of 0.13, without undergoing any optimization. In the following discussion, the tilde symbol, which denotes the grid filtering operation, will be omitted for the sake of simplicity.} The reliability of this {LES} code in accurately simulating turbulent flows has been demonstrated in various configurations, such as rough-wall TBLs \cite{yang2013boundary}, flow over an axisymmetric body of revolution \cite{zhou2020large}, and rotor interactions with thick axisymmetric TBL \cite{zhou2022computational}. A {maximum} Courant–Friedrichs–Lewy (CFL) number of 1.0 is adopted for time advancement in all simulations. To mitigate potential numerical artifacts, each simulation is initially conducted for two flow-through times (FTTs). Subsequently, simulation data is collected over an additional three FTTs to ensure statistically converged results.

\subsection{Simulation Set-up}

This study focuses on the configuration of flow over a two-dimensional Gaussian-shaped bump, adopting the physical conditions identical to the hybrid DNS-WRLES conducted by \citet{uzun2022high} using a compressible flow solver. For brevity, the hybrid DNS-WRLES will henceforth be denoted as DNS. 

\subsubsection{Flow configuration}

\begin{figure}
\centering
\includegraphics[width=\textwidth,trim={0.1cm 0.1cm 0.1cm 0.1cm},clip]{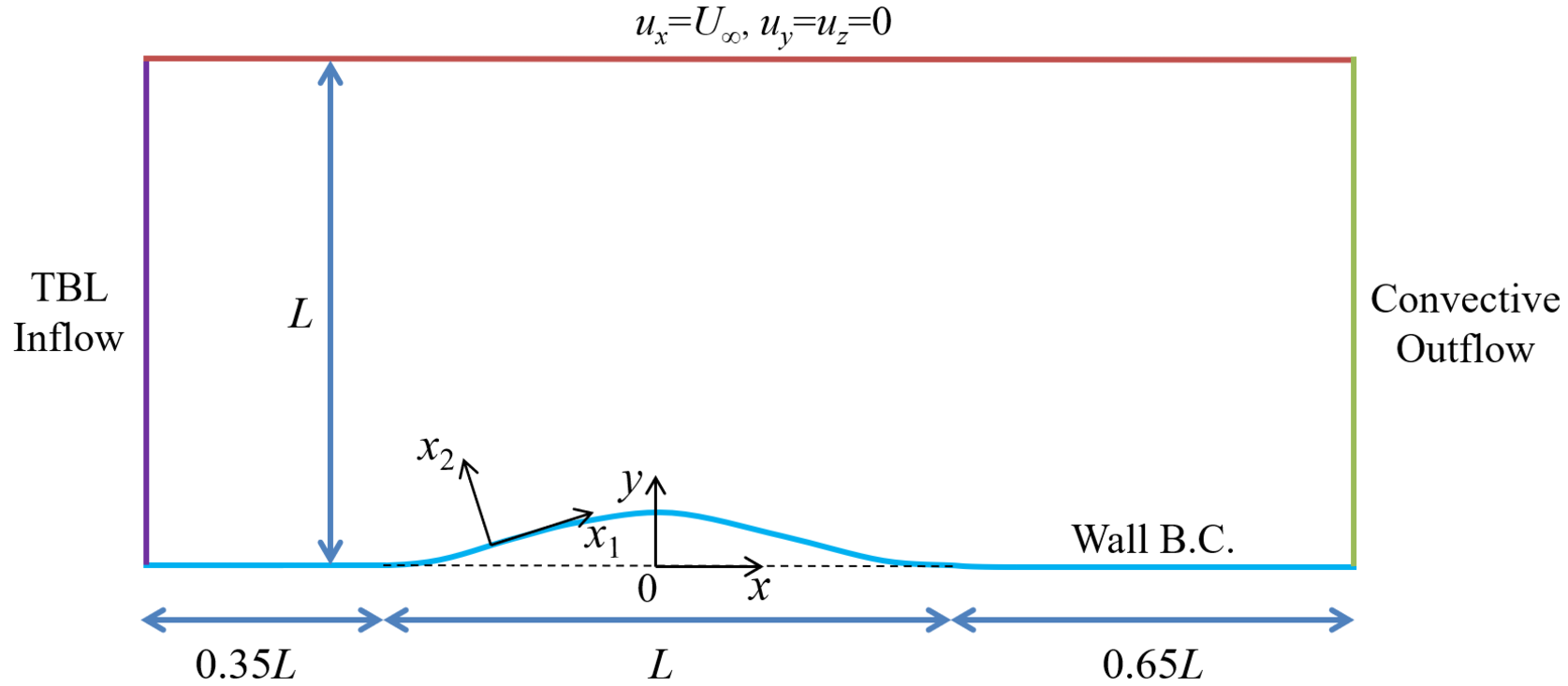}
\caption{Flow configuration and simulation set-up.}
\label{fig:flow_configuration}
\end{figure}

The flow configuration is shown schematically in Fig.~\ref{fig:flow_configuration}. The origin of the $x$-$y$-$z$ coordinate system in the domain is located at the base of the bump peak. In this absolute coordinate system, $x$, $y$, and $z$ denote the horizontal, vertical, and spanwise directions, respectively. For convenience, a localized coordinate system $x_1$-$x_2$-$x_3$ is also introduced at the bottom wall, where $x_1$, $x_2$, and $x_3$ represent the local streamwise, wall-normal, and spanwise directions, respectively. It should be noted that the streamwise direction is pointing towards the positive $x$-direction and the wall-normal direction is oriented towards the interior of the flow field. In the following sections, the components of instantaneous velocity associated with these two coordinate systems are denoted as $(u_x,u_y,u_z)$ and $(u_1,u_2,u_3)$, respectively. 

The geometry of the bump is given by the analytic function $y=f_b(x) = h\exp{[-\left(x/x_0\right)^2]}$, where $f_b$ is the surface representing the geometry of the wall-mounted bump. The length scale $L$, referred to as the bump width, is used to express the other scales of the bump, where $h=0.085L$ is the maximum height of the bump and $x_0=0.195L$. Additionally, the length scale $L$ is used to define the Reynolds number $Re_L = U_\infty L / \nu$, where $U_\infty$ denotes the free-stream velocity and $\nu$ is the kinematic viscosity. In this work, $Re_L$ is fixed to be $2\times10^6$, aligning with the value in the referenced DNS~\citep{uzun2022high}. The simulations are conducted in a cuboidal domain with a length of $2L$, height of $L$, and span of $0.08L$. The dimensions in the vertical and spanwise directions are chosen to be the same values as those in the DNS of \citet{uzun2022high}. 
The domain inlet and the exit are $0.85L$ upstream and $1.15L$ downstream from the bump peak, respectively.

\subsubsection{Boundary conditions}

The boundary conditions consist of a TBL inflow at the inlet, periodic conditions on the spanwise boundaries, convective outflow condition at the exit, and free-stream condition on the top boundary. In WMLES, the inadequately resolved near-wall region on coarse meshes is addressed using modified wall boundary conditions based on wall-stress models. Past investigations \citep{park2017wall, whitmore2020requirements, bae2021effect} indicate that both the wall-stress models and their implementation into the boundary conditions significantly influence the predictions of near-wall turbulence. Hence, for the current study on the sensitivity of WMLES, it is necessary to evaluate the impact of boundary conditions. In order to sidestep the intricacies of modeling wall-shear stress while pinpointing the effects of various factors, the physical no-slip condition at the bottom wall is replaced by three different forms of boundary condition based on the mean wall-shear stress from DNS. The first adopts the Neumann boundary condition given by the form 
\begin{equation}
\left.\left(\frac{\partial u_1}{\partial x_2}\right)\right|_w=\frac{\tau_{w,1}^{\text{DNS}}}{\mu} \hspace{2pt},
\label{Neumann}
\end{equation}
where $\tau_{w,1}^{\text{DNS}}$ is the mean wall-shear stress known \textit{a priori} from the DNS~\citep{uzun2022high}, $\mu$ is the dynamic viscosity, and the subscript $w$ denotes the quantities evaluated at the wall. The boundary condition is time-independent and {does not account for wall-shear stress fluctuations; therefore, it} can be treated as an idealized wall model supplying the exact mean wall-shear stress. Concurrently, a no-penetration condition is enforced for the wall-normal velocity $u_2$. 

The subsequent two boundary conditions are time-dependent and their basic idea is to augment the eddy viscosity at the wall while applying a Dirichlet boundary condition for the velocity components \citep{bae2021effect,zhou2022RLWM}. A previous study \citep{zhou2022RLWM} suggested that using this boundary condition formulation based on the effective wall-eddy viscosity $\nu_{t,w}$ could alleviate the underprediction of separation bubble caused by the directional inconsistency of streamwise velocity at the wall-adjacent control volume and the local wall-shear stress, potentially leading to a more accurate prediction of the velocity field for separated-flow LES. In the current study, $\nu_{t,w}$ is derived from the thin boundary layer equations \citep{bose2018wall, wang2002dynamic} given by
\begin{equation}
\frac{\partial u_i}{\partial t}+\frac{\partial u_i u_j}{\partial x_j}+\frac{1}{\rho}\frac{\partial p}{\partial x_i} = \frac{\partial}{\partial x_2} \left[ \left( \nu+\nu_t \right) \frac{\partial u_i}{\partial x_2} \right], \quad i=1, 3 \hspace{2pt},
\label{TBLE}
\end{equation}
where $\rho$ is the fluid density, $t$ is the time, and $p$ is pressure. The equations are required to satisfy no-slip conditions on the wall and match the velocity at the first off-wall mesh cell center: $u_i=u_{i,c}$ at $x_2=h_c$. The three terms on the left-hand side of Eq.~\ref{TBLE} are related to the effect of unsteadiness, convection, and pressure gradient, respectively. {By neglecting these three terms and integrating the equation from the first off-wall cell center down to the wall, the wall-shear stress components can be approximated by
\begin{equation}
\tau_{w,i} \approx \rho \left(\nu+\nu_{t,w}\right)\frac{u_{i,c}}{h_c}, \quad i=1,3 \hspace{2pt}.
\end{equation}
}Here, we assume the effective wall-eddy viscosity $\nu_{t,w}$ and the velocity $u_{i,c}$ are uncorrelated. Then for the streamwise component, by conducting a running average and incorporating the existing mean wall-shear stress from DNS \citep{uzun2022high}, the effective wall-eddy viscosity {of the second boundary condition} can be given by the form {
\begin{equation}
\overline{\nu_{t,w}} \approx  \left[\tau_{w,1}^{\text{DNS}} \Bigg/ \left(\rho \frac{\overline{u_{1,c}}}{h_c}\right)\right]-\nu \hspace{2pt}.
\label{EQ_nut}
\end{equation} 
}where $\overline{(\cdot)}$ denotes the running-averaged quantity. The running average adopts the approach used by \citet{lund1998generation}, which applies a weighted average that diminishes exponentially over time. Illustrating with the velocity $u_{1,c}$ as an example, the running average takes the subsequent form
\begin{equation}
\overline{u_{1,c}\left(t_{n+1}\right)} = \frac{\Delta t}{T}u_{1,c}\left(t_{n+1}\right)+\left(1-\frac{\Delta t}{T}\right)\overline{u_{1,c}\left(t_{n}\right)} \hspace{2pt},
\label{moving_average}
\end{equation}
where $n$ denotes the time step index, $\Delta t$ represents the computational time step, and $T$ is the characteristic time scale of the averaging interval. In the initial stages of the simulations, to eliminate transients, the averaging interval is set to a smaller value of $0.3L/U_{\infty}$. Once the flow reaches a fully developed state, this interval is increased to $6L/U_{\infty}$ to stabilize the statistics. {From the calculated effective wall-eddy viscosity at each time step, the corresponding local instantaneous wall-shear stress is determined with the formula
\begin{equation}
    \tau_{w,i}=\rho\nu (1+\frac{\overline{\nu_{t,w}}}{\nu})\left.\frac{\partial{u_i}}{\partial{x_2}}\right|_{w}, \quad i=1,3 \hspace{2pt}.
\label{tau_EQ}    
\end{equation} For non-equilibrium wall-bounded turbulent flows, the pressure gradient plays an important role in the flow dynamics. Therefore, in the third boundary condition, the influence of the pressure gradient is included, and the effective wall-eddy viscosity is then expressed by
\begin{equation}
\overline{\nu_{t,w}} \approx \left[\left( \tau_{w,1}^{\text{DNS}} + \frac{h_c}{2}\overline{\frac{\partial p}{\partial x_1}}\right) \Bigg/ \left(\rho \frac{\overline{u_{1,c}}}{h_c}\right)\right]-\nu \hspace{2pt}.
\label{NonEQ_nut}
\end{equation}
The local instantaneous wall-shear stress is determined by
\begin{equation}
    \tau_{w,i}=\rho\nu (1+\frac{\overline{\nu_{t,w}}}{\nu})\left.\frac{\partial{u_i}}{\partial{x_2}}\right|_{w}-\frac{h_c}{2}\frac{\partial p}{\partial x_i}, \quad i=1,3 \hspace{2pt}.
\label{tau_nonEQ}    
\end{equation} It should be noted that the effective wall-eddy viscosity in the current study is constrained to be greater than or equal to zero. This constraint primarily aims to prevent the generation of abnormally large wall-shear stresses around the separation and reattachment points. Given the definitions of these two boundary conditions, they can be regarded as idealized equilibrium and non-equilibrium wall models tailored to the present flow configuration, respectively. For brevity in subsequent discussions, the boundary condition for the bottom wall defined by Eq.~\ref{Neumann} will be termed the velocity Neumann boundary condition, that by Eqs.~\ref{EQ_nut} and \ref{tau_EQ} the equilibrium $\nu_{t,w}$ boundary condition, and by Eqs.~\ref{NonEQ_nut} and \ref{tau_nonEQ} the non-equilibrium $\nu_{t,w}$ boundary condition.}

\subsubsection{Computational meshes}

\begin{figure}
\centering
\includegraphics[width=0.77\textwidth,trim={0.0cm 0.1cm 0.2cm 0.1cm},clip]{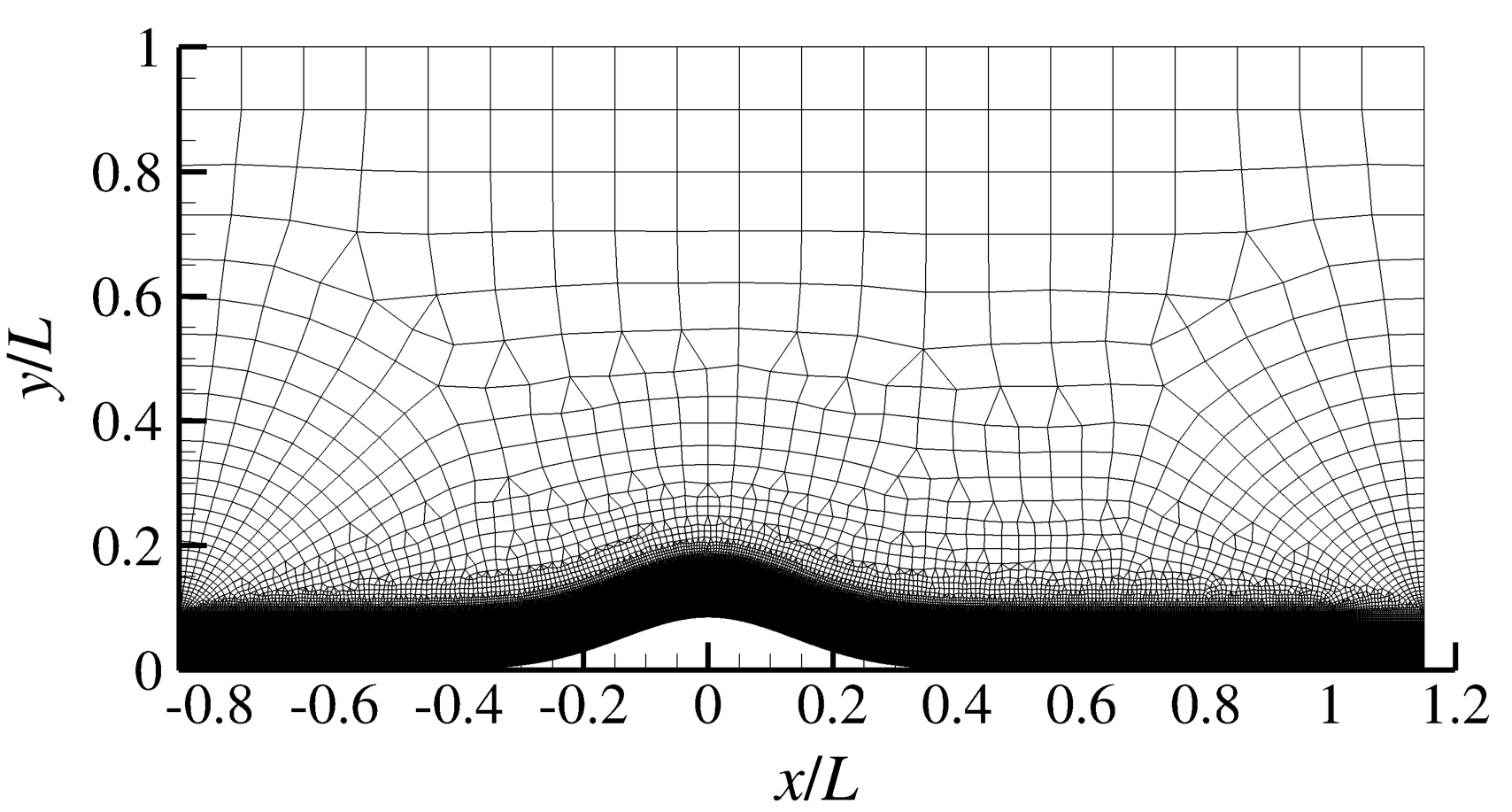}
\caption{Cross section of the baseline mesh in an $x$-$y$ plane.}
\label{fig:mesh}
\end{figure}

In the simulations, hybrid meshes are utilized, with a schematic presented in Fig.~\ref{fig:mesh}. A structured-mesh block is positioned around the near-wall region to cover the entire TBL within the domain, while the remaining regions are occupied by an unstructured-mesh block. To examine the effect of mesh resolution and anisotropy on the simulation results, we utilized both isotropic and anisotropic computational meshes with increasing resolutions within the structured-mesh block. The specifics of these computational meshes are detailed in Table~\ref{tab:table1}. Here, the characteristic size of the mesh cell is denoted as $\Delta_c=\sqrt[3]{V_c}$ where $V_c$ represents the cell volume. Referring to the TBL thickness at $x/L=-0.65$ from the DNS \citep{uzun2022high}, the TBL is resolved by approximately 5 cells in the coarsest mesh, 9 cells in the coarse mesh, 18 in the baseline, and 36 in the fine mesh. {Specifically, the resolution of the fine mesh, determined based on the characteristic size and the mean skin friction from the reference DNS \citep{uzun2022high}, ranges from 10 to 30 wall units within regions of attached flow. While this resolution is comparable to that of the standard WRLES mesh in the streamwise and spanwise directions, it is order of magnitude coarser in the wall-normal direction within the near-wall region.} For the DNS computational mesh, the characteristic cell size is approximately equal to $1.10 \times 10^{-4}L$. This estimation is derived from the mesh resolution at the location of the thickest separation bubble. Moreover, it should be noted that the cells in the anisotropic meshes are uniform. These cells maintain the same spanwise resolution and characteristic length as their isotropic counterparts at equivalent resolutions. The aspect ratio of all anisotropic mesh cells is equal to $\Delta_1:\Delta_2:\Delta_3=4:1:2$, where $\Delta_1$, $\Delta_2$, and $\Delta_3$ denote the mesh-cell size in streamwise, wall-normal and spanwise directions, respectively. For the outer unstructured-mesh blocks, their mesh cell size is smaller than $0.1L$, and the control volumes are refined gradually towards the bottom wall. 

\begin{sidewaystable}
\caption{Parameters of the structured-mesh block in the computational meshes. $\Delta_c$ denotes the characteristic size of the mesh cell.}\label{tab:table1}
\centering
{\small \begin{tabular*}{0.986\linewidth}{c|cc|cc}
\hline
 \rule{0pt}{2.5ex}
\multirow{2}{*}{Mesh label} & \multicolumn{2}{c|}{Isotropic mesh} & \multicolumn{2}{c}{Anisotropic mesh} \\
  & $\Delta_c/L$ & Cell number & $\Delta_c/L$ & Cell number\\ \hline
 \rule{0pt}{2.5ex}
 Coarsest mesh & $1.90\times 10^{-3}$ & $1050 \times 44 \times 42 \approx 1.94$~million & $1.90\times 10^{-3}$ & $525 \times 88 \times 42 \approx 1.94$~million  \\
 Coarse mesh & $9.52\times 10^{-4}$ & $2100 \times 88 \times 84 \approx 15.5$~million & $9.52\times 10^{-4}$ & $1050 \times 176 \times 84 \approx 15.5$~million\\
 Baseline mesh&$4.76\times 10^{-4}$ & $4200 \times 176 \times 168 \approx 124$~million & $4.76\times 10^{-4}$ & $2100 \times 352 \times 168 \approx 124$~million\\
 Fine mesh &$2.38\times 10^{-4}$ & $ 8400 \times 352 \times 336 \approx 993$~million &  & \\ \hline
\end{tabular*}}
\end{sidewaystable}

\subsection{Inflow Generation}

The TBL inflow data for the simulations of flow over Gaussian bump are provided by a separate LES of flat-plate TBL using the rescale-and-recycle method of \citet{lund1998generation}. The size of the computational domain is $0.09L\times 0.15L \times 0.08L$ in the streamwise, wall-normal, and spanwise directions, respectively. A structured mesh is employed, wherein the portion covering the TBL is isotropic, featuring a resolution of $\Delta_0/L=2.19\times10^{-4}$. The outer mesh is gradually coarsened away from the wall. The equilibrium stress-balance wall model \citep{bose2018wall, wang2002dynamic} is applied on the wall and the periodic boundary condition is used in the spanwise direction. The Dirichlet boundary condition $(u_x,u_y,u_z)=(U_\infty, 0, 0)$ is applied at the top of the domain. At the exit, a convective outflow boundary condition is adopted. In the simulation, the DSM \citep{germano1991dynamic,lilly1992proposed} is used, and a {maximum} CFL number of 1.0 is employed. The simulation is run for 100 FTTs ($9L/U_\infty$) at first to pass the transient process. After that, the simulation is run for another 150 FTTs to collect the TBL inflow data.

\begin{figure}
\centering
{\psfrag{a}[][]{{$y^+$}}
\psfrag{b}[][]{{$\left<u_x^+\right>$}}\includegraphics[width=.75\textwidth,trim={0.2cm 5cm 0.2cm 1.7cm},clip]{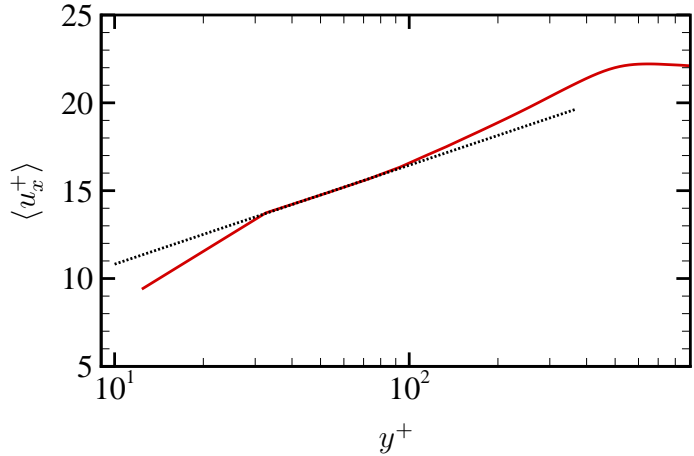}}
\caption{Inner-scaled mean streamwise velocity profile of the collected TBL inflow. Dotted line represents the classical log law of the wall $\left<u_x^+\right>=(1/0.41)\log(y^+)+5.2$.}
\label{inflow_mean}
\end{figure}

The inflow data is collected at $0.025L$ upstream of the domain's exit. The friction Reynolds number $Re_\tau$ at this location is approximately equal to 620 and the TBL thickness is $\delta_{in}/L=0.0061$, which is approximately 10\% larger than that in the DNS of \citet{uzun2022high}. {The momentum thickness Reynolds number of our inflow is $Re_{\theta} \approx 1074$, compared to their slightly smaller value of approximately 1035.} It should be noted that the inflow-generation method used in the study of \citet{uzun2022high} is different from the current method. Figure~\ref{inflow_mean} shows the profile of inner-scaled mean streamwise velocity $\left<u_x^+\right>$ of the collected TBL, with $+$ denoting the quantity in wall unit and $\left<\cdot\right>$ representing the mean quantity derived from both time- and spanwise-averaging. The velocity profile aligns closely with the classical log law of the wall, underscoring the reliability of the captured inflow data. Furthermore, it should be noted that to be compatible with the inlet in the main simulation, inflow data beyond $y/L=0.15$ is adjusted to match the free-stream conditions, characterized by $(u_x,u_y,u_z)=(U_\infty, 0, 0)$.

\section{Results and Discussion}\label{sec:3}

To evaluate the impact of various factors independently, the simulations are divided into three distinct groups, the parameters of which are detailed in Table~\ref{tab:table2}. Specifically, Group 1 concentrates on the effects of the SGS model and mesh resolution, while Group 2 examines the influence of wall boundary conditions, and Group 3 delves into the effect of mesh anisotropy. For brevity, and aside from comparisons among the results from simulations conducted at different mesh resolutions, the subsequent discussions will focus on the baseline-mesh simulation results. Additional demonstrations will be included as needed to clarify specific points. Moreover, the discussion of the velocity field will concentrate on the mean velocity, representing the first-order statistics and standing as the most vital quantity that WMLES needs to predict.

\begin{sidewaystable}
\caption{Parameters of the simulations to evaluate the sensitivities of WMLES with respect to SGS models as well as mesh resolution (Group 1), boundary condition (B.C.) at the wall (Group 2), and mesh anisotropy (Group 3).}\label{tab:table2}
\centering
{\small \begin{tabular*}{0.965\linewidth}{cccc}
\hline
 \rule{0pt}{2.5ex}
 Case group & Mesh & SGS model & wall B.C. \rule{0pt}{2.5ex} \\ 
 \hline
 Group 1 & \begin{tabular}{@{}c@{}} Coarsest mesh (isotropic) \\ Coarse mesh (isotropic) \\ Baseline mesh (isotropic) \\ Fine mesh (isotropic) \end{tabular} & \begin{tabular}{@{}c@{}} \rule{0pt}{2.2ex} None \\ DSM \\ AMD \\ Vreman ($c=0.025$) \\ Vreman ($c=0.07$) \\ {MSM} \end{tabular} & Velocity Neumann boundary condition  \\ \hline
 Group 2 & \begin{tabular}{@{}c@{}} Coarsest mesh (isotropic) \\ Coarse mesh (isotropic) \\ Baseline mesh (isotropic) \end{tabular} & \begin{tabular}{@{}c@{}} \rule{0pt}{2.2ex} None \\ DSM \\ AMD \\ Vreman ($c=0.025$) \\ Vreman ($c=0.07$) \\ {MSM} \end{tabular} & \begin{tabular}{@{}c@{}} Equilibrium $\nu_{t,w}$ boundary condition \\ Non-equilibrium $\nu_{t,w}$ boundary condition \end{tabular} \\ \hline
 Group 3 & \begin{tabular}{@{}c@{}} Coarsest mesh (anisotropic) \\ Coarse mesh (anisotropic) \\ Baseline mesh (anisotropic) \end{tabular} & \begin{tabular}{@{}c@{}} \rule{0pt}{2.2ex} None \\ DSM \\ AMD \\ Vreman ($c=0.025$) \\ Vreman ($c=0.07$) \\ {MSM} \end{tabular} & Velocity Neumann boundary condition \\ \hline
\end{tabular*}}
\end{sidewaystable}

\subsection{Effects of SGS model and mesh resolution}

\begin{figure}
\centering
\includegraphics[width=.7\textwidth,trim={0.0cm 0.0cm 0.0cm 0.0cm},clip]{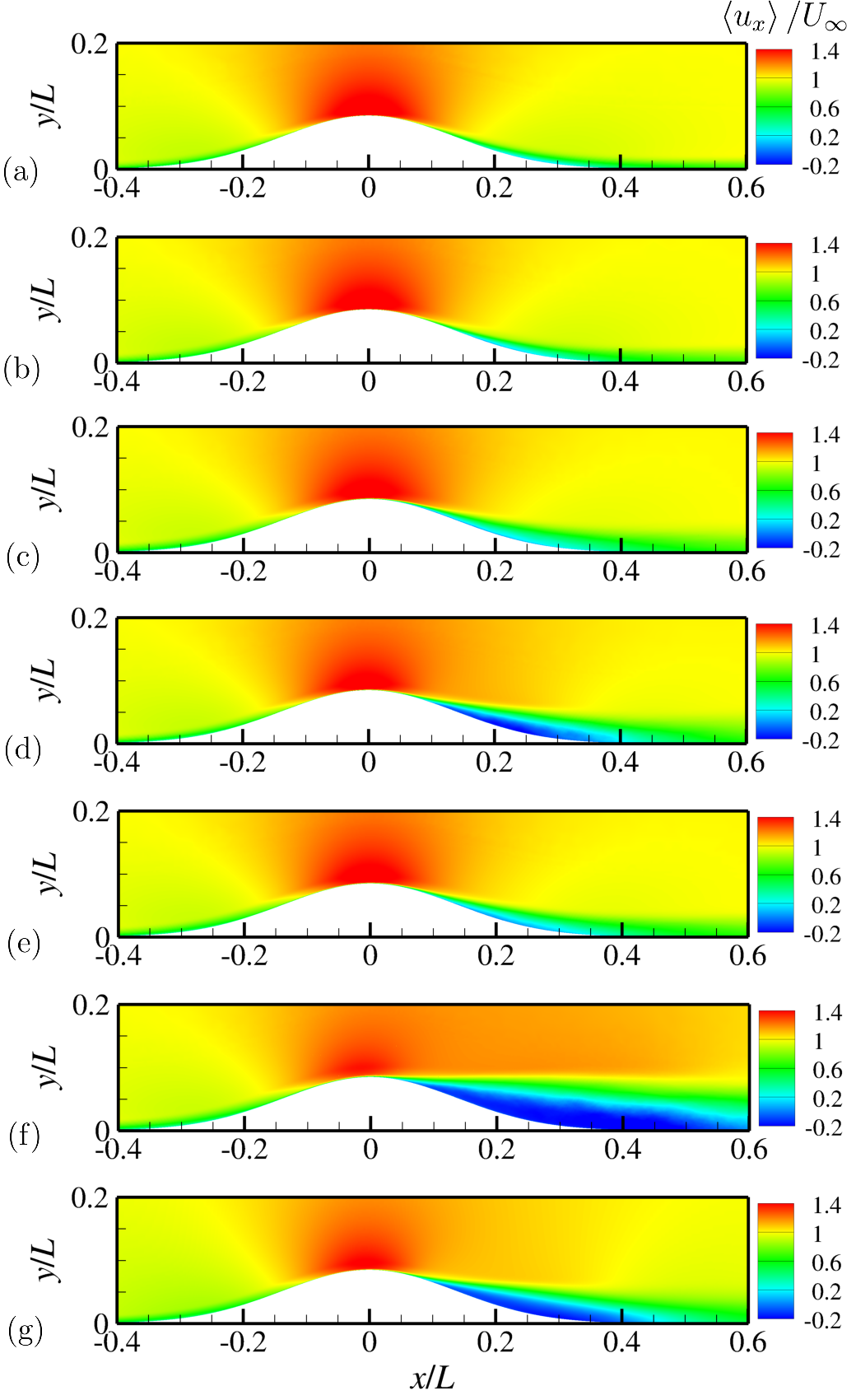}
\caption{{Contours of mean velocity in the $x$ direction, $\left<u_x\right>/U_\infty$, in an $x$-$y$ plane from the baseline-mesh simulations using (a) no SGS model, (b) the DSM, (c) the AMD model, (d) the Vreman model with $c=0.025$, (e) the Vreman model with $c=0.07$, and (f) the MSM that listed in the Group 1 of Table~\ref{tab:table2} and from (g) the DNS \citep{uzun2022high}.}}
\label{fig:ux_ave_xy_SGS}
\end{figure}

First of all, we investigate the effects of SGS model and mesh resolution, focusing on the results of Group 1 simulations. Figure~\ref{fig:ux_ave_xy_SGS} displays contours of the mean velocity in the $x$ direction, $\left< u_x \right>/U_\infty$, in an $x$\nobreakdash-$y$ plane. The results obtained from the baseline-mesh simulations using various SGS models are compared with the referenced DNS results \citep{uzun2022high}. The flow behavior can be characterized by gradual acceleration on the windward side of the bump, reaching its maximum velocity near the bump peak. Downstream from this point, the flow decelerates over the leeward side of the bump, leading to a rapid thickening of the boundary layer. {It is noteworthy that when using the baseline mesh, flow separation on the leeward side of the bump occurs solely in simulations employing the AMD model, the Vreman model ($c=0.025$), and the MSM.} In particular, the simulation utilizing the Vreman model with $c=0.025$ predicts a substantial separation bubble, similar to the DNS results \citep{uzun2022high}. {Meanwhile, the simulation that employs the MSM shows a separation bubble significantly larger than that observed in the DNS.} Contrarily, other simulations predict attached flow over the entire bump surface, underscoring the pronounced influence of SGS models on the prediction of separated turbulent flow.

\begin{figure}
\centering
\includegraphics[width=.7\textwidth,trim={0.0cm 0.3cm 0.0cm 0.0cm},clip]{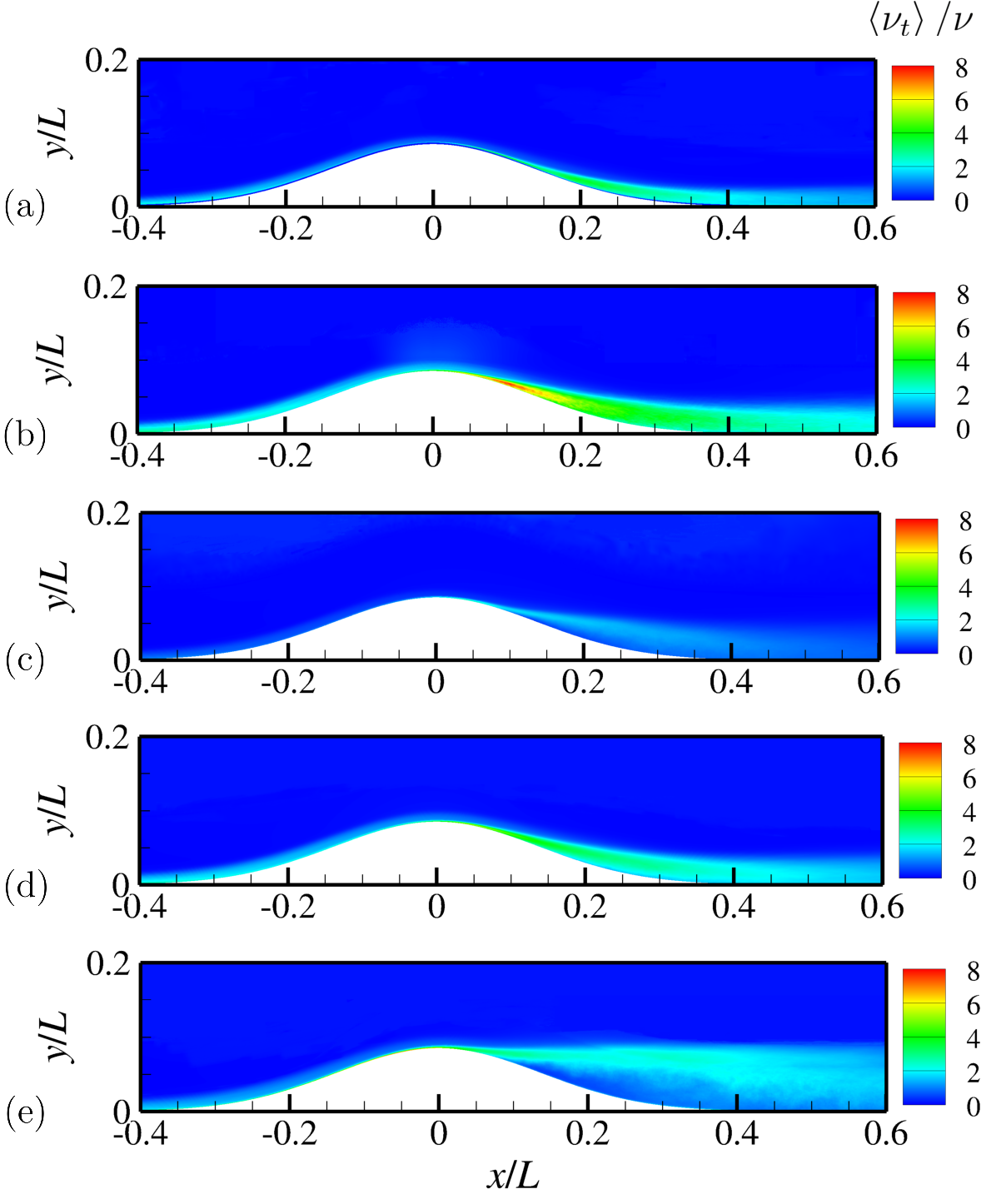}
\caption{{Contours of mean eddy viscosity, $\left< \nu_t \right>/\nu$, from the SGS model in an $x$-$y$ plane from the baseline-mesh simulations of Group 1 using (a) the DSM, (b) the AMD model, (c) the Vreman model with $c=0.025$, (d) the Vreman model with $c=0.07$, and (e) the MSM.}}
\label{fig:nut_ave_xy_SGS}
\end{figure}

Figure~\ref{fig:nut_ave_xy_SGS} presents contours of the mean eddy viscosity, represented by $\left< \nu_t \right>/\nu$, from the SGS model in an $x$\nobreakdash-$y$ plane. {For the MSM, the eddy viscosity is incorporated solely through the term of the Smagorinsky model, as given in Eq.~\ref{MSMeq}.} The magnitude of the eddy viscosity within the TBL is found to be comparable to the fluid viscosity, signifying the importance of SGS stresses in the current simulations. As the flow evolves on the leeward side of the bump, the eddy viscosity within the TBL increases. Moreover, there are noticeable differences in the eddy viscosity magnitudes among the different SGS models. {Specifically, the AMD model produces the highest eddy viscosity within the TBL, while the simulation using the Vreman model with $c=0.025$ results in the smallest eddy viscosity within the TBL. In addition, the MSM demonstrates increased eddy viscosity in the shear layer region downstream of the bump peak.} 

\begin{figure}
\centering
{\psfrag{a}[][]{{$x/L$}}
\psfrag{b}[][]{{$C_p$}}\includegraphics[width=.75\textwidth,trim={0.2cm 6.2cm 0.2cm 1.0cm},clip]{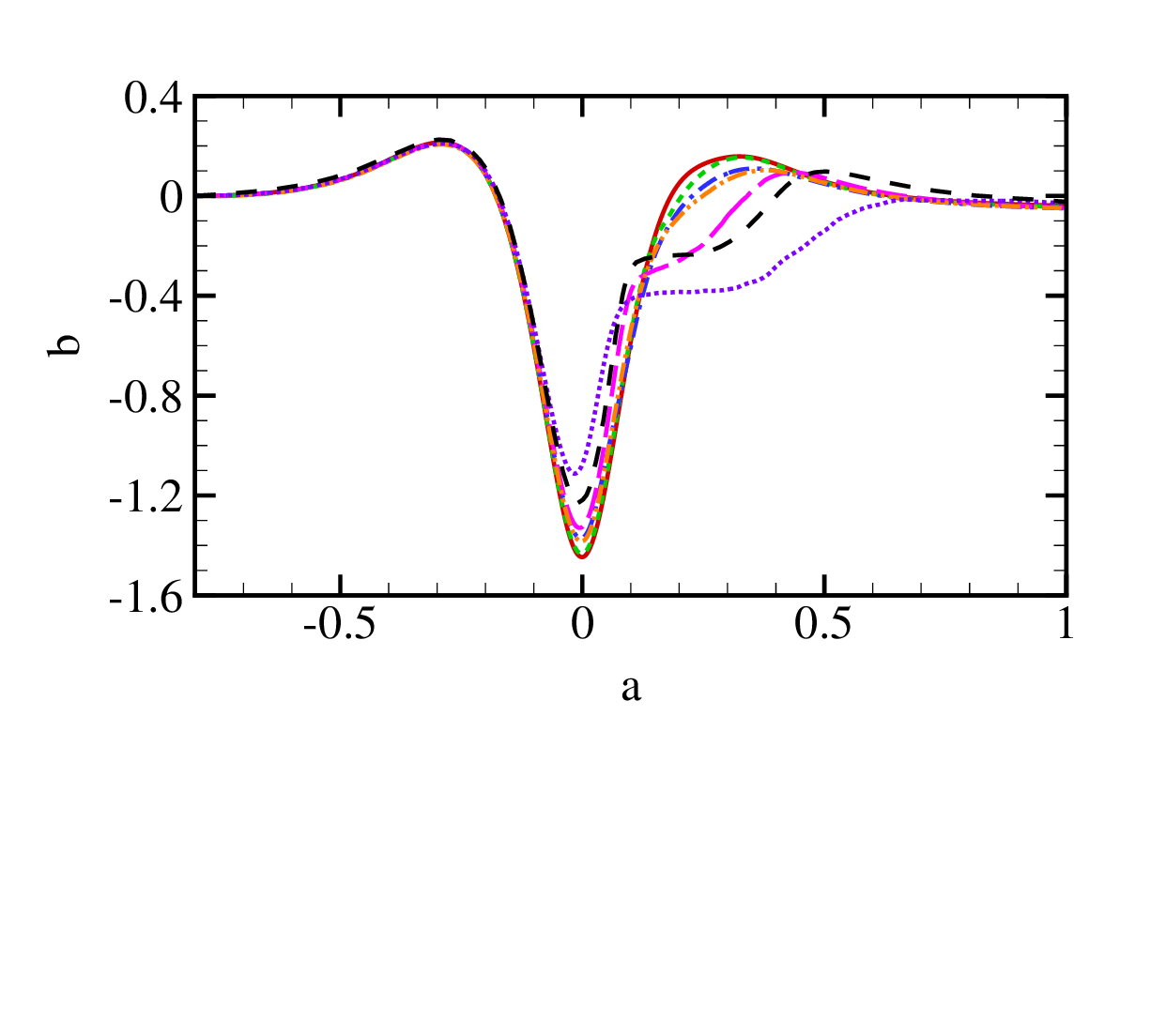}}
\caption{{Comparison of the mean pressure coefficient $C_p$ on the bottom wall from the Group 1 baseline-mesh simulations using different SGS models with the DNS results. Lines indicate {\redsolid}, WMLES without an SGS model; {\greendashed}, WMLES with the DSM; {\bluedashdotted}, WMLES with the AMD model; {\magentadashed}, WMLES with the Vreman model ($c=0.025$); {\orangedashdotdot}, WMLES with the Vreman model ($c=0.07$); {\pupdotdotted}, WMLES with the MSM; {\blackdashed}, DNS~\citep{uzun2022high}.}}
\label{Cp_SGS}
\end{figure}

Figure~\ref{Cp_SGS} compares the distributions of the mean pressure coefficients $C_p=\left<p_w-p_\infty\right>/(0.5\rho U_\infty^2)$ on the bottom wall. Here, $p_w$ represents the local static pressure on the wall, and the reference pressure $p_\infty$ is taken at the inlet near the top boundary. The $C_p$ distribution exhibits a strong favorable pressure gradient (FPG) in front of the bump peak. Past the peak, the flow encounters a strong adverse pressure gradient (APG), followed by a gentler FPG over the downstream flat wall. Furthermore, the results from different simulations are closely aligned upstream of the bump peak and within the region of the downstream flat wall. However, distinctions become marked around the peak and on the leeward side of the bump. {Particularly, the simulation using the Vreman model with $c=0.025$ exhibits a better agreement with the DNS results \citep{uzun2022high}.} 

\begin{figure}
\centering
{\psfrag{a}[][]{{$x/L$}}
\psfrag{b}[][]{{$\left<u_1 \right>/U_{\infty}$}}\includegraphics[width=.75\textwidth,trim={0.4cm 6.6cm 0.2cm 0.2cm},clip]{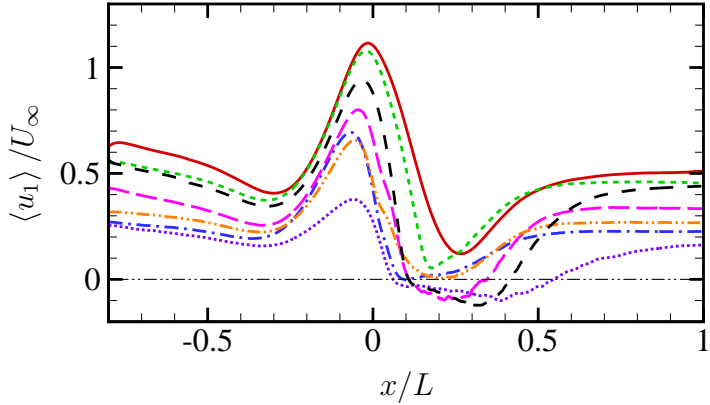}}
\caption{{The profiles of mean streamwise velocity at $x_2/L=2.38\times 10^{-4}$ (the wall-normal location of the first off-wall cell center) from the baseline-mesh simulations of Group 1, alongside a reference profile from the DNS \citep{uzun2022high} at the same wall-normal location. Lines indicate {\redsolid}, WMLES without an SGS model; {\greendashed}, WMLES with the DSM; {\bluedashdotted}, WMLES with the AMD model; {\magentadashed}, WMLES with the Vreman model ($c=0.025$); {\orangedashdotdot}, WMLES with the Vreman model ($c=0.07$); {\pupdotdotted}, WMLES with the MSM; {\blackdashed}, DNS~\citep{uzun2022high}; {\blackdashdotdot}, $\left<u_1 \right>/U_{\infty}=0$.}}
\label{mean_vel_1st_SGS}
\end{figure}

{Since every simulation employs a velocity Neumann boundary condition at the bottom wall, which consistently applies a wall-shear stress matching the local mean wall-shear stress from the reference DNS \citep{uzun2022high}, the conventional method that relies on skin friction measurements to determine the size of separation bubble cannot be utilized.} To circumvent this limitation, we estimate the length of the predicted separation bubble using the mean streamwise velocity at the first off-wall cell center, as depicted in Fig.~\ref{mean_vel_1st_SGS}. A close examination of the distributions of the mean streamwise velocity reveals significant differences in the near-wall velocity among these baseline-mesh simulations. It can be found that the separation bubble is tiny in the simulation with the AMD model, and the flow nearly separates in the simulation with the Vreman model ($c=0.07$). Nevertheless, these differences are not confined to the leeward side of the bump but span the entire domain length. Compared to other simulations, the one employing the DSM aligns more closely with the DNS \citep{uzun2022high} in the region with small pressure-gradient effects. {In contrast, the simulation utilizing the MSM exhibits a lower magnitude of streamwise velocity, suggesting that significant dissipation occurs in the near-wall region.} Moreover, it is essential to recognize that the near-wall field of the TBL is crucial to WMLES, as the wall model relies on the data within this region to make predictions of wall quantities. The current findings imply that the performance of wall models can be sensitive to the choice of the SGS model.

\begin{figure}
\centering
{\psfrag{a}[][]{{$\Delta_c/L\times10^3$}}
\psfrag{b}[][]{{$L_s/L$}}\includegraphics[width=.75\textwidth,trim={0.2cm 7.7cm 0.2cm 0.2cm},clip]{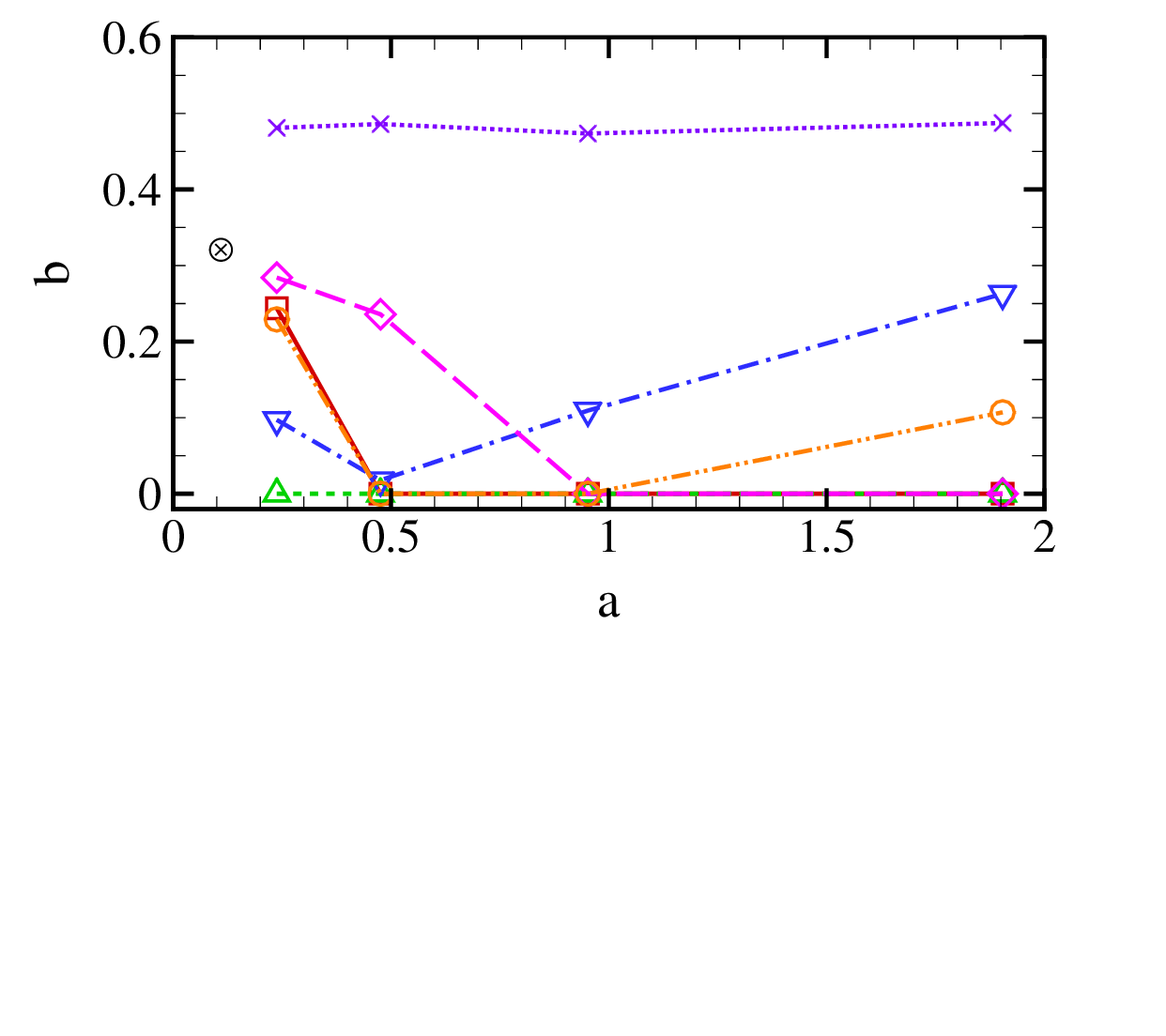}}
\vspace{0.1cm}
\caption{{The horizontal length of separation bubble ($L_s/L$) as a function of mesh-cell characteristic length ($\Delta_c/L$) from the simulations of Group 1. Lines indicate {\redsolid} with \protect\marksymbol{square}{red}, WMLES without an SGS model; {\greendashed} with \protect\marksymbol{triangle}{green}, WMLES using the DSM; {\bluedashdotted} with \protect\marksymbol{triangle}{blue,style={rotate=180}}, WMLES using the AMD model; {\magentadashed} with \protect\marksymbol{diamond}{magenta}, WMLES using the Vreman model ($c=0.025$); {\orangedashdotdot} with \protect\marksymbol{o}{orange}, WMLES using the Vreman model ($c=0.07$); {\pupdotdotted} with \textcolor{purp2}{$\times$}, WMLES using the MSM. The symbol \textcolor{black}{$\otimes$} represents the DNS results where $\Delta_c/L\times 10^3=0.11$ and $L_s/L=0.32$.}}
\label{Bump_size_SGS}
\end{figure}

Figure~\ref{Bump_size_SGS} illustrates the horizontal length ($L_s/L$) of the predicted separation bubble as a function of the characteristic mesh resolution ($\Delta_c/L$). In the DNS conducted by \citet{uzun2022high}, the separation bubble's horizontal length measures $0.32L$, and the approximate characteristic mesh resolution is $1.10\times10^{-4}$. The DNS results are included in the figure for reference. The results indicate that the fine-mesh simulations approach grid convergence, especially in the cases using the AMD model and the Vreman model. Nevertheless, the variations in separation bubble length with respect to mesh resolution present a complex pattern. In the simulations employing the Vreman model with $c=0.025$, convergence toward the DNS results appears to be monotonic. Conversely, for the simulations involving the AMD model and the Vreman model with $c=0.07$, convergence is non-monotonic, leading to a spurious diminishment or shrinking of the separation bubble upon mesh refinement. This non-monotonic convergence of WMLES toward DNS or experimental results aligns with observations from previous studies \citep{whitmore2021large,agrawal2022non} and reinforces the idea that the non-monotonic behavior of WMLES solutions is connected to the particular behavior of the SGS model. {Although the simulation using the MSM predicts a larger separation bubble compared to others, its results remain consistent across varying mesh resolutions. This consistency is a promising feature, suggesting that the inclusion of SGS stress anisotropy could have a positive influence \citep{inagaki2023analysis}. Given that the current MSM has not undergone any optimization and employs a fixed Smagorinsky coefficient, it is reasonable to anticipate that the MSM's performance could be further enhanced through either a dynamically calculated model coefficient or other optimization techniques.} In addition, the performance of WMLES without an SGS model notably improves with the use of the fine mesh. This improvement can be anticipated since the contribution from the SGS model decreases as the mesh is refined. The resolution of the current fine mesh approaches that of WRLES, especially in the region of the separation bubble. In this area, the boundary layer thickness is covered by over 150 cells, further underscoring the reduced dependence on the SGS model.

\begin{figure}
\centering
\begin{subfigure}{0.325\textwidth}
\subcaption{~~~~~~~~~~~~~~~~~~~~~~~~~~~~~~~~~~~~~~~~~}
\vspace{-1pt}
{\psfrag{b}[][]{{$x_2/L$}}
\psfrag{a}[][]{{$\left<u_1 \right>/U_{\infty}$}}
\includegraphics[width=\textwidth,trim={2.2 1.8cm 8.2cm 0.4cm},clip]{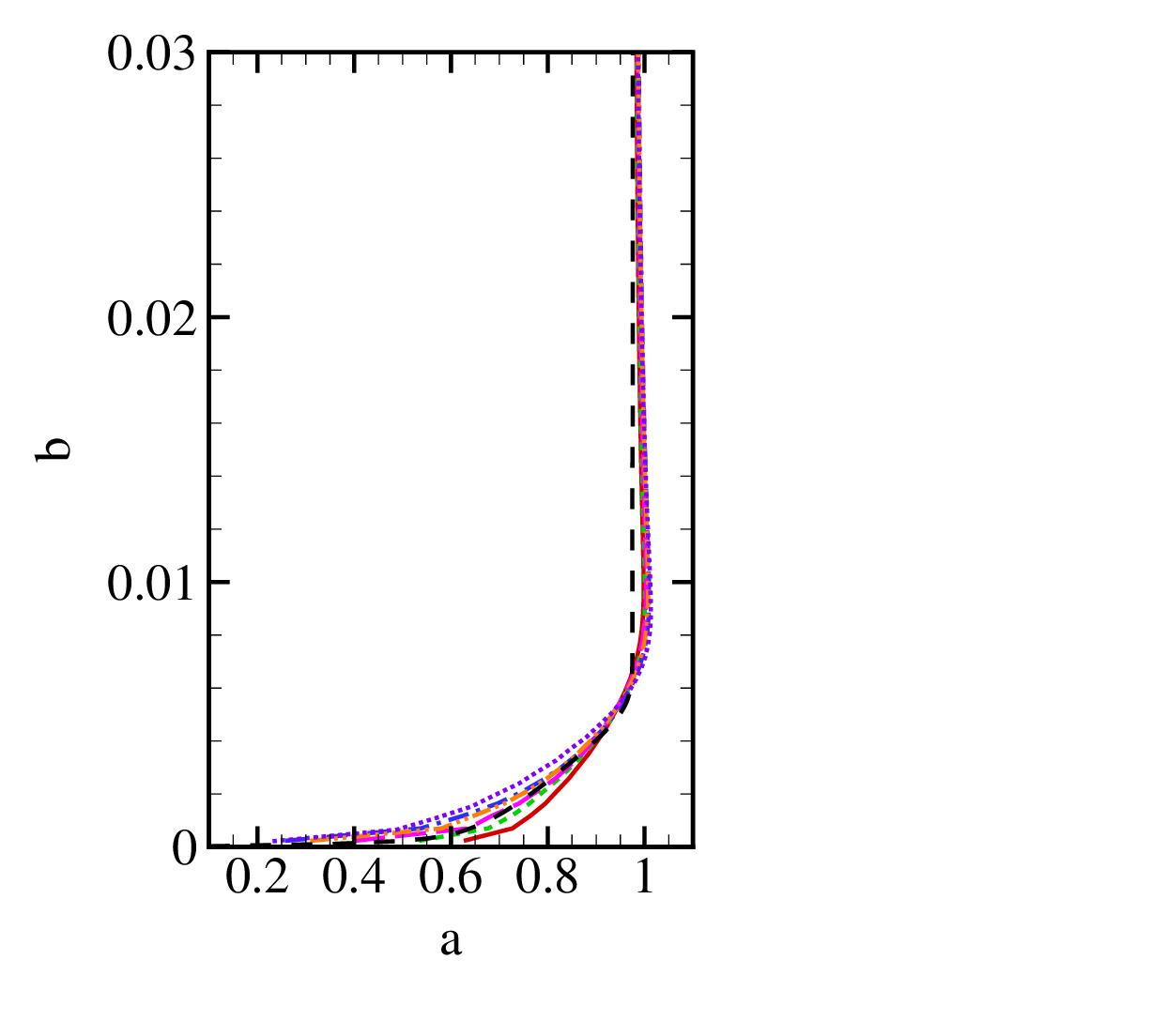}
}
\end{subfigure}
\begin{subfigure}{0.325\textwidth}
\subcaption{~~~~~~~~~~~~~~~~~~~~~~~~~~~~~~~~~~~~~~~~~}
\vspace{-1pt}
{\psfrag{b}[][]{{$x_2/L$}}
\psfrag{a}[][]{{$\left<u_1 \right>/U_{\infty}$}}
\includegraphics[width=\textwidth,trim={2.2 1.8cm 8.2cm 0.4cm},clip]{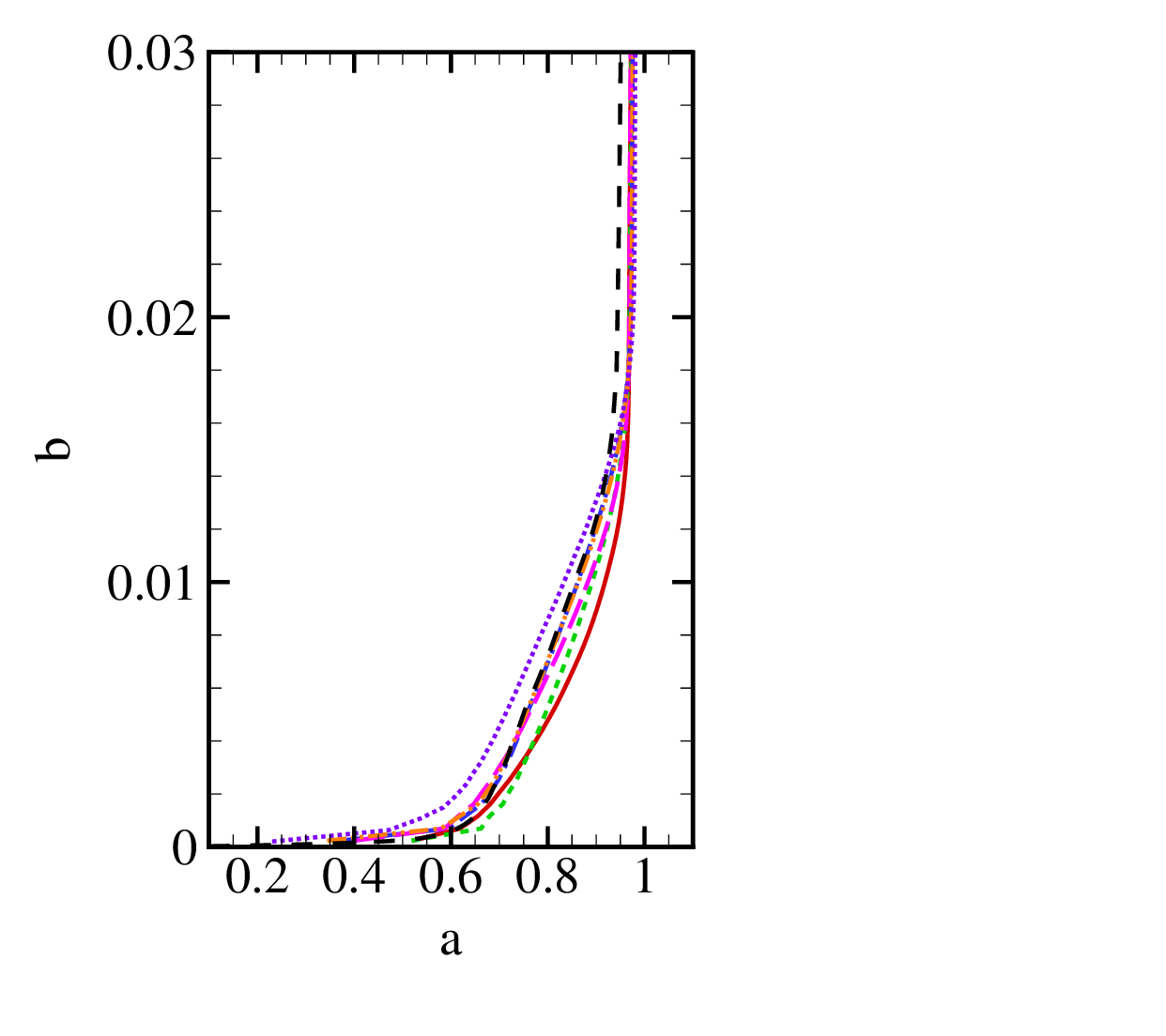}
}
\end{subfigure}
\begin{subfigure}{0.325\textwidth}
\subcaption{~~~~~~~~~~~~~~~~~~~~~~~~~~~~~~~~~~~~~~~~~}
\vspace{-1pt}
{\psfrag{b}[][]{{$x_2/L$}}
\psfrag{a}[][]{{$\left<u_1 \right>/U_{\infty}$}}
\includegraphics[width=\textwidth,trim={2.2 1.8cm 8.2cm 0.4cm},clip]{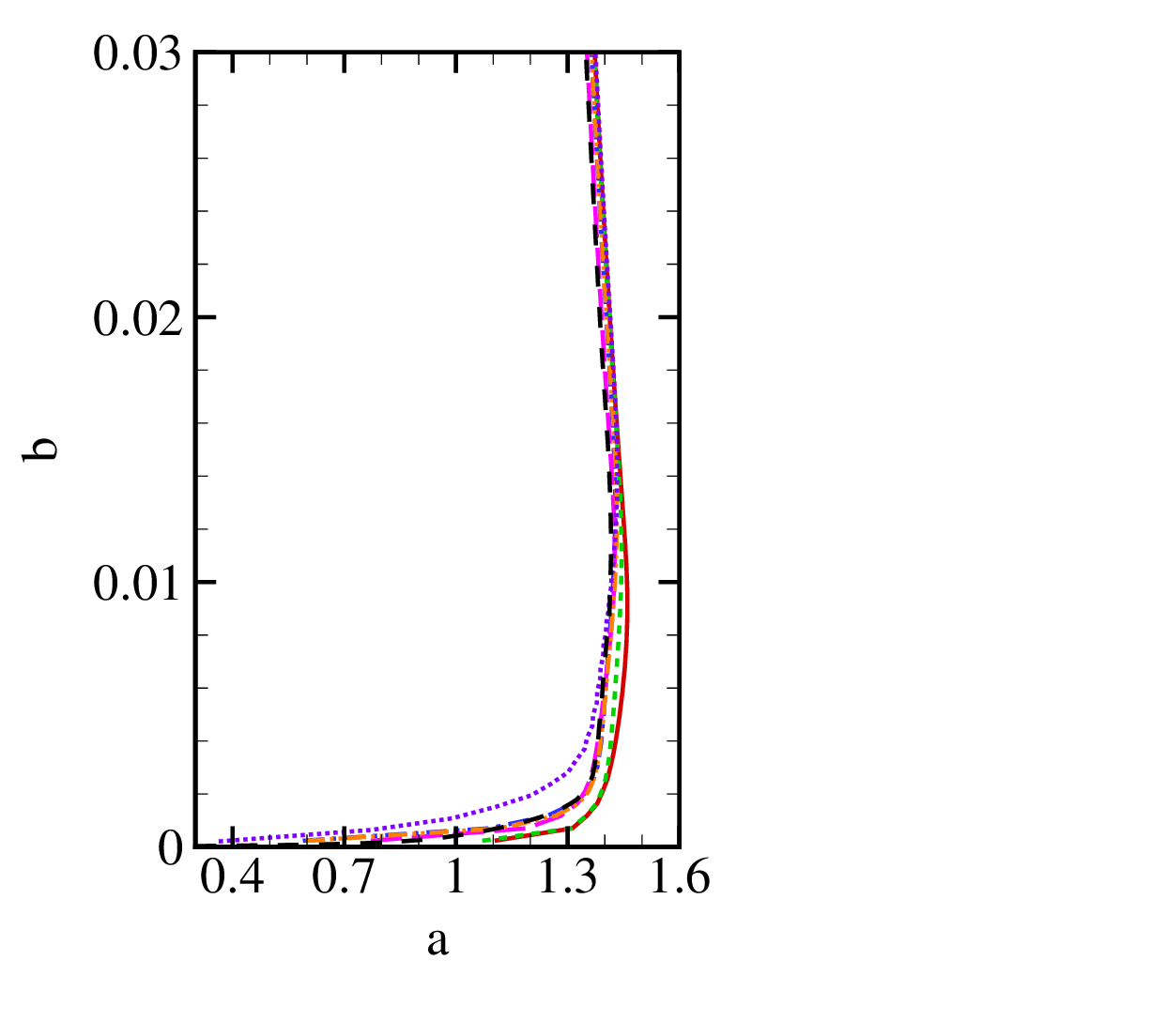}
}
\end{subfigure}\\[-10pt]
\begin{subfigure}{0.325\textwidth}
\subcaption{~~~~~~~~~~~~~~~~~~~~~~~~~~~~~~~~~~~~~~~~~}
\vspace{-1pt}
{\psfrag{b}[][]{{$x_2/L$}}
\psfrag{a}[][]{{$\left<u_1 \right>/U_{\infty}$}}
\includegraphics[width=\textwidth,trim={2.2 1.8cm 8.2cm 0.4cm},clip]{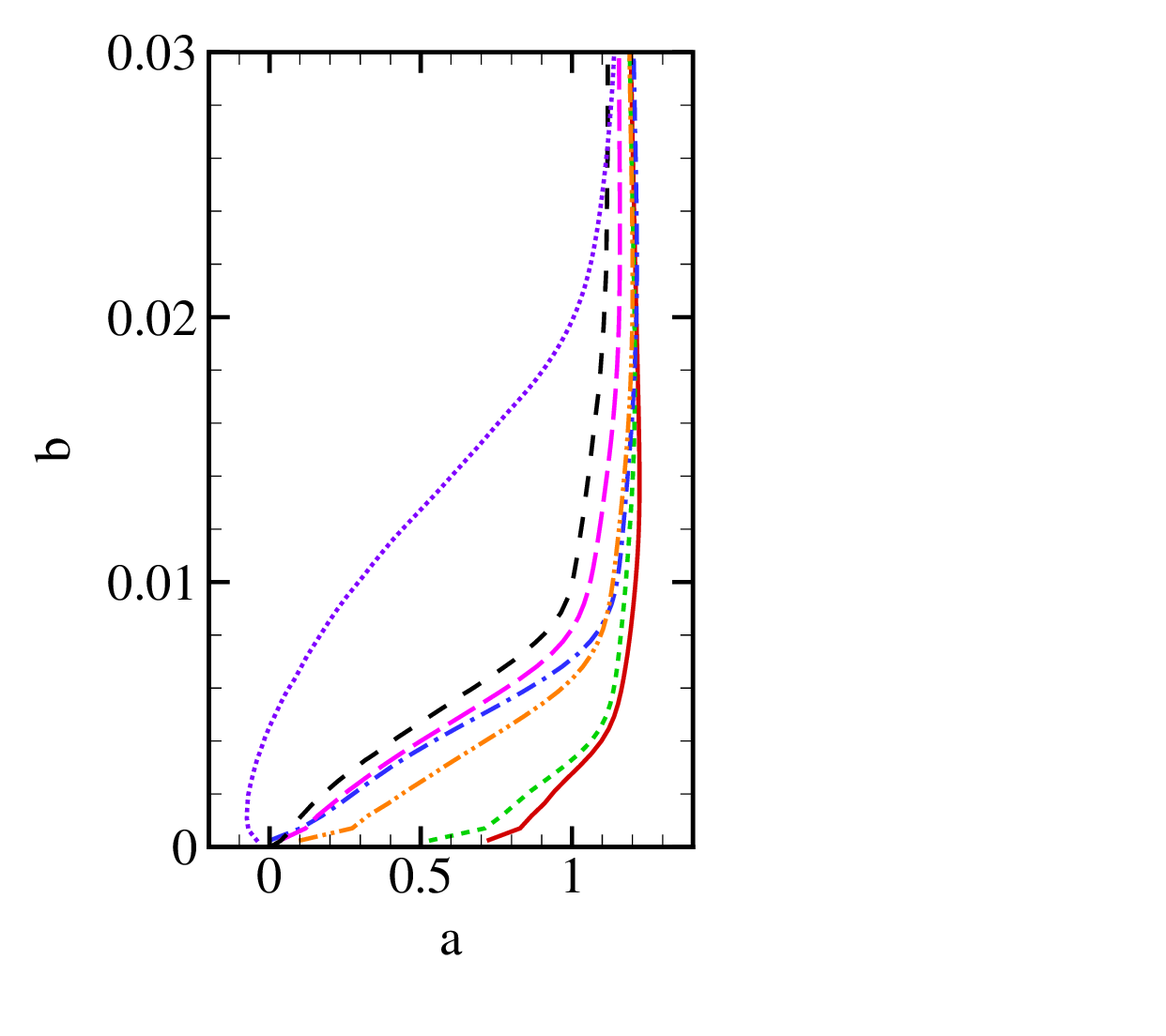}
}
\end{subfigure}
\begin{subfigure}{0.325\textwidth}
\subcaption{~~~~~~~~~~~~~~~~~~~~~~~~~~~~~~~~~~~~~~~~~}
\vspace{-1pt}
{\psfrag{b}[][]{{$x_2/L$}}
\psfrag{a}[][]{{$\left<u_1 \right>/U_{\infty}$}}
\includegraphics[width=\textwidth,trim={2.2 1.8cm 8.2cm 0.4cm},clip]{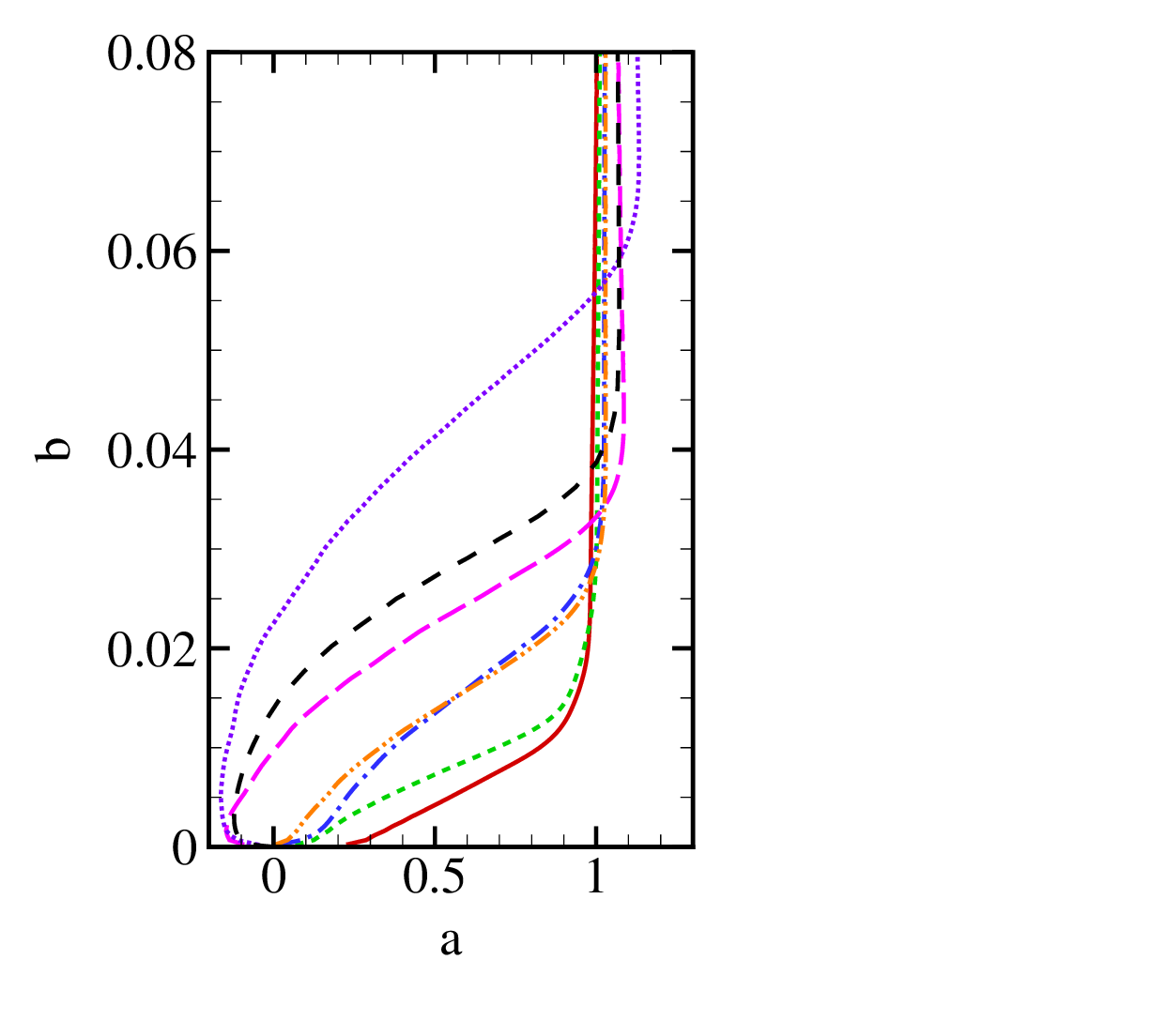}
}
\end{subfigure}
\begin{subfigure}{0.325\textwidth}
\subcaption{~~~~~~~~~~~~~~~~~~~~~~~~~~~~~~~~~~~~~~~~~}
\vspace{-1pt}
{\psfrag{b}[][]{{$x_2/L$}}
\psfrag{a}[][]{{$\left<u_1 \right>/U_{\infty}$}}
\includegraphics[width=\textwidth,trim={2.2 1.8cm 8.2cm 0.4cm},clip]{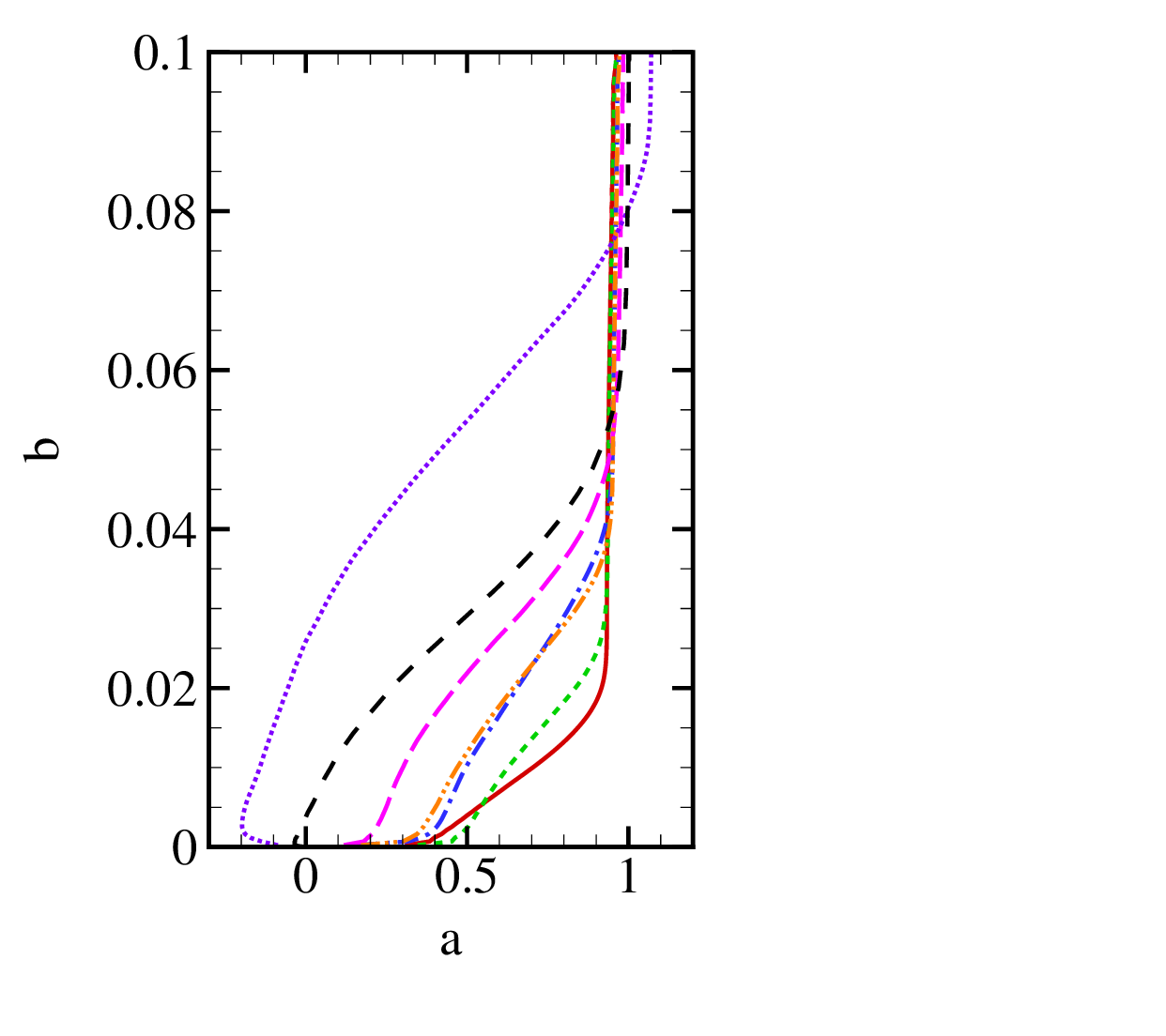}
}
\end{subfigure}
\vskip -0.3cm
\caption{{Comparison of the mean streamwise velocity profiles from the Group 1 baseline-mesh simulations using different SGS models with the DNS results at (a) $x/L=-0.709$, (b) $x/L=-0.2$, (c) $x/L=-0.025$, (d) $x/L=0.1$, (e) $x/L=0.2$ and (f) $x/L=0.4$. Lines indicate {\redsolid}, WMLES without an SGS model; {\greendashed}, WMLES with the DSM; {\bluedashdotted}, WMLES with the AMD model; {\magentadashed}, WMLES with the Vreman model ($c=0.025$); {\orangedashdotdot}, WMLES with the Vreman model ($c=0.07$); {\pupdotdotted}, WMLES with the MSM; {\blackdashed}, DNS~\citep{uzun2022high}.}}
\label{streamwise_cuts_base_SGS}
\end{figure}

To further evaluate the simulation results, we compare boundary layer profiles with the DNS results of \citet{uzun2022high} at six stations along the $x$ direction. Figure~\ref{streamwise_cuts_base_SGS} illustrates the profiles of mean streamwise velocity $\left< u_1 \right>/U_\infty$ along the wall-normal direction. At the stations positioned in front of the bump peak, where $x/L<0$, the discrepancies among the velocity profiles can be noticed, and the profiles from simulations using the AMD model and the two Vreman models align more closely with the DNS results. Conversely, on the leeward side of the bump, where the flow is subject to a strong APG, the mean velocity results demonstrate marked differences. {The simulation employing the MSM significantly overpredicts the thickness of the TBL, whereas the TBL thickness predicted by other simulations is undersized compared to the DNS results.} Of these simulations, the one using the Vreman model ($c=0.025$) provides the most accurate prediction of velocity.

\begin{figure}
\centering
\begin{subfigure}{0.325\textwidth}
\subcaption{~~~~~~~~~~~~~~~~~~~~~~~~~~~~~~~~~~~~~~~~~}
\vspace{-1pt}
{\psfrag{b}[][]{{$x_2/L$}}
\psfrag{a}[][]{{$\left<\epsilon_t \right>\nu/u^4_{\tau}$}}
\includegraphics[width=\textwidth,trim={2.2 1.8cm 8.2cm 0.4cm},clip]{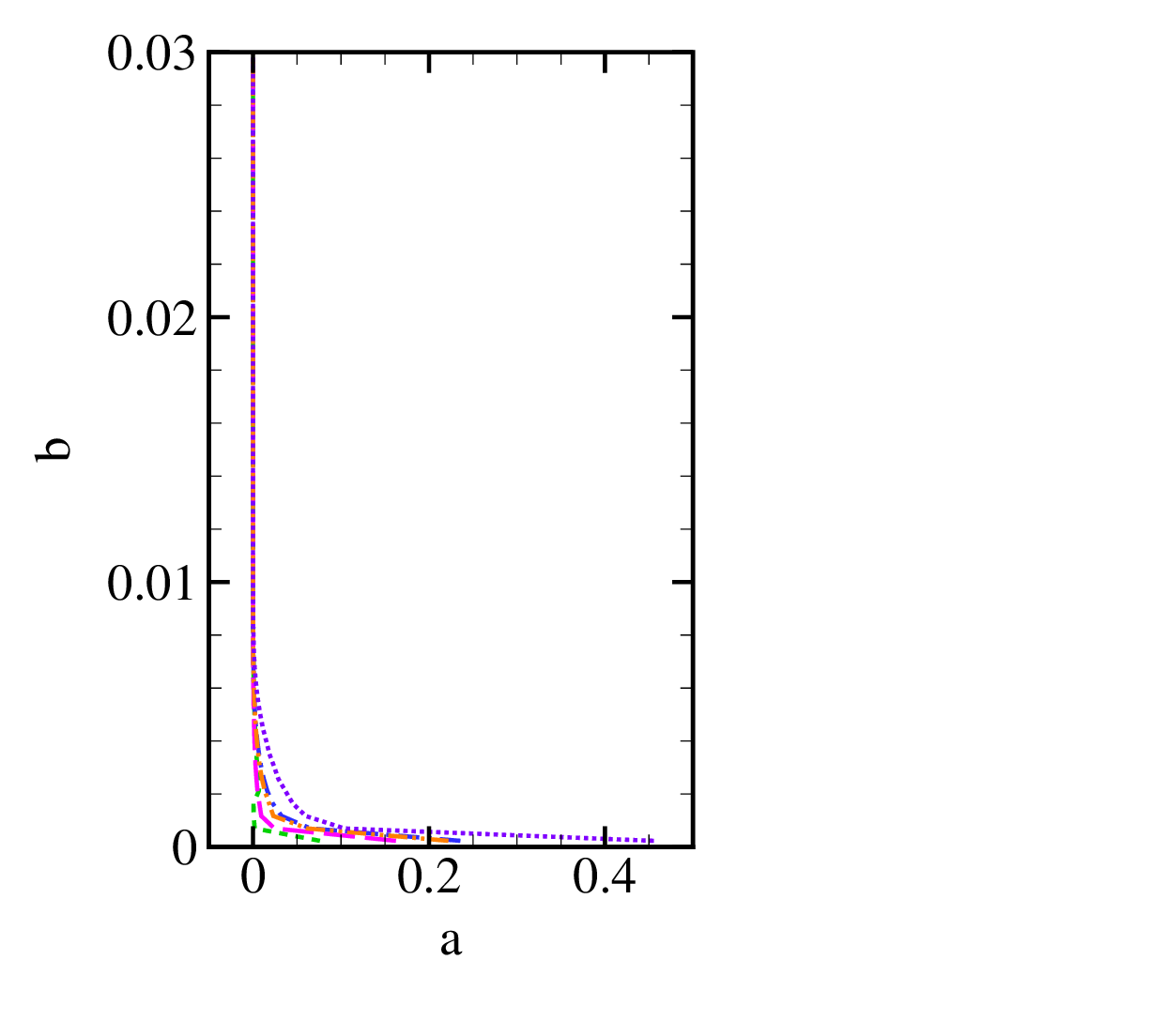}
}
\end{subfigure}
\begin{subfigure}{0.325\textwidth}
\subcaption{~~~~~~~~~~~~~~~~~~~~~~~~~~~~~~~~~~~~~~~~~}
\vspace{-1pt}
{\psfrag{b}[][]{{$x_2/L$}}
\psfrag{a}[][]{{$\left<\epsilon_t \right>\nu/u^4_{\tau}$}}
\includegraphics[width=\textwidth,trim={2.2 1.8cm 8.2cm 0.4cm},clip]{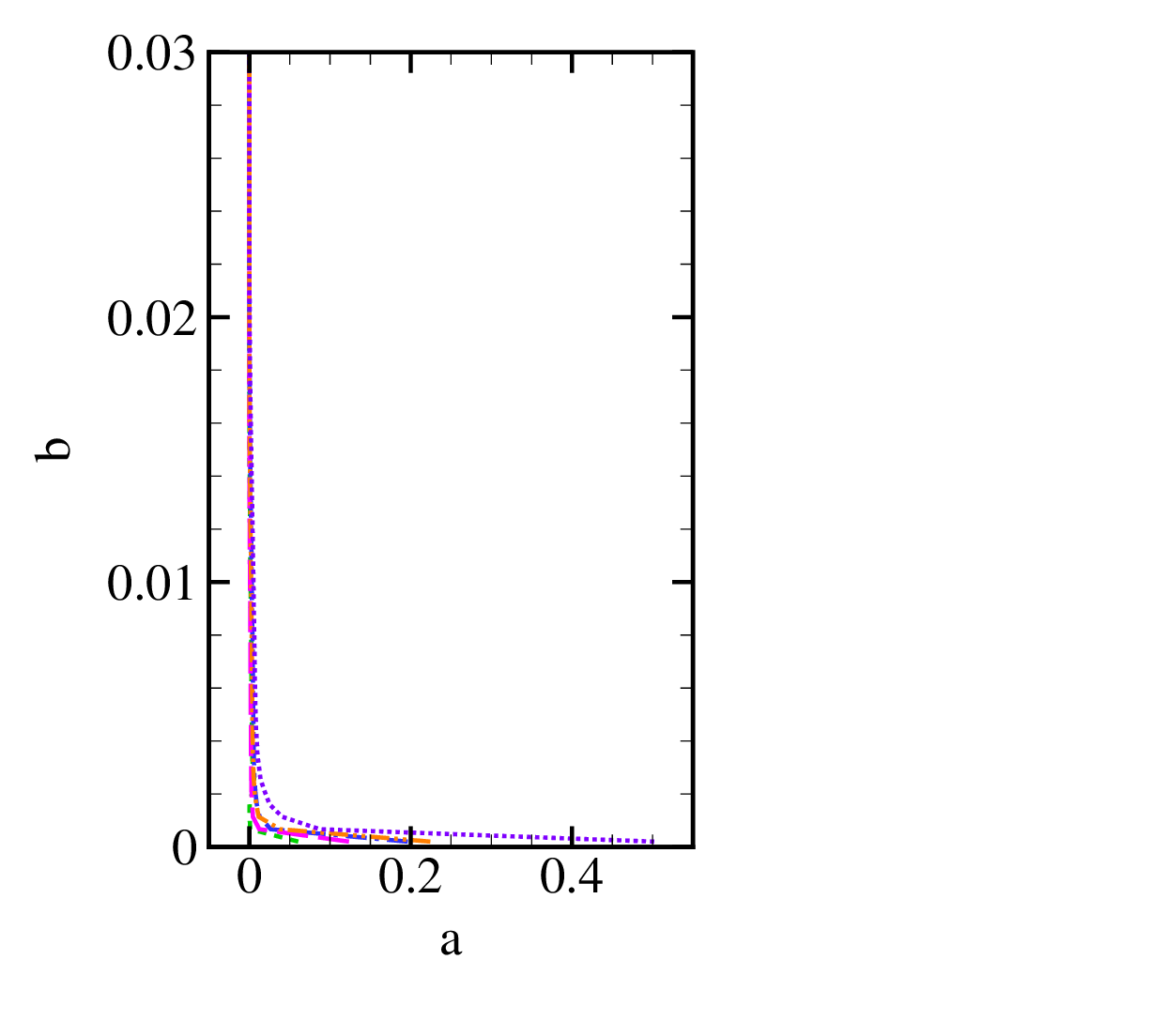}
}
\end{subfigure}
\begin{subfigure}{0.325\textwidth}
\subcaption{~~~~~~~~~~~~~~~~~~~~~~~~~~~~~~~~~~~~~~~~~}
\vspace{-1pt}
{\psfrag{b}[][]{{$x_2/L$}}
\psfrag{a}[][]{{$\left<\epsilon_t \right>\nu/u^4_{\tau}$}}
\includegraphics[width=\textwidth,trim={2.2 1.8cm 8.2cm 0.4cm},clip]{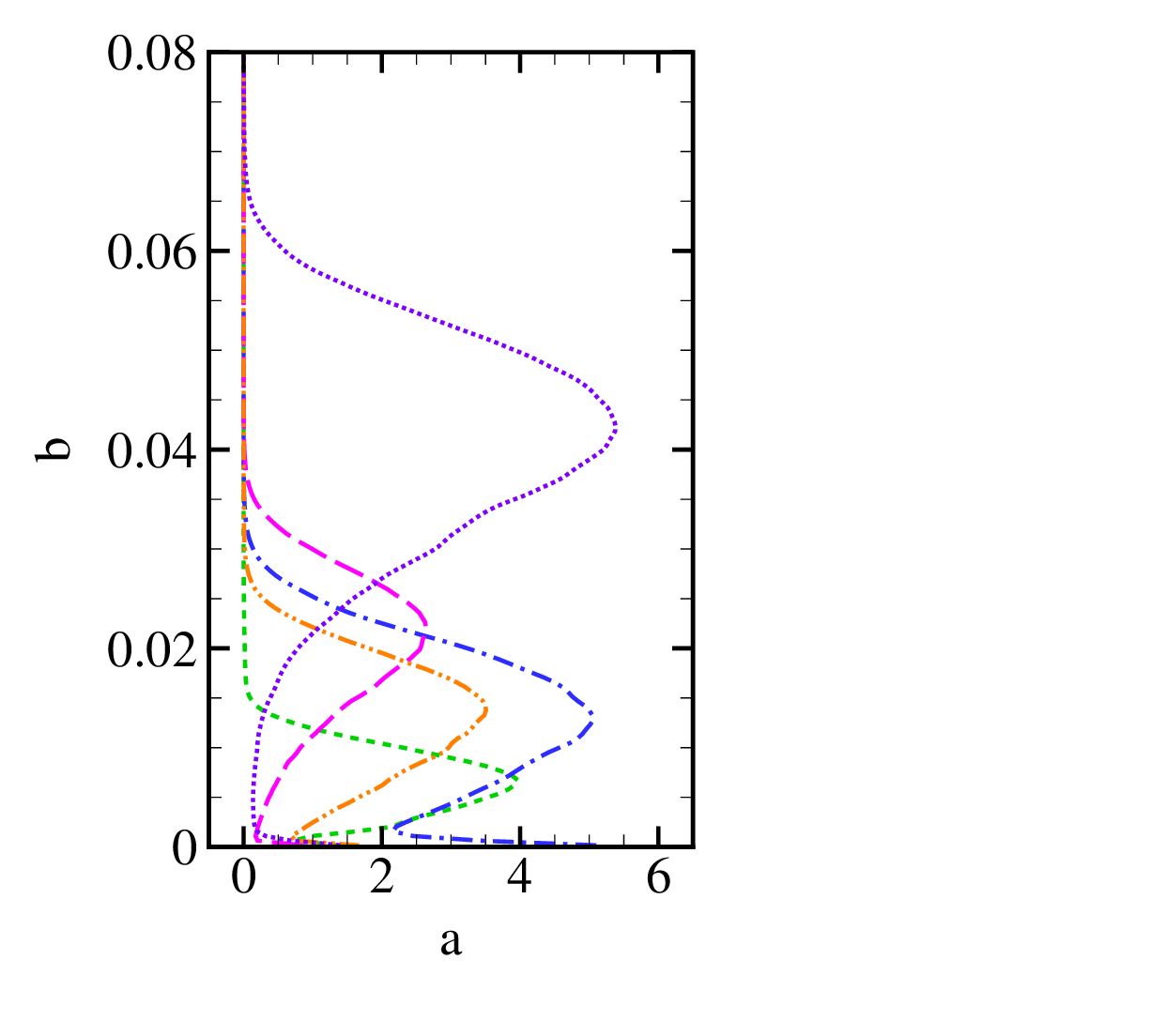}
}
\end{subfigure}
\vskip -0.3cm
\caption{{The profiles of the inner-scaled mean SGS dissipation from the Group 1 baseline-mesh simulations using different SGS models at (a) $x/L=-0.709$, (b) $x/L=-0.2$, (c) $x/L=0.2$. Lines indicate {\greendashed}, WMLES with the DSM; {\bluedashdotted}, WMLES with the AMD model; {\magentadashed}, WMLES with the Vreman model ($c=0.025$); {\orangedashdotdot}, WMLES with the Vreman model ($c=0.07$); {\pupdotdotted}, WMLES with the MSM.}}
\label{streamwise_cuts_SGSdp_base_SGS}
\end{figure}

\begin{figure}
\centering
\begin{subfigure}{0.325\textwidth}
\subcaption{~~~~~~~~~~~~~~~~~~~~~~~~~~~~~~~~~~~~~~~~~}
\vspace{-1pt}
{\psfrag{b}[][]{{$x_2/L$}}
\psfrag{a}[][]{{$s_a$}}
\includegraphics[width=\textwidth,trim={2.2 1.8cm 8.2cm 0.4cm},clip]{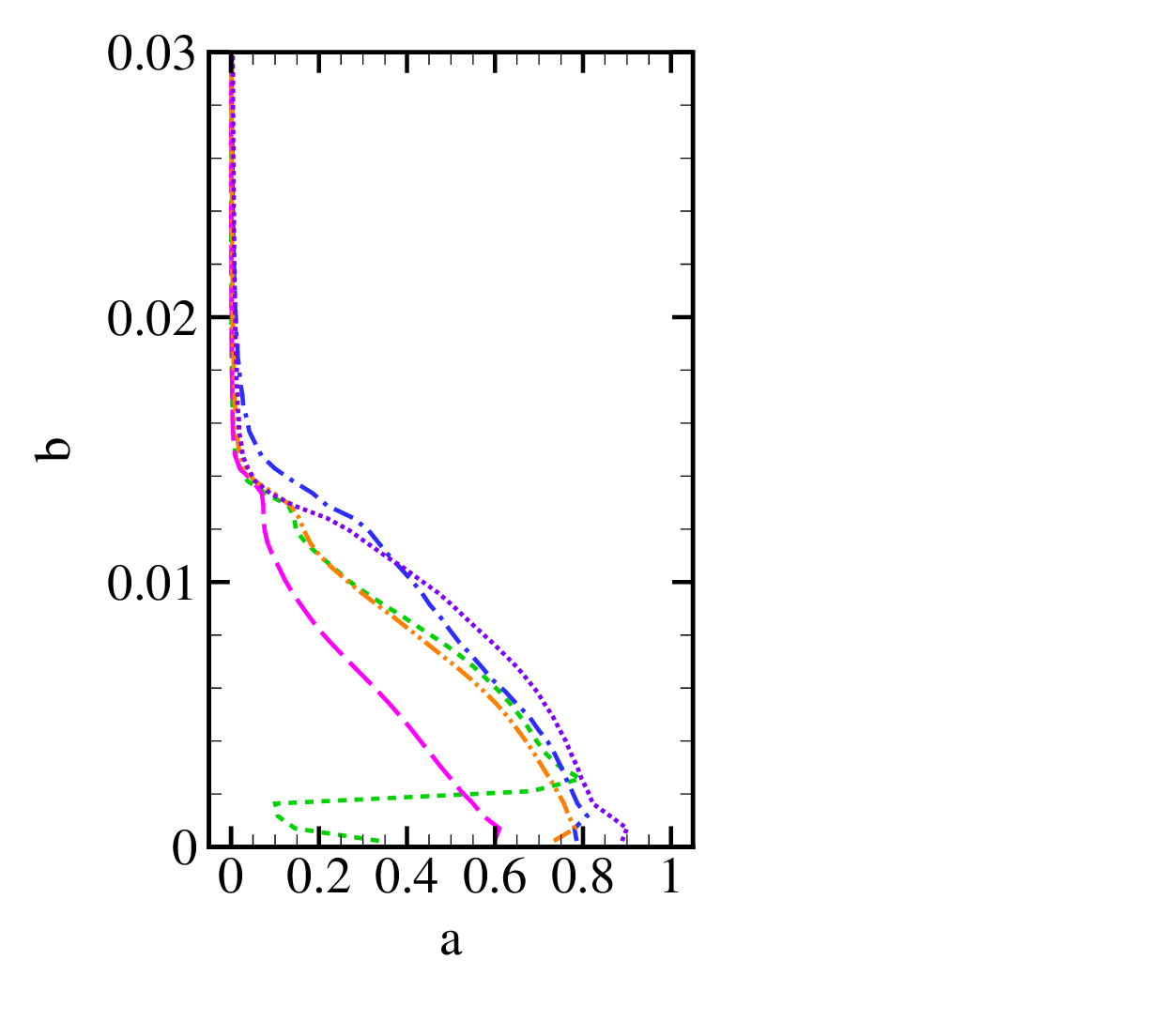}
}
\end{subfigure}
\begin{subfigure}{0.325\textwidth}
\subcaption{~~~~~~~~~~~~~~~~~~~~~~~~~~~~~~~~~~~~~~~~~}
\vspace{-1pt}
{\psfrag{b}[][]{{$x_2/L$}}
\psfrag{a}[][]{{$s_a$}}
\includegraphics[width=\textwidth,trim={2.2 1.8cm 8.2cm 0.4cm},clip]{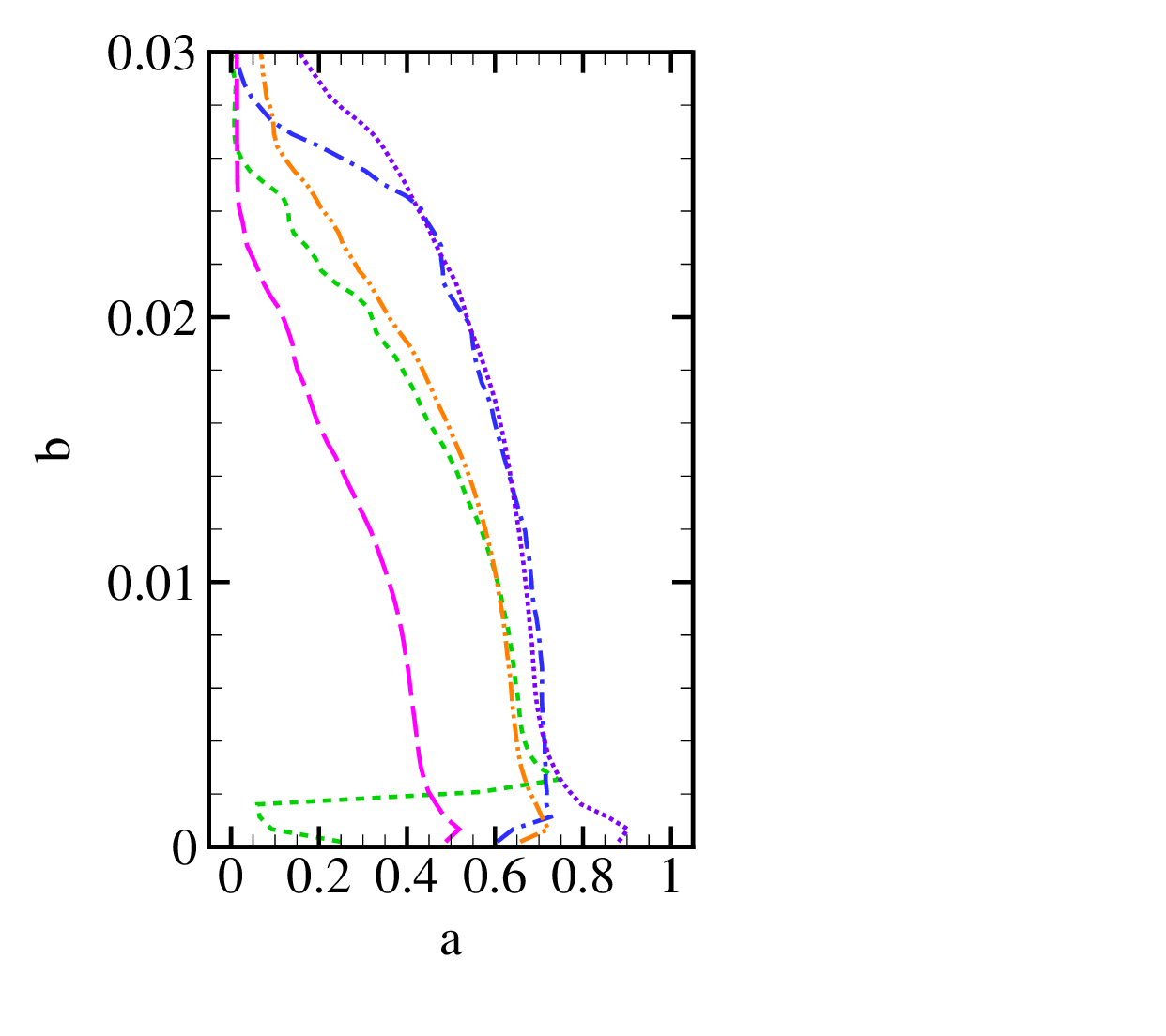}
}
\end{subfigure}
\begin{subfigure}{0.325\textwidth}
\subcaption{~~~~~~~~~~~~~~~~~~~~~~~~~~~~~~~~~~~~~~~~~}
\vspace{-1pt}
{\psfrag{b}[][]{{$x_2/L$}}
\psfrag{a}[][]{{$s_a$}}
\includegraphics[width=\textwidth,trim={2.2 1.8cm 8.2cm 0.4cm},clip]{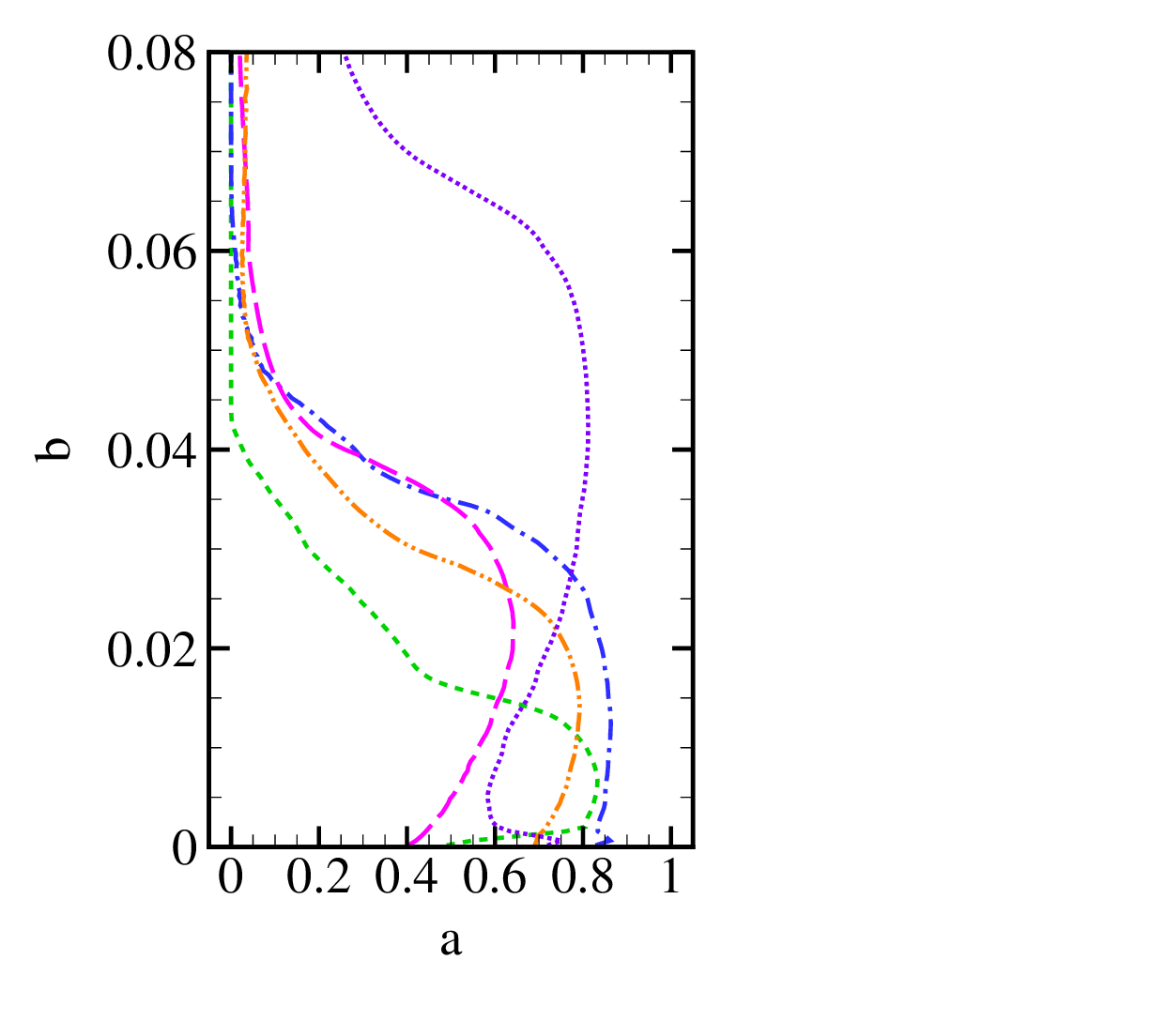}
}
\end{subfigure}
\vskip -0.3cm
\caption{{The profiles of the SGS activity parameter $s_a$ from the Group 1 baseline-mesh simulations using different SGS models at (a) $x/L=-0.709$, (b) $x/L=-0.2$, (c) $x/L=0.2$. Lines indicate {\greendashed}, WMLES with the DSM; {\bluedashdotted}, WMLES with the AMD model; {\magentadashed}, WMLES with the Vreman model ($c=0.025$); {\orangedashdotdot}, WMLES with the Vreman model ($c=0.07$); {\pupdotdotted}, WMLES with the MSM.}}
\label{streamwise_cuts_SGSact_base_SGS}
\end{figure}

{In Fig.~\ref{streamwise_cuts_SGSdp_base_SGS}, the inner-scaled mean SGS dissipation profiles at three stations along the $x$ direction are presented. SGS dissipation is defined as $\epsilon_t = -\tau_{ij}S_{ij}$, with the local friction velocity $u_\tau$ calculated based on the mean skin friction data from the reference DNS \citep{uzun2022high}. At the first two stations, where the flow remains attached, the variation in SGS dissipation relative to streamwise location is small. However, on the leeward side of the bump, where a separation bubble is expected, there is a significant increase in dissipation. Across all stations, the MSM yields higher dissipation compared to other models. In front of the bump peak, strong dissipation in the near-wall region results in the smaller mean streamwise velocity observed in Figs.~\ref{streamwise_cuts_base_SGS}a and \ref{streamwise_cuts_base_SGS}b. Downstream of the bump peak, the MSM generates substantial dissipation in the outer region of the TBL, where shear is pronounced. For other eddy-viscosity models, a noticeable difference in SGS dissipation can also be observed, particularly at the station on the leeward side of the bump. The AMD model produces relatively higher dissipation, consistent with the results of eddy viscosity presented in the previous Fig.~\ref{fig:nut_ave_xy_SGS}. Additionally, to highlight the relative importance of the SGS model in each simulation, the profiles of the SGS activity parameter \citep{lozano2019error, geurts2002framework} are depicted in Fig.~\ref{streamwise_cuts_SGSact_base_SGS}. This parameter is defined as $s_a=\left<\epsilon_t\right> / \left(\left<\epsilon_t\right> + \left<\epsilon_\nu\right>\right)$, where $\epsilon_\nu = 2\nu S_{ij}S_{ij}$ denotes the viscous dissipation. By design, the parameter $s_a$ is constrained to values between 0 and 1, with higher values indicating a more important role of SGS dissipation. The results illustrate that as the location moves downstream, the relative importance of the SGS model within the TBL changes. Furthermore, significant differences among the simulations are evident. In the simulations employing the MSM and the AMD model, SGS dissipation is predominant throughout the boundary layer thickness, whereas the SGS dissipation generated by the Vreman model ($c=0.025$) is comparable to the local viscous dissipation. In addition, it is noteworthy that the SGS activity of the DSM decreases to lower values in the near-wall cells, a trend that is also reflected in the eddy viscosity results shown in Fig.~\ref{fig:nut_ave_xy_SGS}. This reduction leads to diminished SGS dissipation within this specific region, as depicted in Fig.~\ref{streamwise_cuts_SGSdp_base_SGS}. Consequently, it results in the amplified mean streamwise velocity at the first off-wall cell, as shown in Fig.~\ref{mean_vel_1st_SGS}, and aligns with findings from previous WMLES studies \citep{whitmore2020requirements, bae2018investigation}.} A promising feature of DSM in WRLES is its proper asymptotic behavior close to the wall, realized without requiring damping functions \citep{germano1991dynamic}. However, the extension of this property to WMLES of inhomogeneous wall-bounded turbulence, which employs a more coarse mesh, remains an open question. Both the current study and prior research indicate the need for further investigation to clarify its validity in such scenarios. In addition, the results presented in Fig.~\ref{Bump_size_SGS} reveal that the simulations using the DSM did not predict any separation bubble. However, it is crucial to recognize that this conclusion is confined to the specific flow solver and simulation set-up employed in this study. While earlier WMLES by \citet{iyer2022wall, Iyer2021wall} for the same two-dimensional bump flow configuration also did not capture flow separation with the DSM, the WMLES studies \citep{whitmore2021large,agrawal2022non} have demonstrated the potential of the DSM to identify separation on the leeward side of the bump. The discrepancies in the DSM's ability to predict separation might arise from variations in \textit{ad hoc} clipping, the use of filtering, the topology of computational mesh, and other intricate numerical aspects.

\begin{figure}
\centering
\begin{subfigure}{0.325\textwidth}
\subcaption{~~~~~~~~~~~~~~~~~~~~~~~~~~~~~~~~~~~~~~~~~}
\vspace{-1pt}
{\psfrag{b}[][]{{$x_2/L$}}
\psfrag{a}[][]{{$\left<u_1 \right>/U_{\infty}$}}
\includegraphics[width=\textwidth,trim={2.2 1.8cm 8.2cm 0.4cm},clip]{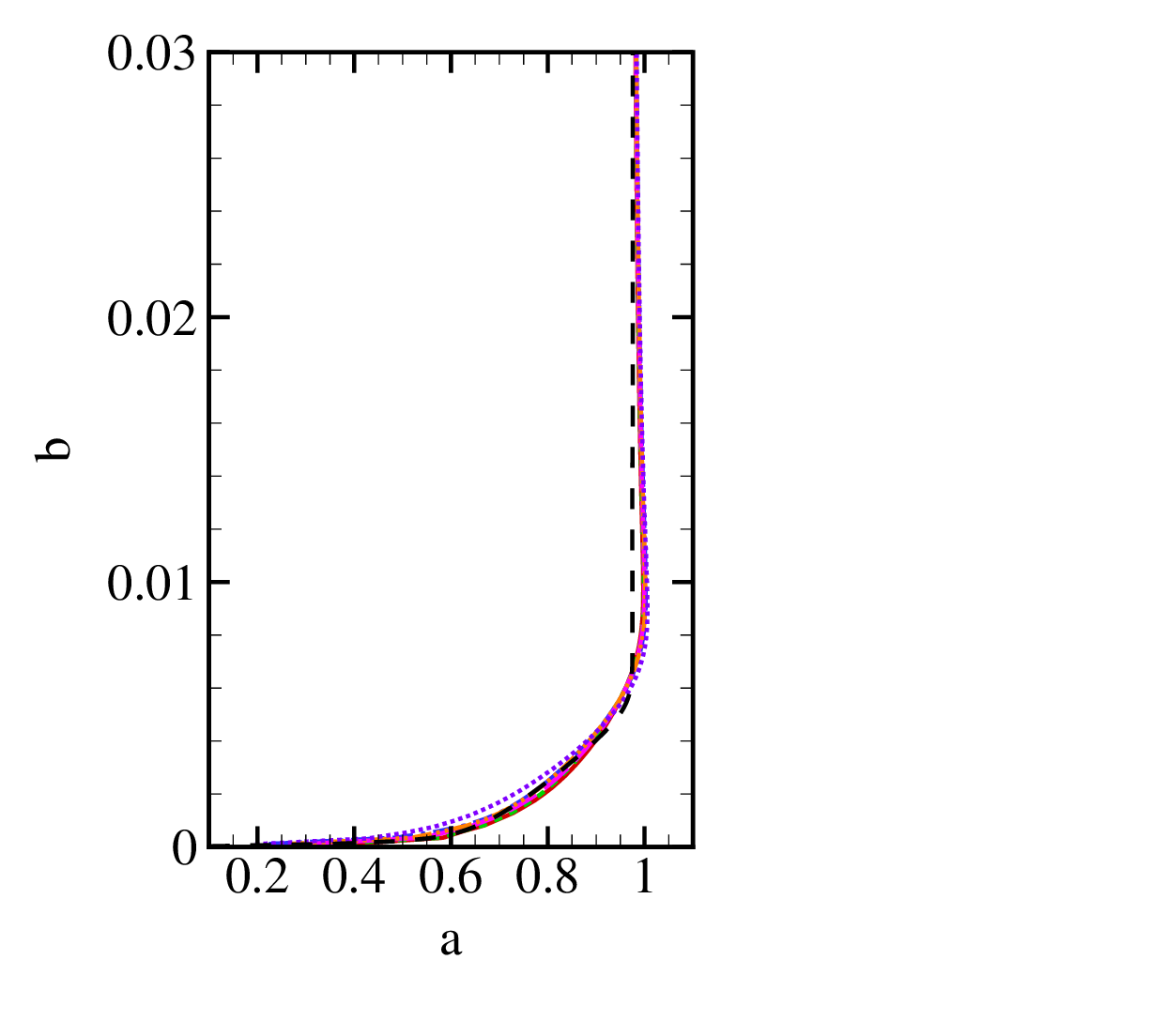}
}
\end{subfigure}
\begin{subfigure}{0.325\textwidth}
\subcaption{~~~~~~~~~~~~~~~~~~~~~~~~~~~~~~~~~~~~~~~~~}
\vspace{-1pt}
{\psfrag{b}[][]{{$x_2/L$}}
\psfrag{a}[][]{{$\left<u_1 \right>/U_{\infty}$}}
\includegraphics[width=\textwidth,trim={2.2 1.8cm 8.2cm 0.4cm},clip]{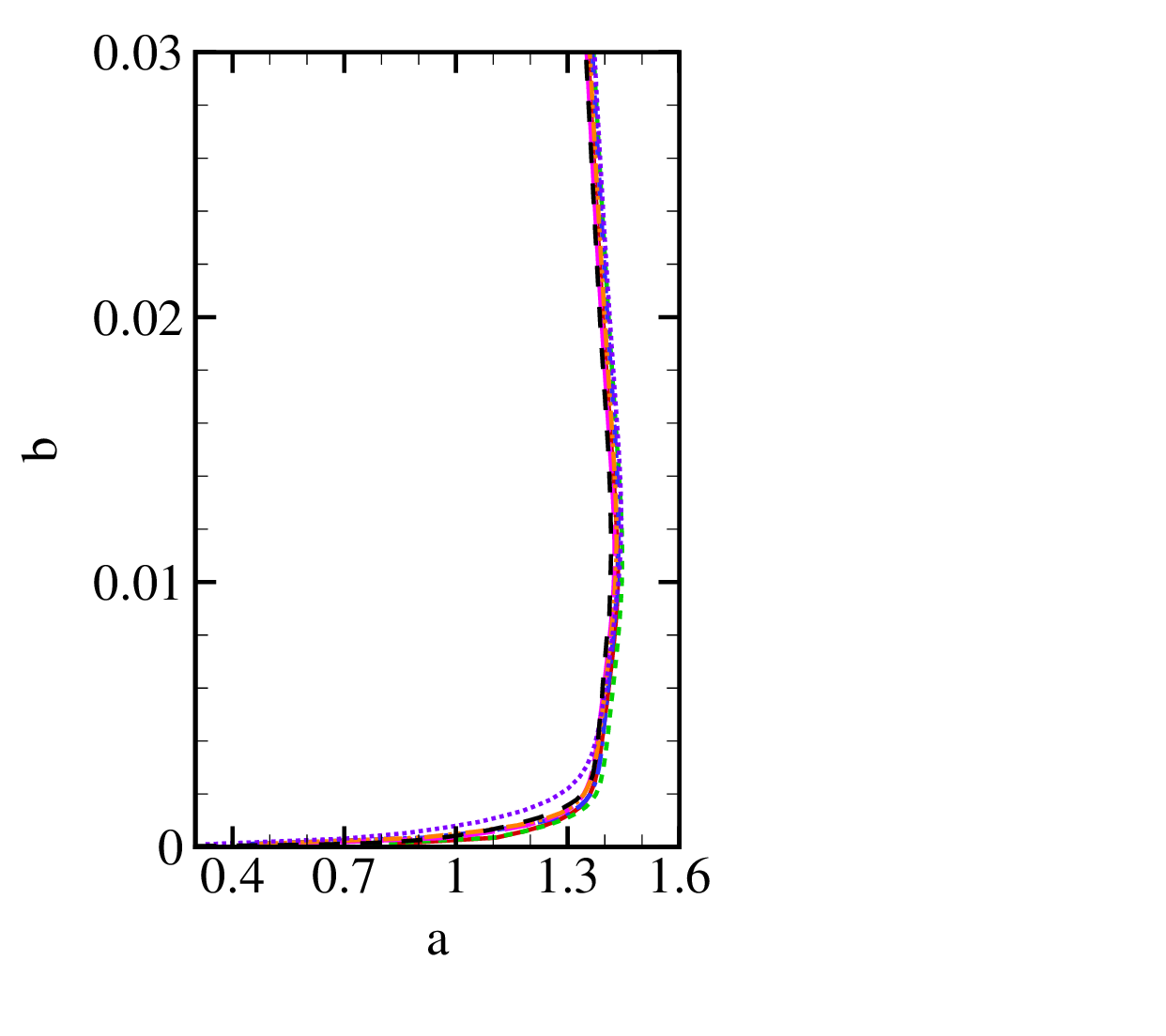}
}
\end{subfigure}
\begin{subfigure}{0.325\textwidth}
\subcaption{~~~~~~~~~~~~~~~~~~~~~~~~~~~~~~~~~~~~~~~~~}
\vspace{-1pt}
{\psfrag{b}[][]{{$x_2/L$}}
\psfrag{a}[][]{{$\left<u_1 \right>/U_{\infty}$}}
\includegraphics[width=\textwidth,trim={2.2 1.8cm 8.2cm 0.4cm},clip]{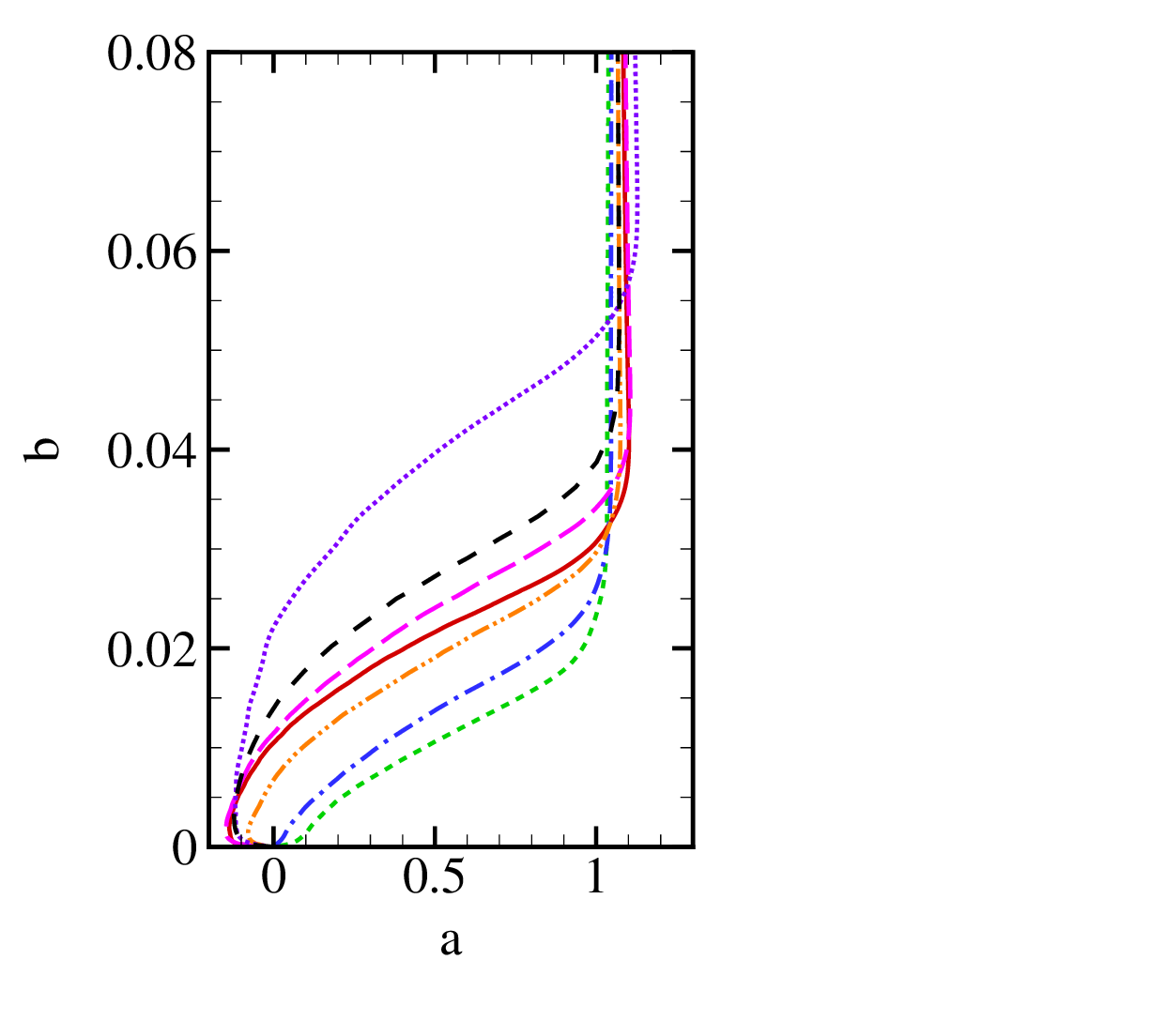}
}
\end{subfigure}
\vskip -0.3cm
\caption{{The profiles of mean streamwise velocity from the fine-mesh simulations of Group 1 at (a) $x/L=-0.709$, (b) $x/L=-0.025$ and (c) $x/L=0.2$. Lines indicate {\redsolid}, WMLES without an SGS model; {\greendashed}, WMLES with the DSM; {\bluedashdotted}, WMLES with the AMD model; {\magentadashed}, WMLES with the Vreman model ($c=0.025$); {\orangedashdotdot}, WMLES with the Vreman model ($c=0.07$); {\pupdotdotted}, WMLES with the MSM; {\blackdashed}, DNS~\citep{uzun2022high}.}}
\label{streamwise_cuts158_fine_SGS}
\end{figure}

Based on the discussions presented earlier, it can be inferred that the SGS model plays a pivotal role in the simulations. However, to reinforce this conclusion, further exploration must isolate the history effect coming from the upstream flow. Let us direct our attention to the results obtained from the fine-mesh simulations. Figure~\ref{streamwise_cuts158_fine_SGS} illustrates the profiles of the mean streamwise velocity at three distinct stations along the $x$ direction. {It is evident that at the stations located far upstream of the bump peak ($x/L=-0.709$) and near the peak ($x/L=-0.025$), the results from different simulations align more closely with each other and with the DNS results. This improved alignment suggests that the simulations are approaching grid convergence within the attached-flow region.} Although very similar upstream flow conditions are observed, the velocity predictions at the station on the leeward side of the bump ($x/L=0.2$) exhibit significant differences. These results, even after isolating the history effect of upstream flow, underscore the important role that the SGS model plays in predicting flow separation. Besides, to confirm that the behaviors of the SGS models are independent of the TBL inflow condition, we conducted the simulations with a different TBL inflow. The results demonstrate that the inflow condition only has a negligible effect on the velocity field. {More details can be referred to \ref{append}.} 

\subsection{Effect of wall boundary condition}

\begin{figure}
\centering
\begin{subfigure}{0.495\textwidth}
\subcaption{~~~~~~~~~~~~~~~~~~~~~~~~~~~~~~~~~~~~~~~~~~~~~~~~~~~~~~~~~~~~~~~~~}
\vspace{-1pt}
{\psfrag{a}[][]{{$x/L$}}
\psfrag{b}[][]{{$C_f\times10^3$}}
\includegraphics[width=\textwidth,trim={1.5 6cm 2.0cm 1.3cm},clip]{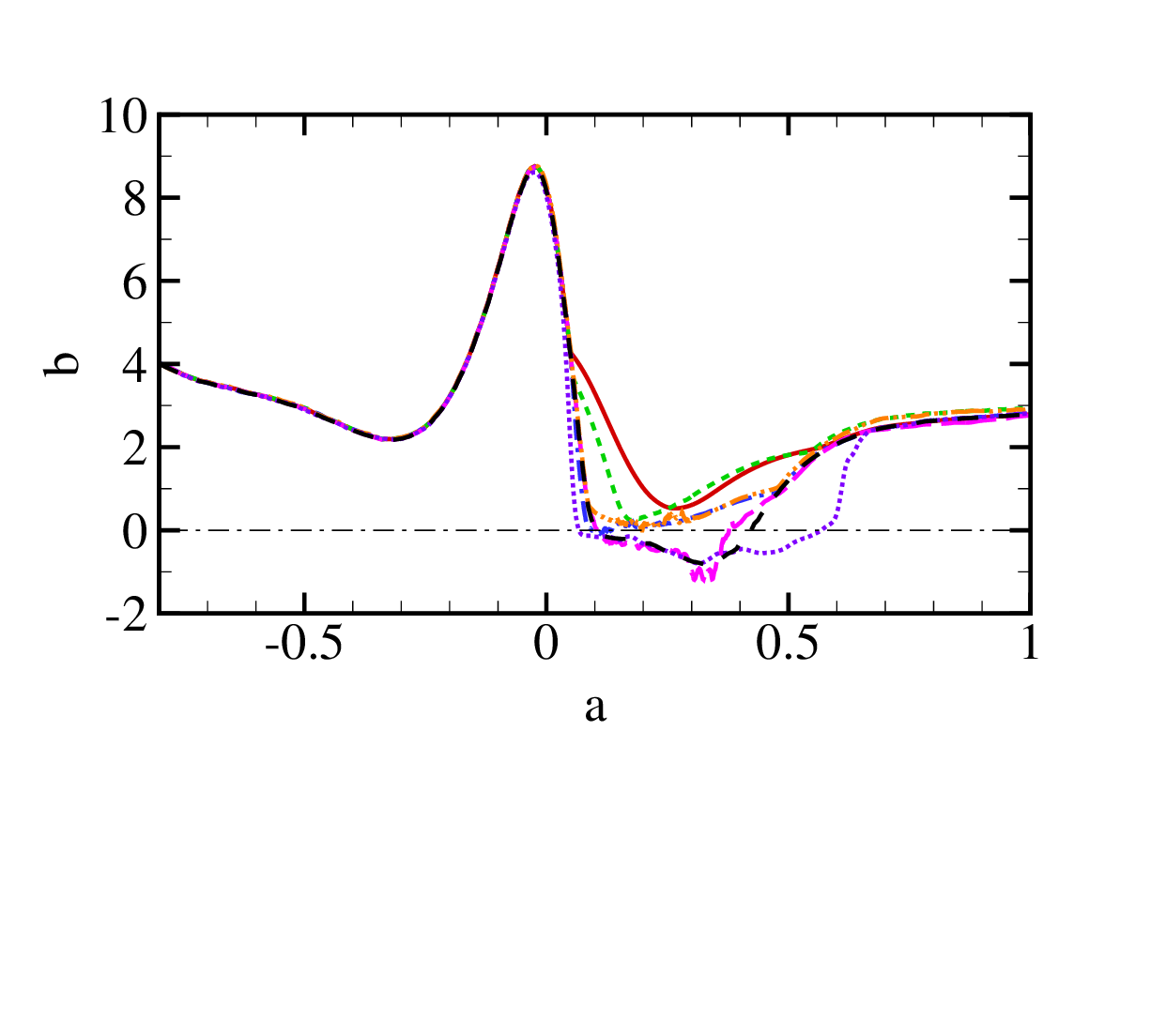}
}
\end{subfigure}
\begin{subfigure}{0.495\textwidth}
\subcaption{~~~~~~~~~~~~~~~~~~~~~~~~~~~~~~~~~~~~~~~~~~~~~~~~~~~~~~~~~~~~~~~~~}
\vspace{-1pt}
{\psfrag{a}[][]{{$x/L$}}
\psfrag{b}[][]{{$C_f\times10^3$}}
\includegraphics[width=\textwidth,trim={1.5 6cm 2.0cm 1.3cm},clip]{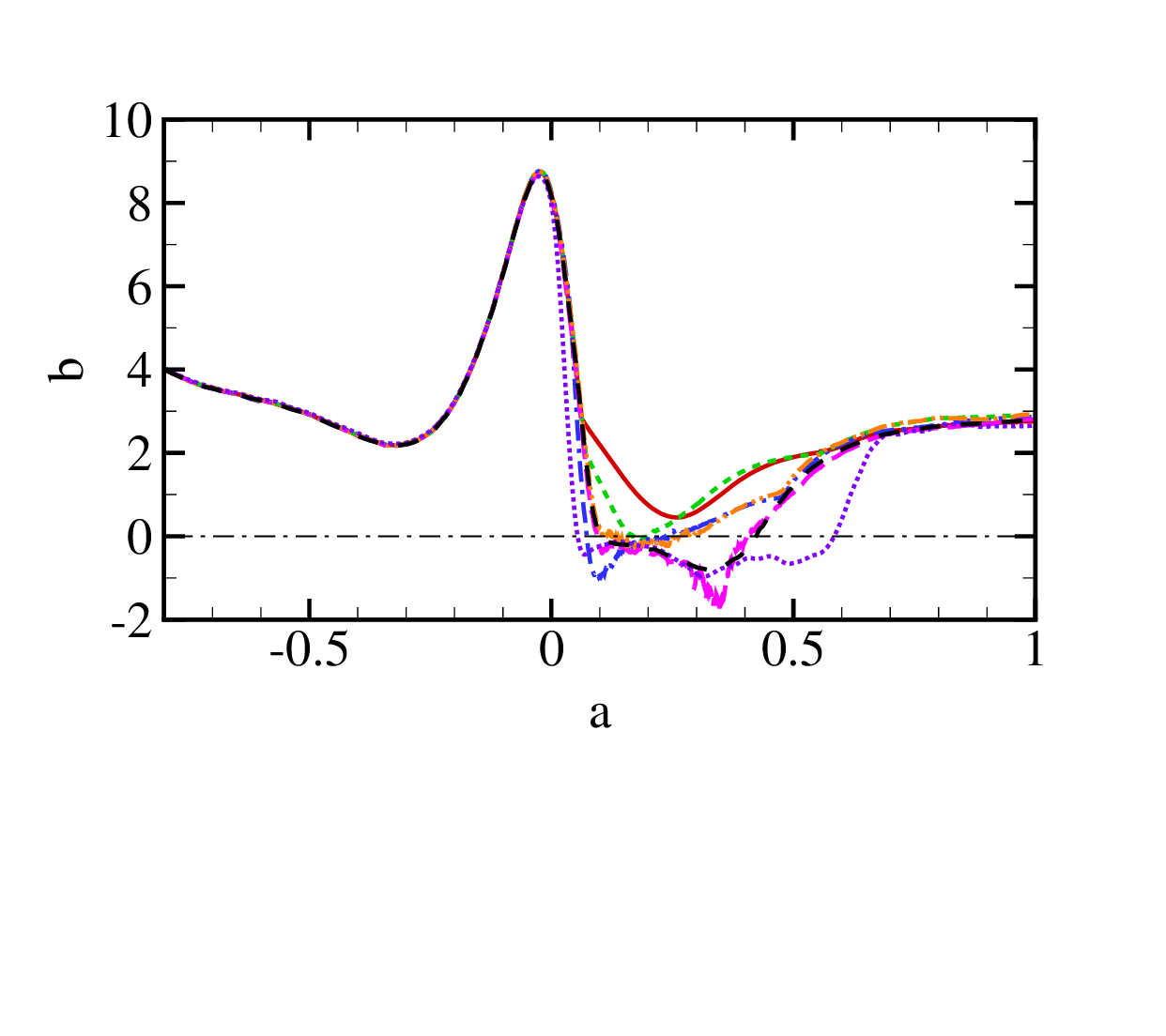}
}
\end{subfigure}
\vskip -0.4cm
\caption{{Comparison of the mean skin friction coefficient $C_f$ on the bottom wall from the Group 2 baseline-mesh simulations using (a) the equilibrium $\nu_{t,w}$ boundary condition and (b) the non-equilibrium $\nu_{t,w}$ boundary condition with the DNS results. Lines indicate {\redsolid}, WMLES without an SGS model; {\greendashed}, WMLES with the DSM; {\bluedashdotted}, WMLES with the AMD model; {\magentadashed}, WMLES with the Vreman model ($c=0.025$); {\orangedashdotdot}, WMLES with the Vreman model ($c=0.07$); {\pupdotdotted}, WMLES with the MSM; {\blackdashed}, DNS~\citep{uzun2022high}; {\blackdashdot}, $C_f=0$.}}
\label{bump_Cf_BC}
\end{figure}

\begin{figure}
\begin{subfigure}{0.495\textwidth}
\subcaption{~~~~~~~~~~~~~~~~~~~~~~~~~~~~~~~~~~~~~~~~~~~~~~~~~~~~~~~~~~~~~~}
\vspace{-1pt}
{\psfrag{a}[][]{{$\Delta_c/L\times10^3$}}
\psfrag{b}[][]{{$L_s/L$}}
\includegraphics[width=\textwidth,trim={0.3 6cm 2.0cm 1.6cm},clip]{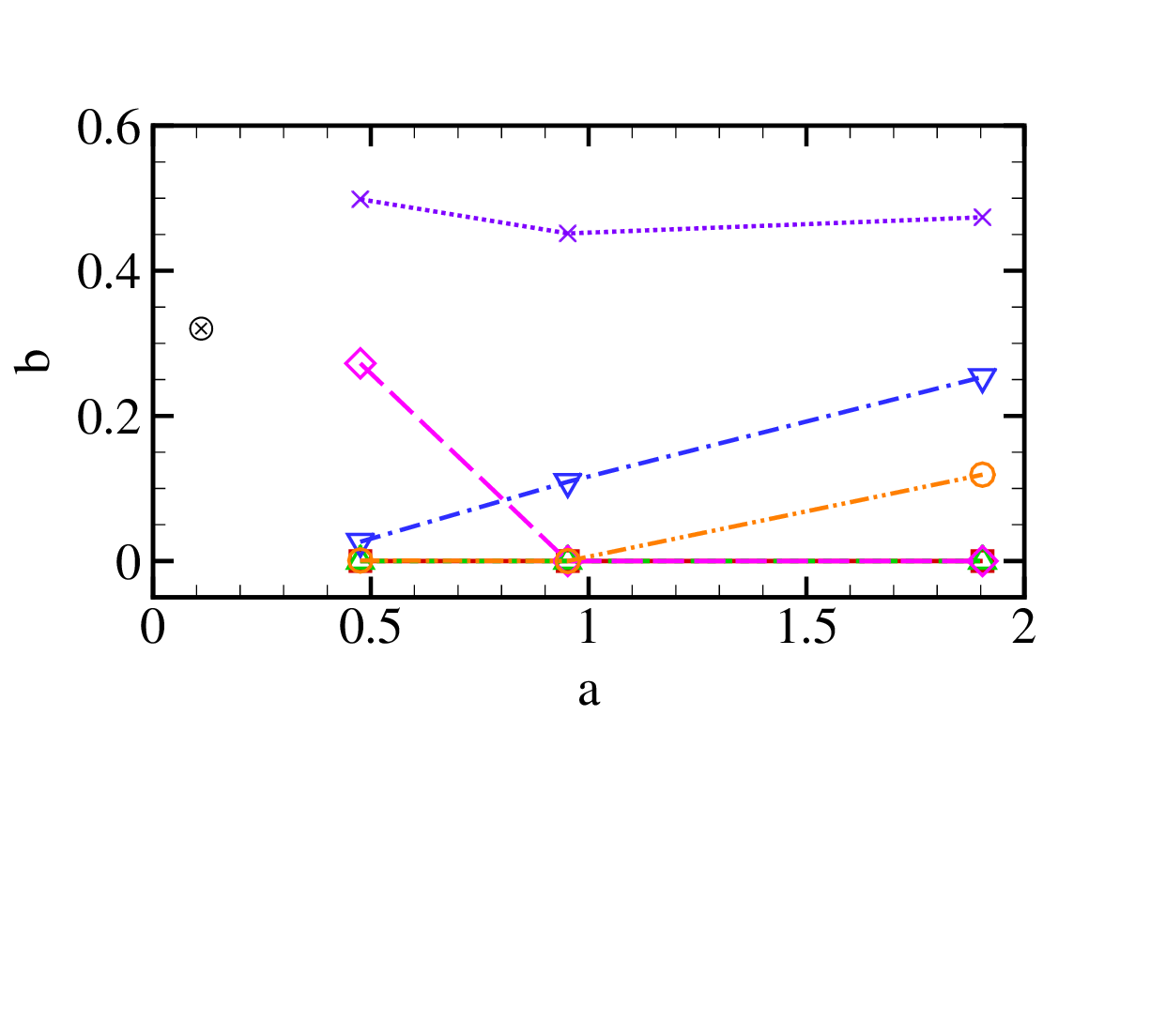}
}
\end{subfigure}
\begin{subfigure}{0.495\textwidth}
\subcaption{~~~~~~~~~~~~~~~~~~~~~~~~~~~~~~~~~~~~~~~~~~~~~~~~~~~~~~~~~~~~~~}
\vspace{-1pt}
{\psfrag{a}[][]{{$\Delta_c/L\times10^3$}}
\psfrag{b}[][]{{$L_s/L$}}
\includegraphics[width=\textwidth,trim={0.3 6cm 2.0cm 1.6cm},clip]{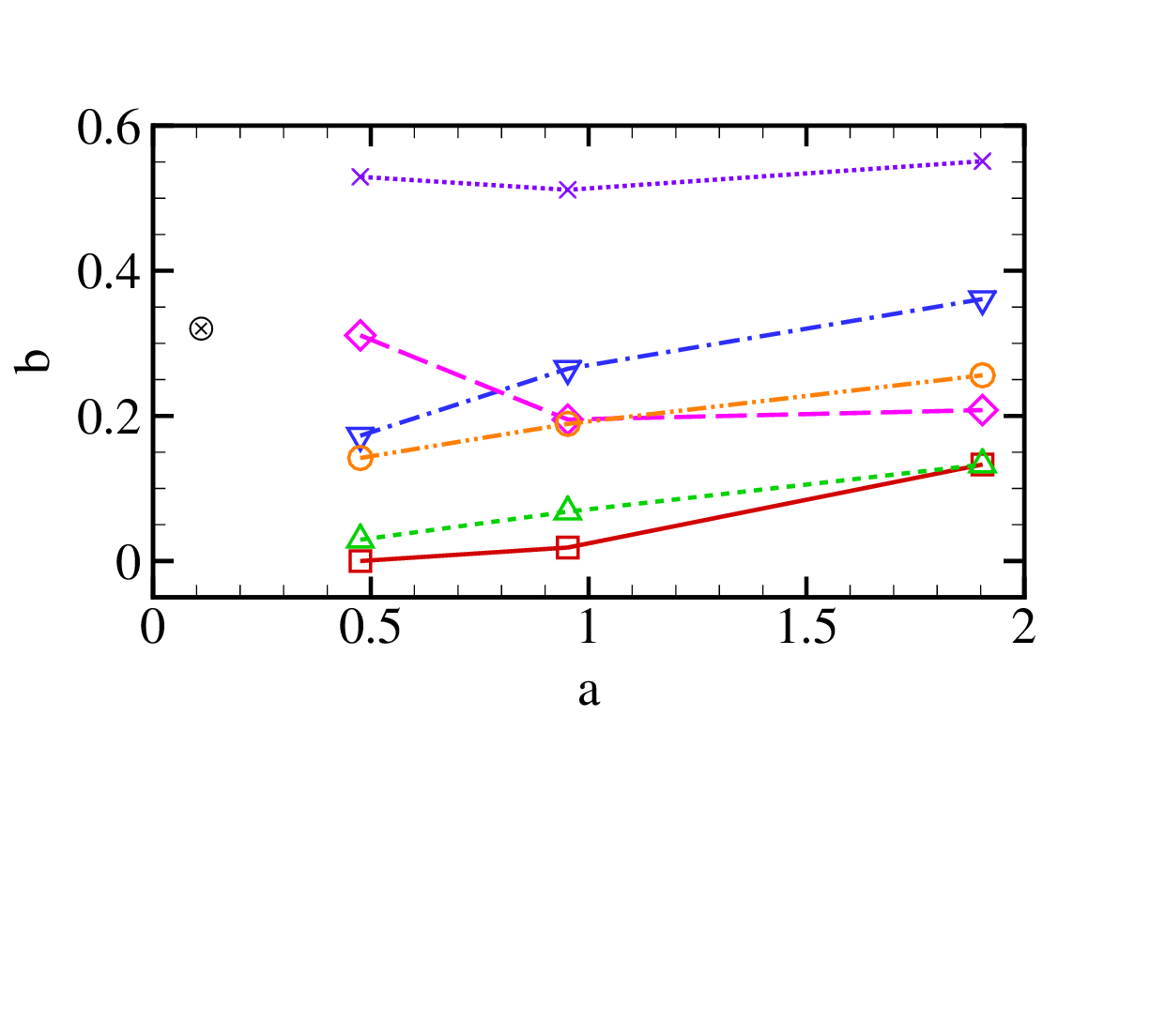}
}
\end{subfigure}\\[-15pt]
\begin{subfigure}{0.495\textwidth}
\subcaption{~~~~~~~~~~~~~~~~~~~~~~~~~~~~~~~~~~~~~~~~~~~~~~~~~~~~~~~~~~~~~~}
\vspace{-1pt}
{\psfrag{a}[][]{{$\Delta_c/L\times10^3$}}
\psfrag{b}[][]{{$L_s/L$}}
\includegraphics[width=\textwidth,trim={0.3 6cm 2.0cm 1.6cm},clip]{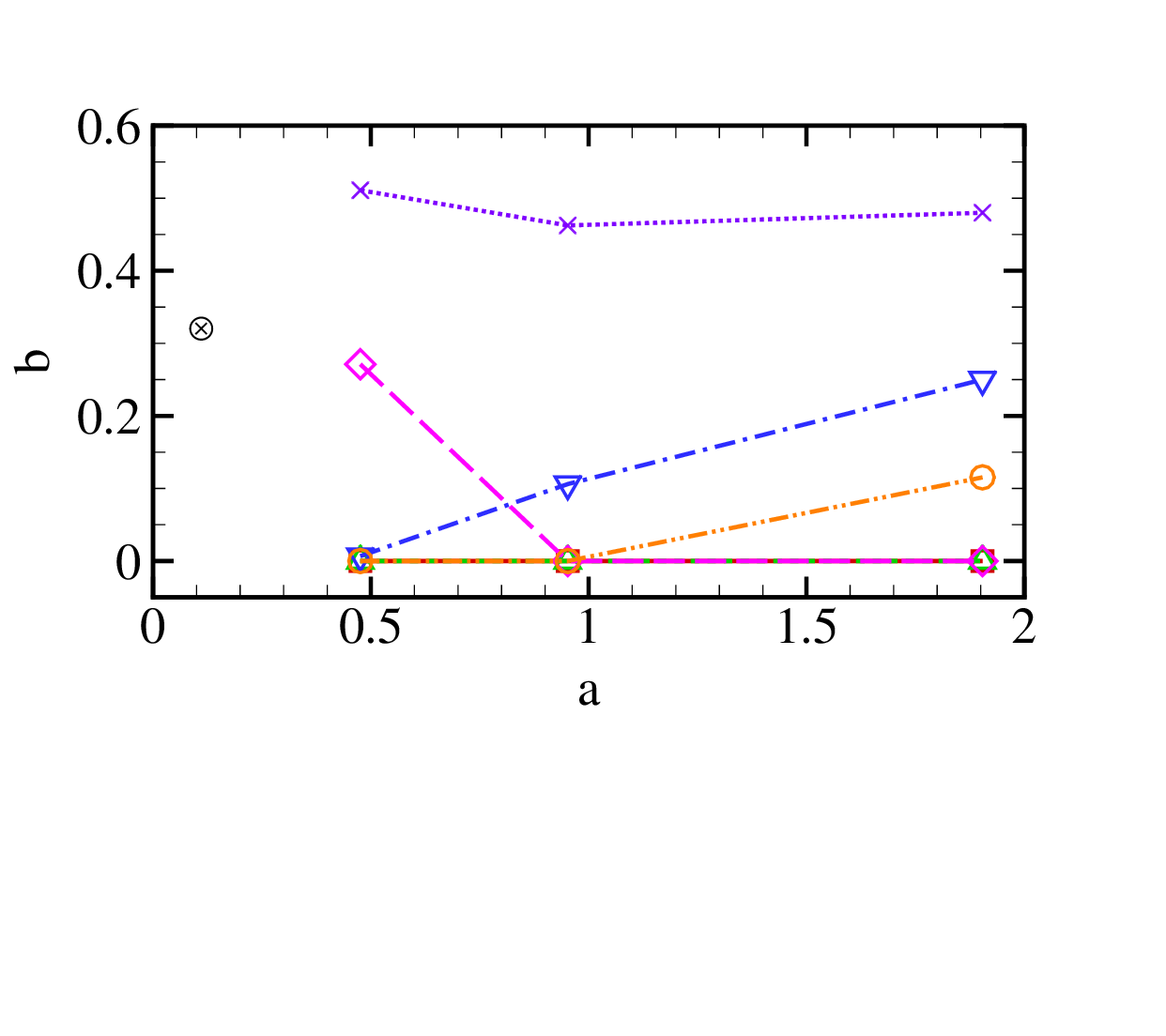}
}
\end{subfigure}
\begin{subfigure}{0.495\textwidth}
\subcaption{~~~~~~~~~~~~~~~~~~~~~~~~~~~~~~~~~~~~~~~~~~~~~~~~~~~~~~~~~~~~~~}
\vspace{-1pt}
{\psfrag{a}[][]{{$\Delta_c/L\times10^3$}}
\psfrag{b}[][]{{$L_s/L$}}
\includegraphics[width=\textwidth,trim={0.3 6cm 2.0cm 1.6cm},clip]{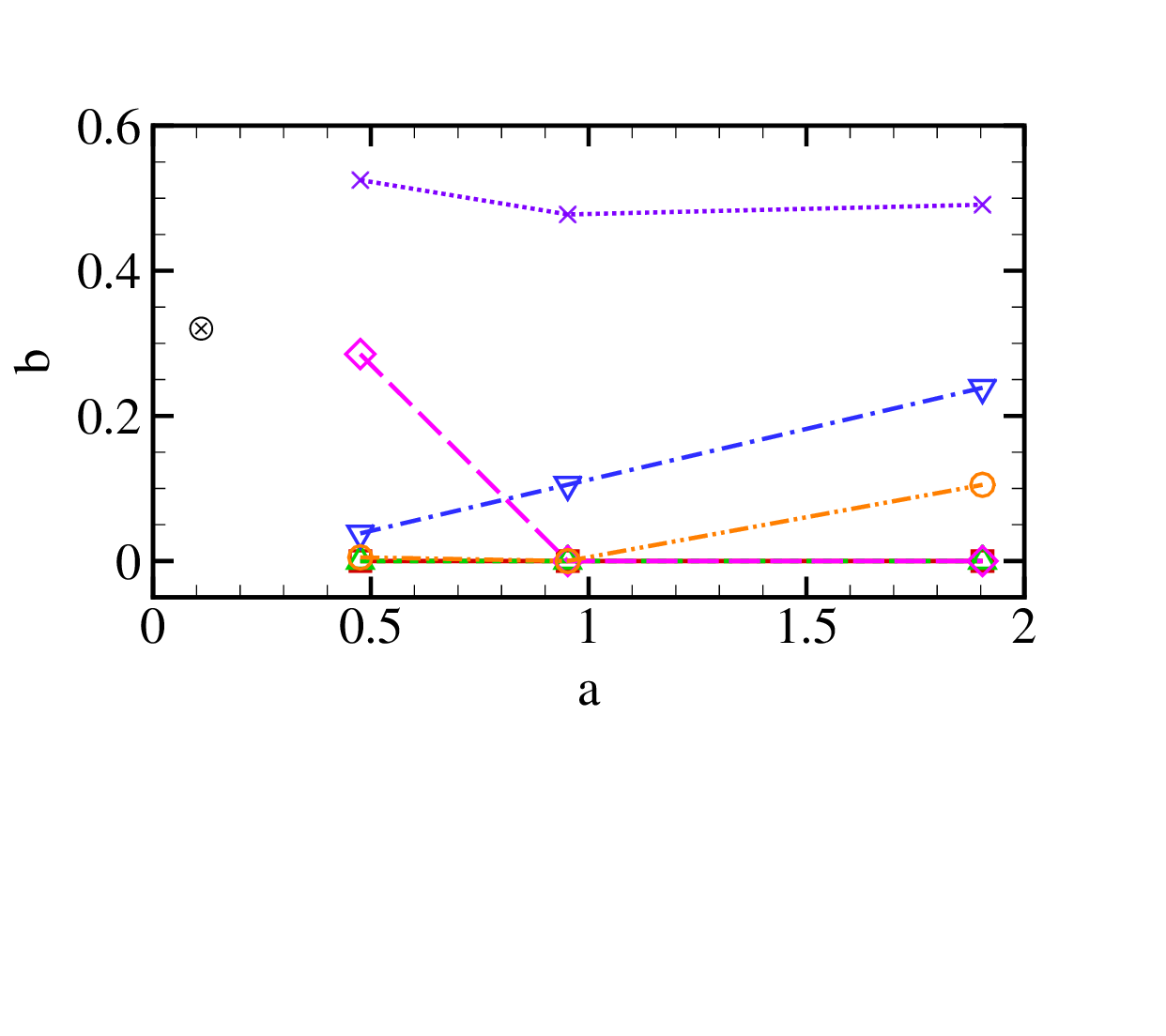}
}
\end{subfigure}
\vskip -0.4cm
\caption{{The horizontal length of separation bubble ($L_s/L$) estimated based on the results of (a,b) $C_f$ and (c,d) the mean streamwise velocity at the first off-wall cell center from the Group 2 simulations using (a,c) the equilibrium $\nu_{t,w}$ boundary condition and (b,d) the non-equilibrium $\nu_{t,w}$ boundary condition. Lines indicate {\redsolid} with \protect\marksymbol{square}{red}, WMLES without an SGS model; {\greendashed} with \protect\marksymbol{triangle}{green}, WMLES using the DSM; {\bluedashdotted} with \protect\marksymbol{triangle}{blue,style={rotate=180}}, WMLES using the AMD model; {\magentadashed} with \protect\marksymbol{diamond}{magenta}, WMLES using the Vreman model ($c=0.025$); {\orangedashdotdot} with  \protect\marksymbol{o}{orange}, WMLES using the Vreman model ($c=0.07$); {\pupdotdotted} with \textcolor{purp2}{$\times$}, WMLES using the MSM. The symbol \textcolor{black}{$\otimes$} represents the DNS results where $\Delta_c/L\times 10^3=0.11$ and $L_s/L=0.32$.}}
\label{bump_size_BC_Cf&vel}
\end{figure}

In order to examine the effect of different wall boundary conditions, a series of simulations in Group 2 of Table~\ref{tab:table2} were conducted using both the aforementioned equilibrium and non-equilibrium $\nu_{t,w}$ boundary conditions. Figure~\ref{bump_Cf_BC} shows the distributions of the mean skin friction coefficient $C_f$ on the bottom wall from the baseline-mesh simulations with these two boundary conditions. The skin friction coefficient is defined as $C_f=\left<\tau_{w,1}\right>/(0.5\rho U_\infty^2)$, and for comparison, the mean skin friction coefficient from the DNS~\citep{uzun2022high} is also included. Noticeable differences among the profiles from the simulations using different SGS models can be observed in the region of {$x/L\in \left( 0,0.7\right)$}, which is where the separation bubble is expected to appear. In other locations, the agreement among the profiles, as well as with the DNS results, is satisfactory. {Part of the reason for the discrepancy observed on the leeward side of the bump is related to the clipping of effective wall-eddy viscosity, which occurs only within this region. The occurrence of this clipping varies over time and spatial location, with an average occurrence rate below 13\% in each case.} Based on the results of $C_f$, the size of the separation bubble can be quantified. Figures~\ref{bump_size_BC_Cf&vel}(a) and \ref{bump_size_BC_Cf&vel}(b) illustrate the measured separation bubble's horizontal length as a function of the characteristic mesh resolution. From this, it is evident that both the SGS model and the mesh resolution still largely influence the prediction of the separation bubble. {Additionally, the previously mentioned non-monotonic convergence is observed in simulations employing eddy viscosity models, such as the AMD model and the Vreman model with $c=0.07$. In contrast, simulations using the MSM still demonstrate improved consistency in the size of the separation bubble across different mesh resolutions. These observations suggest that the application of both equilibrium and non-equilibrium $\nu_{t,w}$ boundary conditions has a negligible impact on the convergence behavior. Comparing results from simulations that employ these two distinct wall boundary conditions reveals that the non-equilibrium boundary condition notably improves the prediction of the separation bubble in simulations using eddy viscosity models.} However, an assessment of the predicted separation bubble's length, based on the mean streamwise velocity at the first off-wall cell center (as depicted in Figs.~\ref{bump_size_BC_Cf&vel}(c) and \ref{bump_size_BC_Cf&vel}(d)), reveals that the disparity in performance between these two boundary conditions is relatively small. The predictions from the simulations using the equilibrium boundary condition closely resemble the results in Fig.~\ref{Bump_size_SGS}, which are obtained using the velocity Neumann boundary condition. While the implementation of the non-equilibrium boundary condition results in some improvements, the magnitude of this enhancement is not substantial.


\begin{figure}
\centering
\begin{subfigure}{0.495\textwidth}
\subcaption{~~~~~~~~~~~~~~~~~~~~~~~~~~~~~~~~~~~~~~~~~~~~~~~~~~~~~~~~~~~~~~~~~}
\vspace{-1pt}
{\psfrag{a}[][]{{$x/L$}}
\psfrag{b}[][]{{$\left<u_1 \right>/U_{\infty}$}}
\includegraphics[width=\textwidth,trim={1.5 6.6cm 2.0cm 1.3cm},clip]{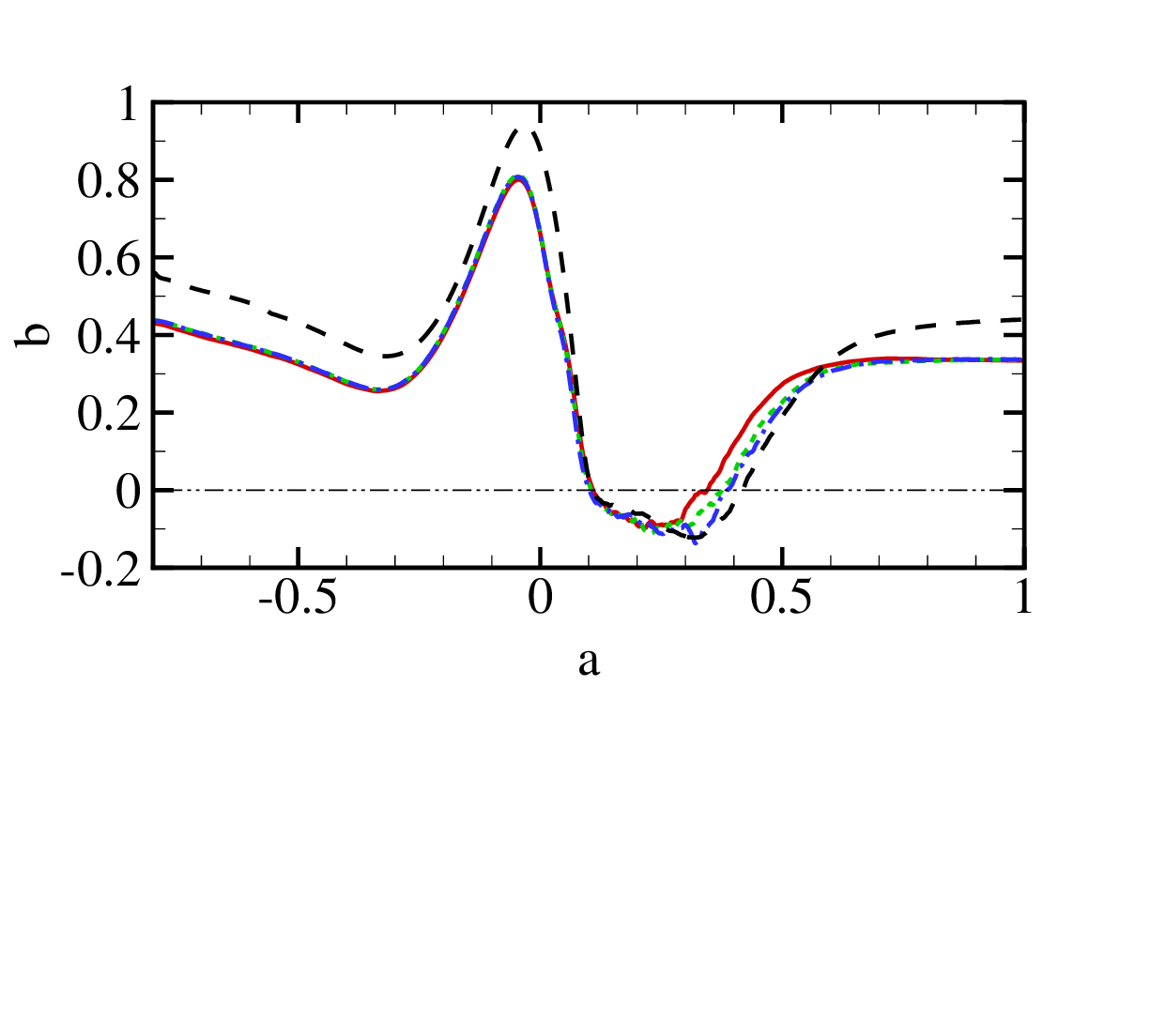}
}
\end{subfigure}
\begin{subfigure}{0.495\textwidth}
\subcaption{~~~~~~~~~~~~~~~~~~~~~~~~~~~~~~~~~~~~~~~~~~~~~~~~~~~~~~~~~~~~~~~~~}
\vspace{-1pt}
{\psfrag{a}[][]{{$x/L$}}
\psfrag{b}[][]{{$\left<u_1 \right>/U_{\infty}$}}
\includegraphics[width=\textwidth,trim={1.5 6.6cm 2.0cm 1.3cm},clip]{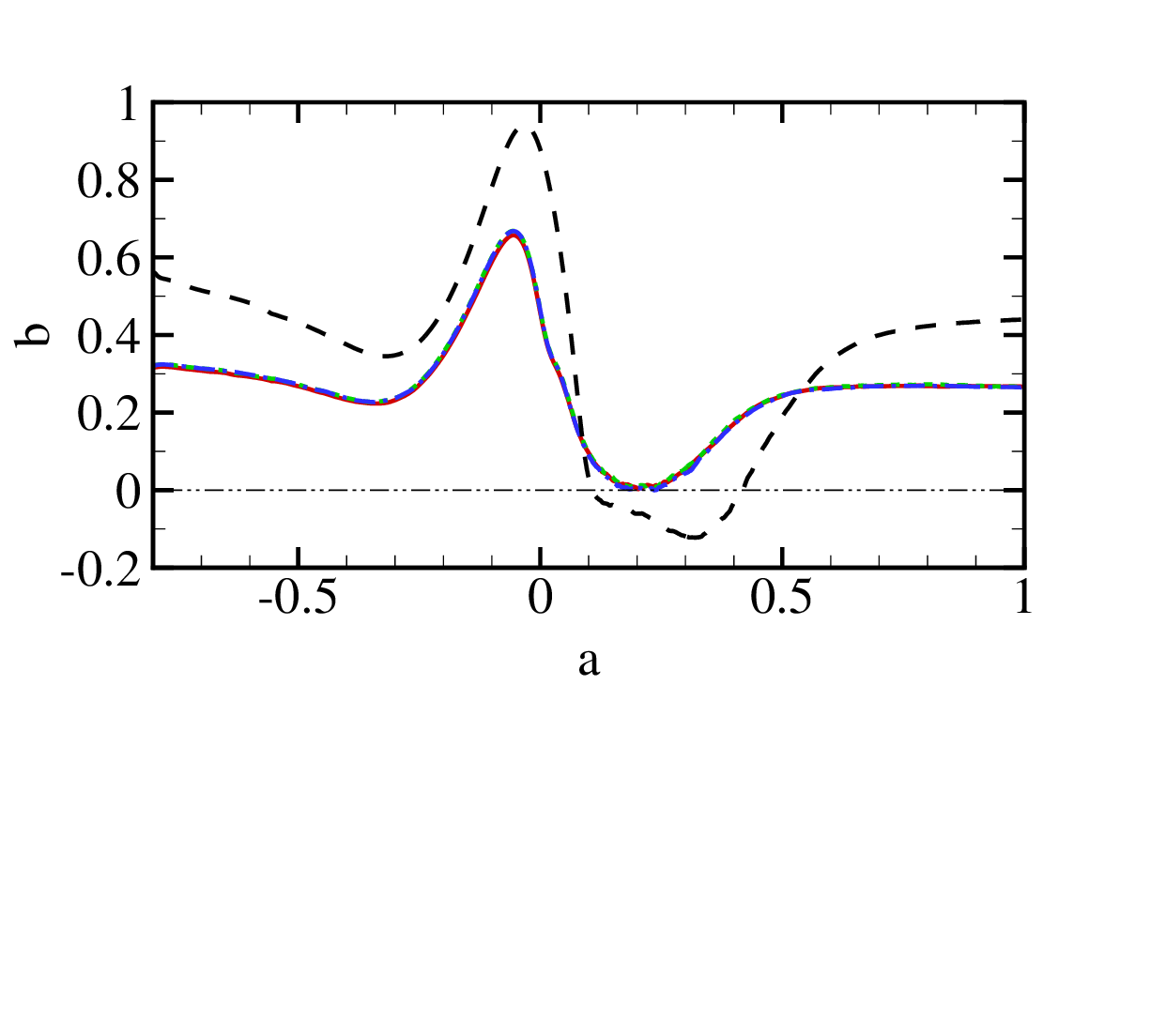}
}
\end{subfigure}
\begin{subfigure}{0.495\textwidth}
\subcaption{~~~~~~~~~~~~~~~~~~~~~~~~~~~~~~~~~~~~~~~~~~~~~~~~~~~~~~~~~~~~~~~~~}
\vspace{-1pt}
{\psfrag{a}[][]{{$x/L$}}
\psfrag{b}[][]{{$\left<u_1 \right>/U_{\infty}$}}
\includegraphics[width=\textwidth,trim={1.5 6.6cm 2.0cm 1.3cm},clip]{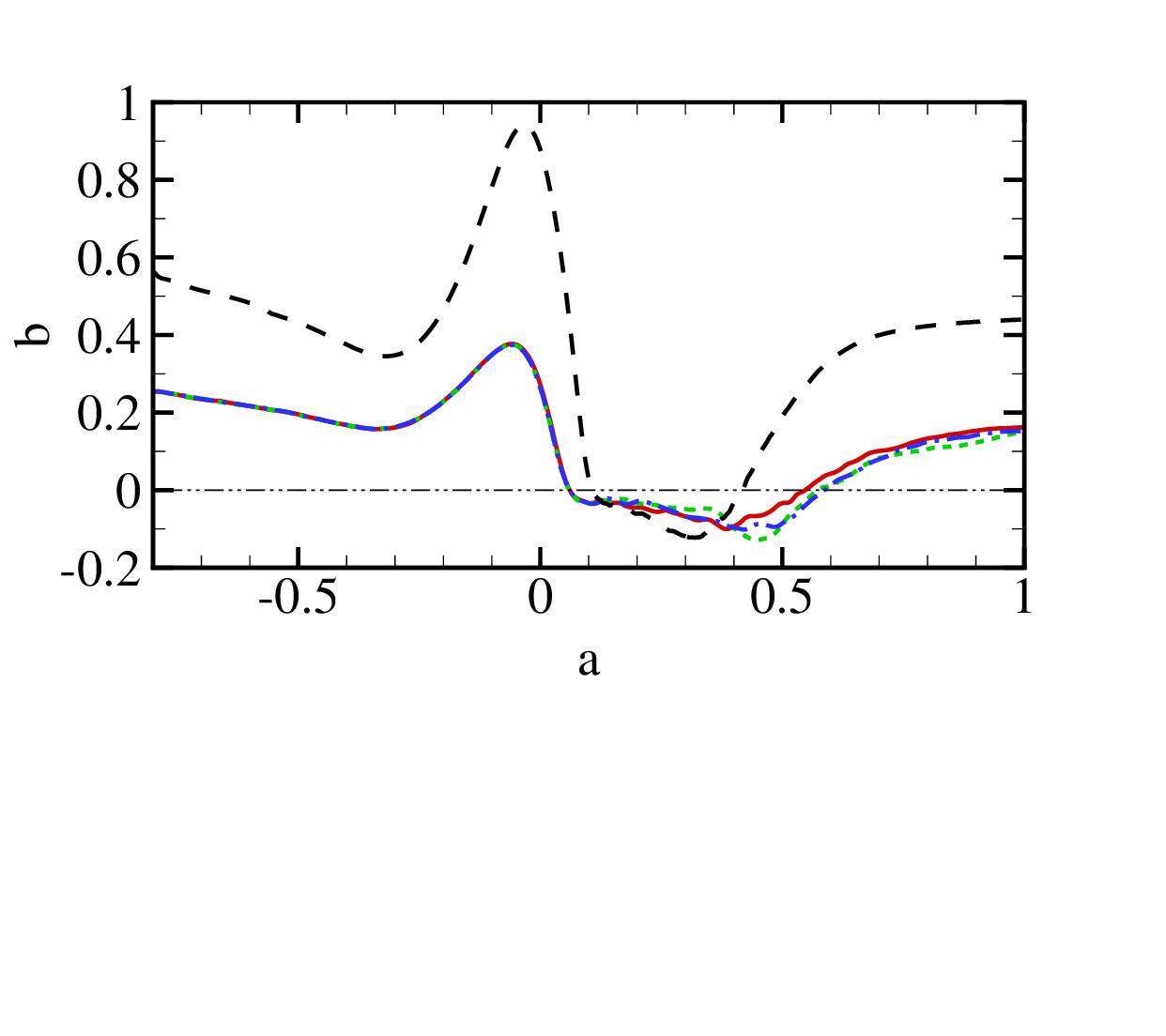}
}
\end{subfigure}
\vskip -0.4cm
\caption{{The profiles of mean streamwise velocity at $x_2/L=2.38\times 10^{-4}$ (the wall-normal location of the first off-wall cell center) from the baseline-mesh simulations using (a) the Vreman model ($c=0.025$), (b) the Vreman model ($c=0.07$), and (c) the MSM with different wall boundary conditions, alongside a reference profile from the DNS \citep{uzun2022high} at the same wall-normal location. Lines indicate {\redsolid}, WMLES with the velocity Neumann boundary condition on the bottom wall; {\greendashed}, WMLES with the equilibrium $\nu_{t,w}$ boundary condition on the bottom wall; {\bluedashdotted}, WMLES with the non-equilibrium $\nu_{t,w}$ boundary condition on the bottom wall; {\blackdashed}, DNS~\citep{uzun2022high}; {\blackdashdotdot}, $\left<u_1 \right>/U_{\infty}=0$.}}
\label{wallnormal_cut_vel_base_BC}
\end{figure}

A detailed quantitative comparison of the velocity field was conducted to investigate the influence of boundary conditions on velocity-field predictions further. Figure~\ref{wallnormal_cut_vel_base_BC} illustrates the mean streamwise velocity at the first off-wall cell center, derived from baseline-mesh simulations employing different wall boundary conditions. {Specifically, the simulation results using the two previously introduced Vreman models and the MSM are shown. It is observable that the profiles, although obtained using the same wall boundary condition, vary when different SGS models are applied, underscoring that the effect of wall boundary conditions is intertwined with the choice of the SGS model.} For instance, using the Vreman model with $c=0.07$, the velocity results obtained under different wall boundary conditions agree well with each other. {Conversely, in the cases of the MSM and the Vreman model with $c=0.025$, the velocity predictions within the attached-flow region are virtually identical, but there is a noticeable variation in the velocity prediction within the separation bubble when the equilibrium and non-equilibrium $\nu_{t,w}$ boundary conditions are employed. Specifically, these two boundary conditions yield a further downstream reattachment location and a more extended separation bubble.} Moreover, the results from these two Vreman models, each with a different model constant $c$, further demonstrate that the performance of the Vreman model is sensitive to this constant.

\begin{figure}
\centering
\begin{subfigure}{0.325\textwidth}
\subcaption{~~~~~~~~~~~~~~~~~~~~~~~~~~~~~~~~~~~~~~~~~~~~~~}
\vspace{-1pt}
{\psfrag{b}[][]{{$x_2/L$}}
\psfrag{a}[][]{{$\left<u_1 \right>/U_{\infty}$}}
\includegraphics[width=\textwidth,trim={2 2.1cm 8cm 0.1cm},clip]{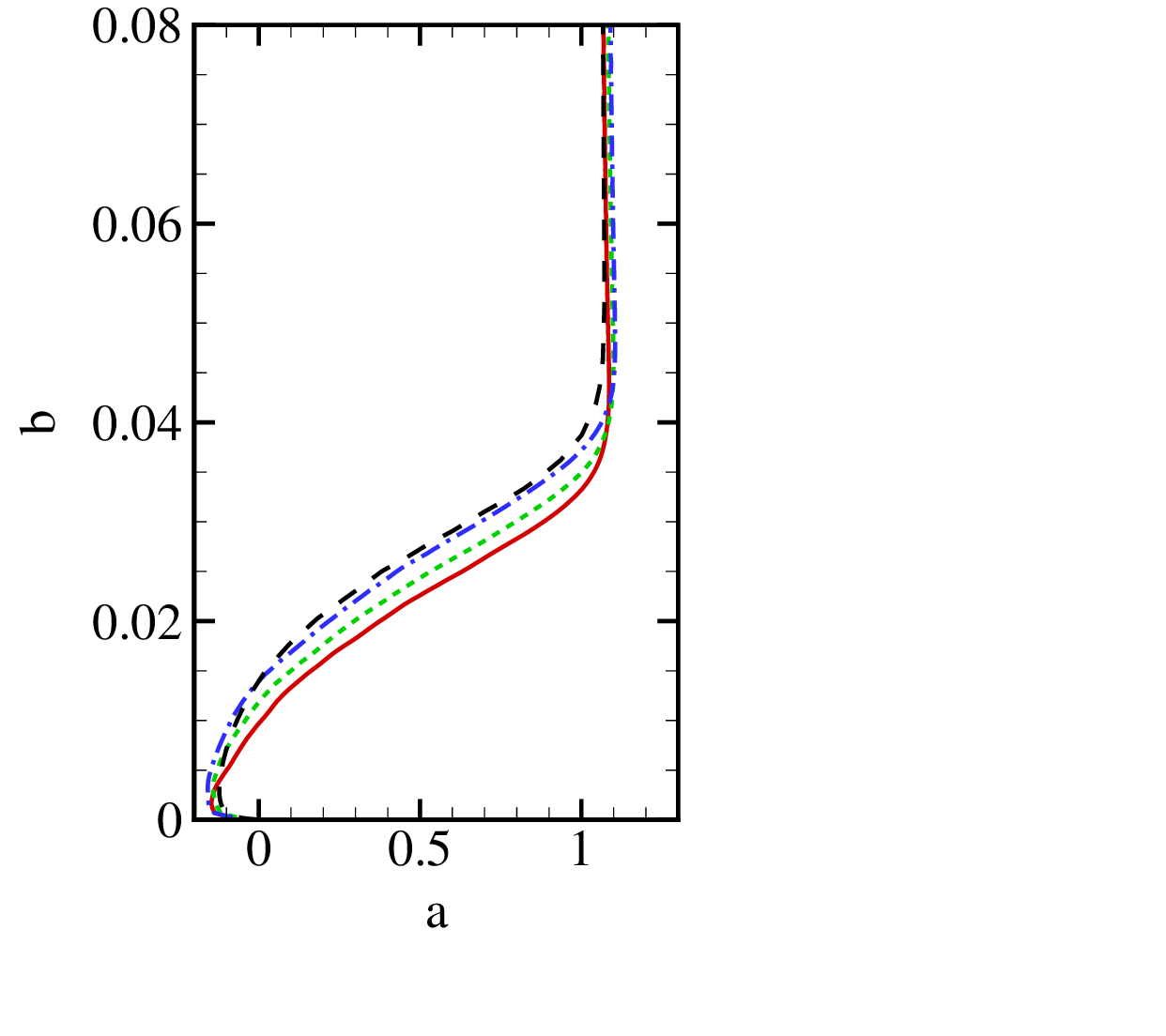}
}
\end{subfigure}
\begin{subfigure}{0.325\textwidth}
\subcaption{~~~~~~~~~~~~~~~~~~~~~~~~~~~~~~~~~~~~~~~~~~~~~~}
\vspace{-1pt}
{\psfrag{b}[][]{{$x_2/L$}}
\psfrag{a}[][]{{$\left<u_1 \right>/U_{\infty}$}}
\includegraphics[width=\textwidth,trim={2 2.1cm 8cm 0.1cm},clip]{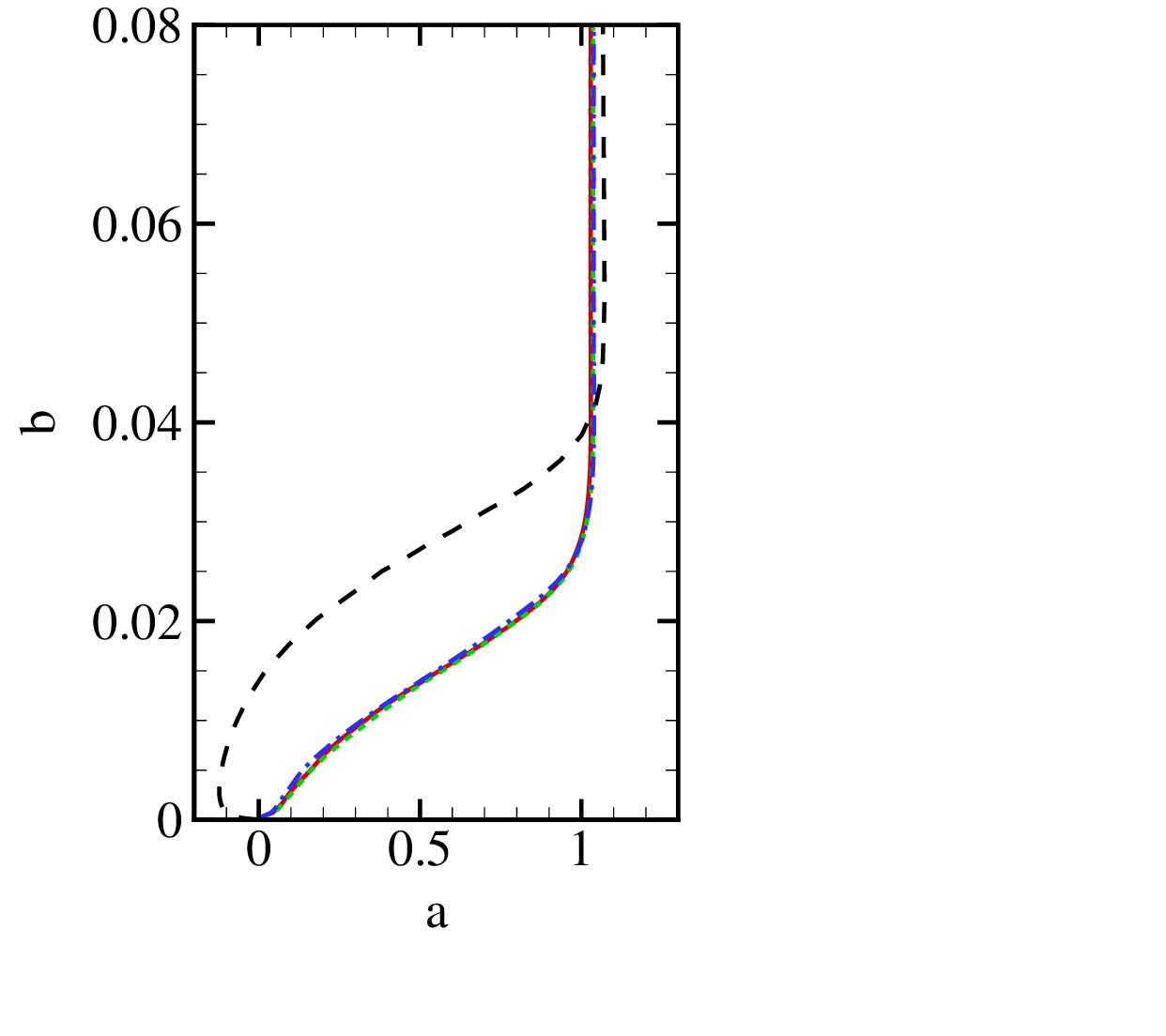}
}
\end{subfigure}
\begin{subfigure}{0.325\textwidth}
\subcaption{~~~~~~~~~~~~~~~~~~~~~~~~~~~~~~~~~~~~~~~~~~~~~~}
\vspace{-1pt}
{\psfrag{b}[][]{{$x_2/L$}}
\psfrag{a}[][]{{$\left<u_1 \right>/U_{\infty}$}}
\includegraphics[width=\textwidth,trim={2 2.1cm 8cm 0.1cm},clip]{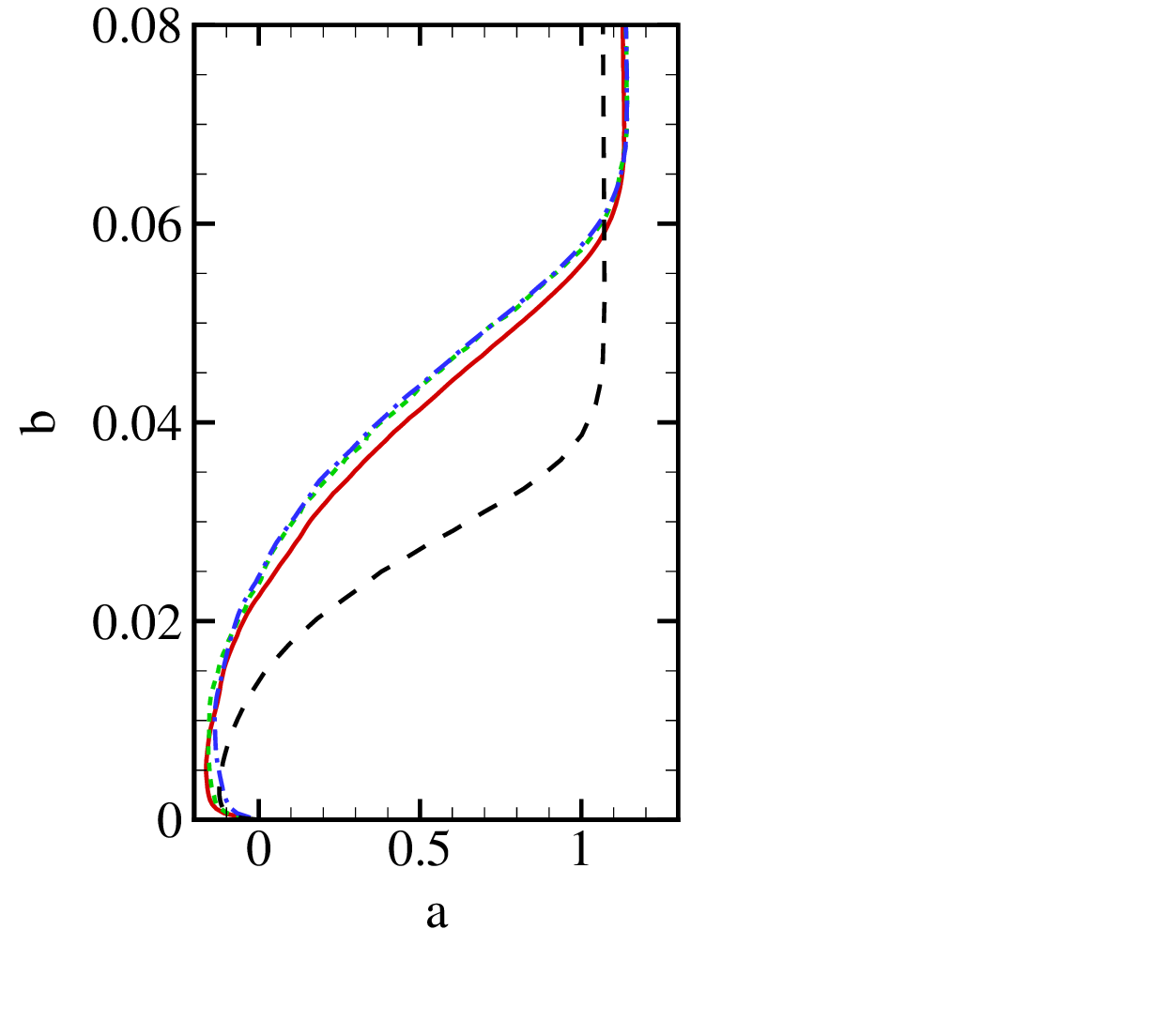}
}
\end{subfigure}
\vskip -0.4cm
\caption{{The mean streamwise velocity profiles at $x/L=0.2$ from the baseline-mesh simulations using (a) the Vreman model ($c=0.025$), (b) the Vreman model ($c=0.07$), and (c) the MSM with different wall boundary conditions. Lines indicate {\redsolid}, WMLES with the velocity Neumann boundary condition on the bottom wall; {\greendashed}, WMLES with the equilibrium $\nu_{t,w}$ boundary condition on the bottom wall; {\bluedashdotted}, WMLES with the non-equilibrium $\nu_{t,w}$ boundary condition on the bottom wall; {\blackdashed}, DNS~\citep{uzun2022high}.}}
\label{streamwise_cut8_base_BCs}
\end{figure}

In Fig.~\ref{streamwise_cut8_base_BCs}, the mean streamwise velocity profiles at $x/L=0.2$, a location expected to be within the separation bubble, are exhibited alongside the DNS results for reference. {From the comparison of velocity profiles, similar conclusions can be drawn. Specifically, employing the non-equilibrium $\nu_{t,w}$ boundary condition, the simulation using the Vreman model with $c=0.025$ yields an enhanced prediction of the streamwise velocity at this location, aligning more closely with the DNS results.} Additionally, similar behavior can be observed in the results of mean wall pressure, although these are omitted here for conciseness. Overall, these test simulations suggest that while the wall boundary condition can enhance the prediction of flow separation, it has a less dominant effect on the simulation compared to the SGS model.

\subsection{Effect of mesh anisotropy}

\begin{figure}
\centering
{\psfrag{a}[][]{{$\Delta_c/L\times10^3$}}
\psfrag{b}[][]{{$L_s/L$}}\includegraphics[width=.75\textwidth,trim={0.2cm 6.2cm 0.2cm 1.5cm},clip]{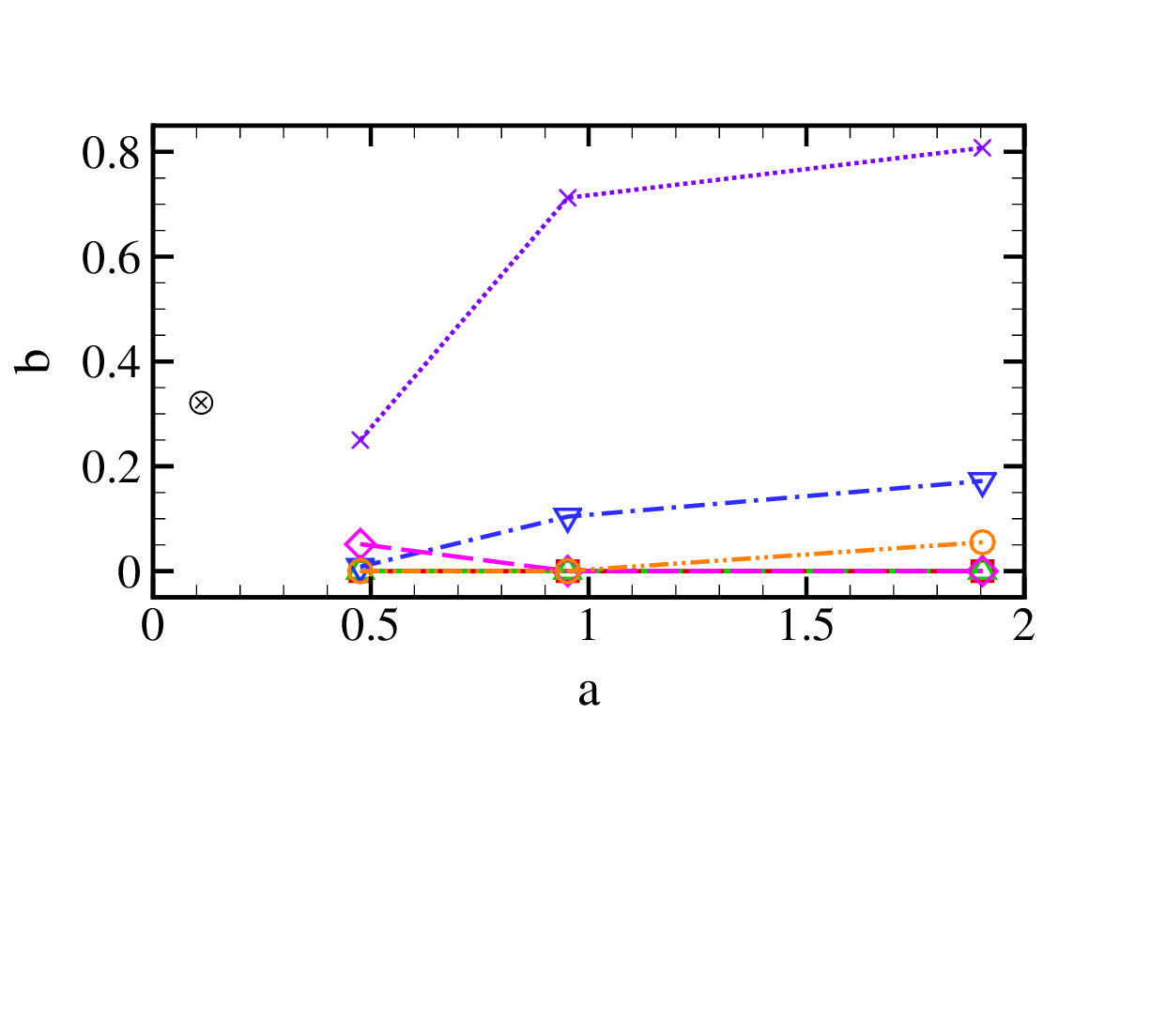}}
\vspace{0.1cm}
\caption{{The horizontal length of separation bubble ($L_s/L$) from the simulations of Group 3. Lines indicate {\redsolid} with \protect\marksymbol{square}{red}, WMLES without an SGS model; {\greendashed} with \protect\marksymbol{triangle}{green}, WMLES using the DSM; {\bluedashdotted} with \protect\marksymbol{triangle}{blue,style={rotate=180}}, WMLES using the AMD model; {\magentadashed} with \protect\marksymbol{diamond}{magenta}, WMLES using the Vreman model ($c=0.025$); {\orangedashdotdot} with  \protect\marksymbol{o}{orange}, WMLES using the Vreman model ($c=0.07$); {\pupdotdotted} with \textcolor{purp2}{$\times$}, WMLES using the MSM. The symbol \textcolor{black}{$\otimes$} represents the DNS results where $\Delta_c/L\times 10^3=0.11$ and $L_s/L=0.32$.}}
\label{Bump_size_SGS_aniso}
\end{figure}

\begin{figure}
\centering
\begin{subfigure}{0.495\textwidth}
\subcaption{~~~~~~~~~~~~~~~~~~~~~~~~~~~~~~~~~~~~~~~~~~~~~~~~~~~~~~~~~~~~~~}
\vspace{-1pt}
{\psfrag{a}[][]{{$x/L$}}
\psfrag{b}[][]{{$C_p$}}
\includegraphics[width=\textwidth,trim={1.5 6.9cm 1.6cm 1.2cm},clip]{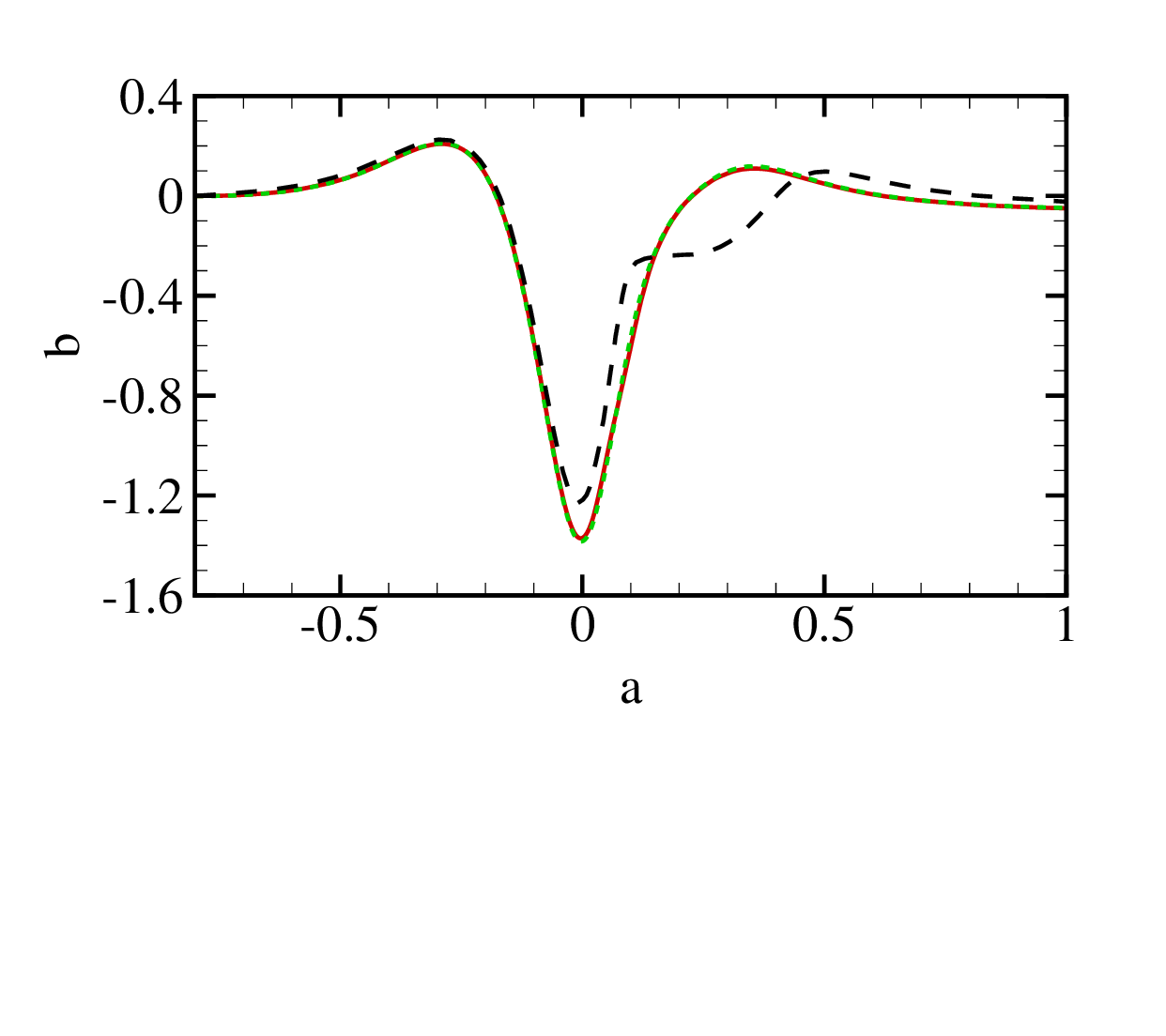}
}
\end{subfigure}
\begin{subfigure}{0.495\textwidth}
\subcaption{~~~~~~~~~~~~~~~~~~~~~~~~~~~~~~~~~~~~~~~~~~~~~~~~~~~~~~~~~~~~~~}
\vspace{-1pt}
{\psfrag{a}[][]{{$x/L$}}
\psfrag{b}[][]{{$C_p$}}
\includegraphics[width=\textwidth,trim={1.5 6.9cm 1.6cm 1.2cm},clip]{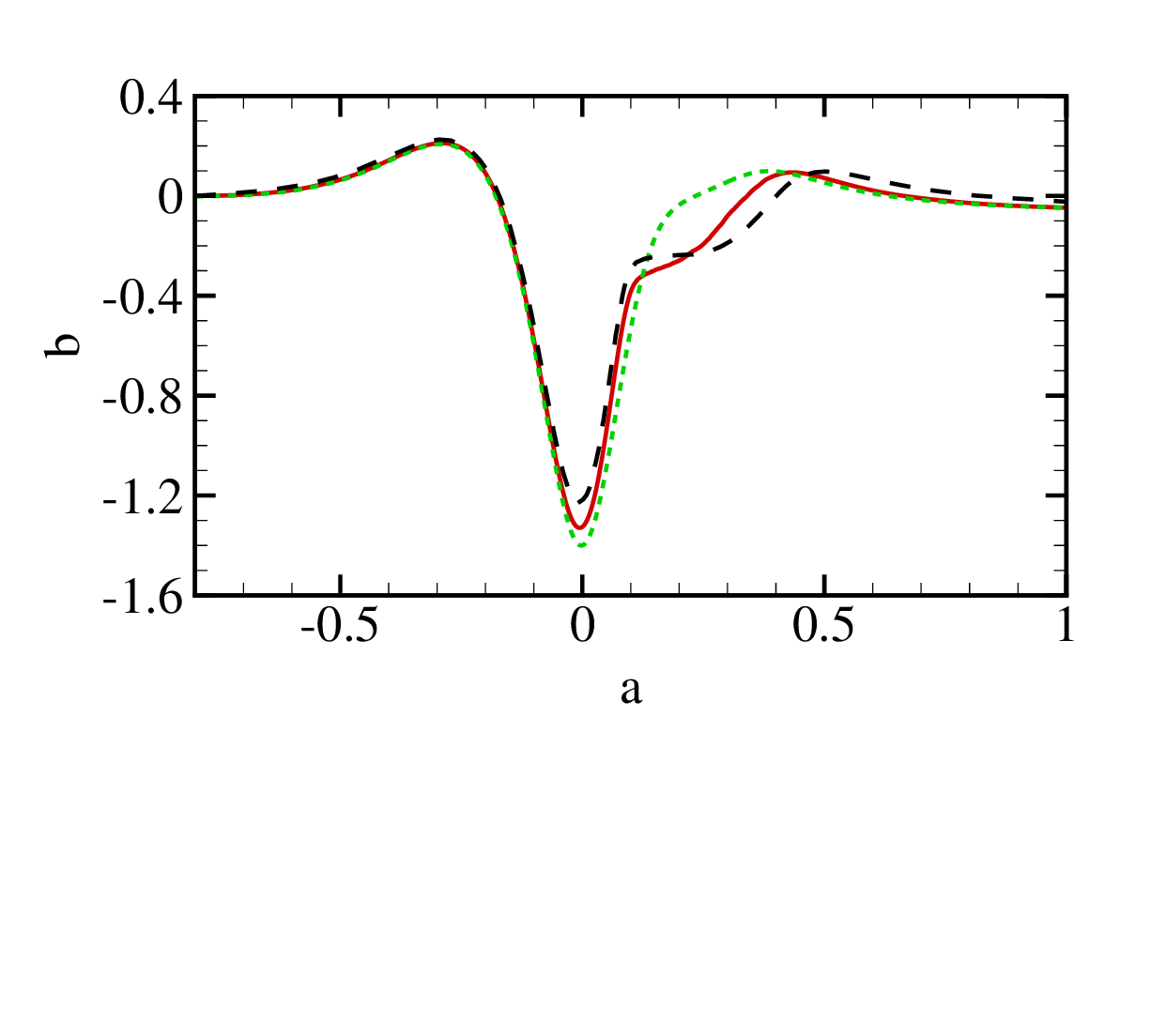}
}
\end{subfigure}
\begin{subfigure}{0.495\textwidth}
\subcaption{~~~~~~~~~~~~~~~~~~~~~~~~~~~~~~~~~~~~~~~~~~~~~~~~~~~~~~~~~~~~~~}
\vspace{-1pt}
{\psfrag{a}[][]{{$x/L$}}
\psfrag{b}[][]{{$C_p$}}
\includegraphics[width=\textwidth,trim={1.5 6.9cm 1.6cm 1.2cm},clip]{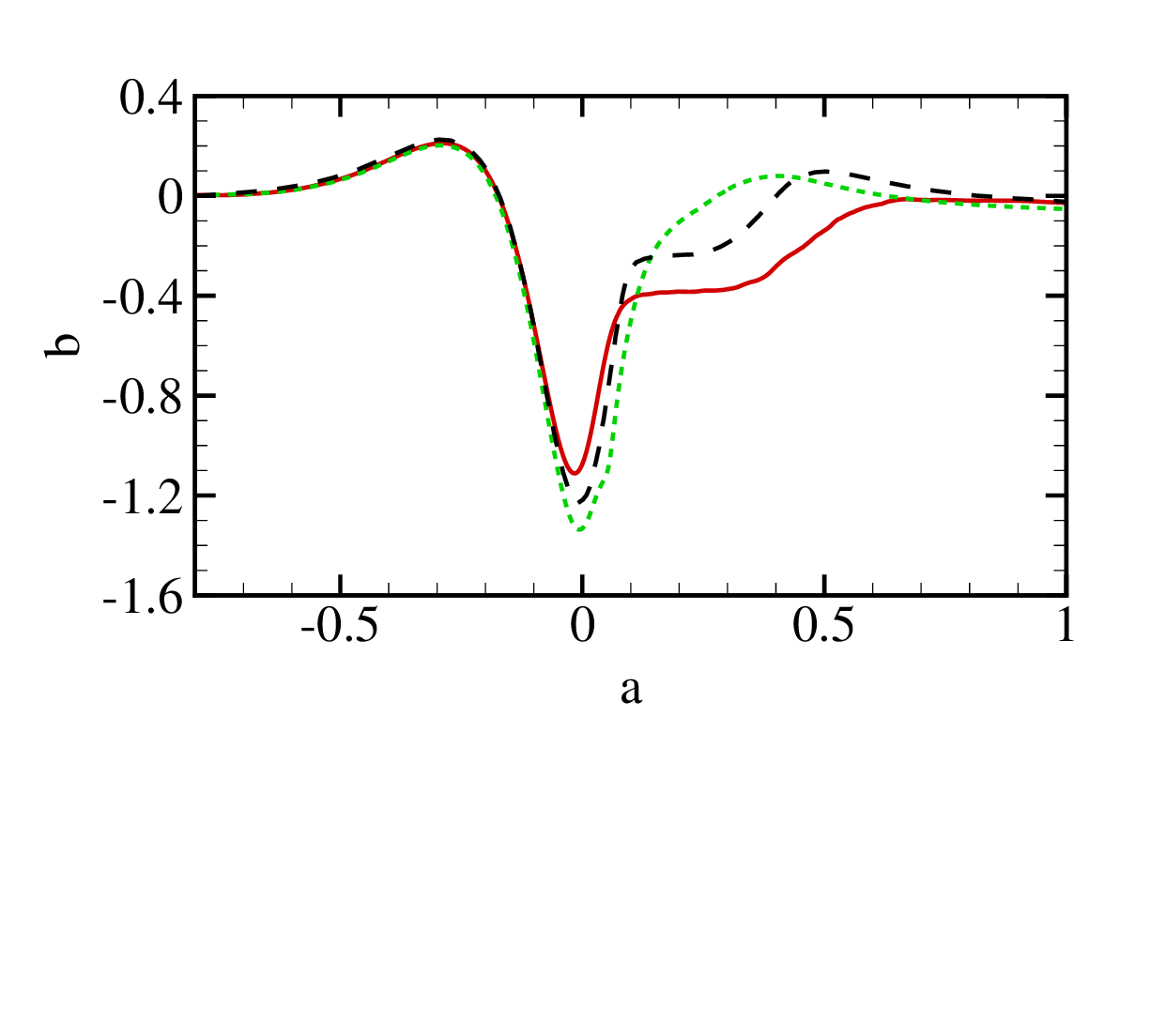}
}
\end{subfigure}
\vskip -0.2cm
\caption{{The mean pressure coefficient $C_p$ on the bottom wall from the baseline-mesh simulations using (a) the AMD model, (b) the Vreman model ($c=0.025$), and (c) the MSM with different types of mesh. Lines indicate {\redsolid}, WMLES with the isotropic mesh; {\greendashed}, WMLES with the anisotropic mesh; {\blackdashed}, DNS~\citep{uzun2022high}.}}
\label{Cp_base_aniso}
\end{figure}

The examination of the mesh anisotropy's influence on flow prediction necessitates a comparison between simulations using meshes with isotropic and anisotropic cells. It is worth noting that these simulations solely utilize the velocity Neumann boundary condition based on the mean wall-shear stress from the DNS \citep{uzun2022high} at the bottom wall. Figure~\ref{Bump_size_SGS_aniso} illustrates the measured separation bubble size $L_s/L$ derived from simulations with an anisotropic mesh as shown in Group 3 of Table~\ref{tab:table2}. This size is assessed through the mean streamwise velocity at the first off-wall cell center. {When compared with the results from isotropic-mesh simulations in Fig.~\ref{Bump_size_SGS}, a degradation in the prediction of the separation bubble by using an anisotropic mesh is evident, especially in case employing the Vreman model ($c=0.025$). Furthermore, in the cases using the MSM, the size of the separation bubble is enlarged in simulations with the coarsest and coarse anisotropic meshes. However, the size is significantly reduced in the baseline-mesh simulation, disrupting the consistency observed in the aforementioned simulations with isotropic meshes. Figure~\ref{Cp_base_aniso} compares the distributions of $C_p$ on the bottom wall from both isotropic and anisotropic baseline-mesh simulations using the AMD model, the Vreman model ($c=0.025$), and the MSM to the DNS results.} A reasonable agreement between these simulations and the DNS is observed on the windward side of the bump and the flat wall. On the leeward side of the bump, the prediction of $C_p$ from simulations using the AMD model is barely affected by the mesh type. {In contrast, the prediction from simulations using the Vreman model ($c=0.025$) and the MSM is negatively influenced by the adoption of an anisotropic mesh.}

\begin{figure}
\centering
\begin{subfigure}{0.495\textwidth}
\subcaption{~~~~~~~~~~~~~~~~~~~~~~~~~~~~~~~~~~~~~~~~~~~~~~~~~~~~~~~~~~~~~~~~~}
\vspace{-1pt}
{\psfrag{a}[][]{{$x/L$}}
\psfrag{b}[][]{{$\left<u_1 \right>/U_{\infty}$}}
\includegraphics[width=\textwidth,trim={1.5 6.6cm 1.6cm 1.3cm},clip]{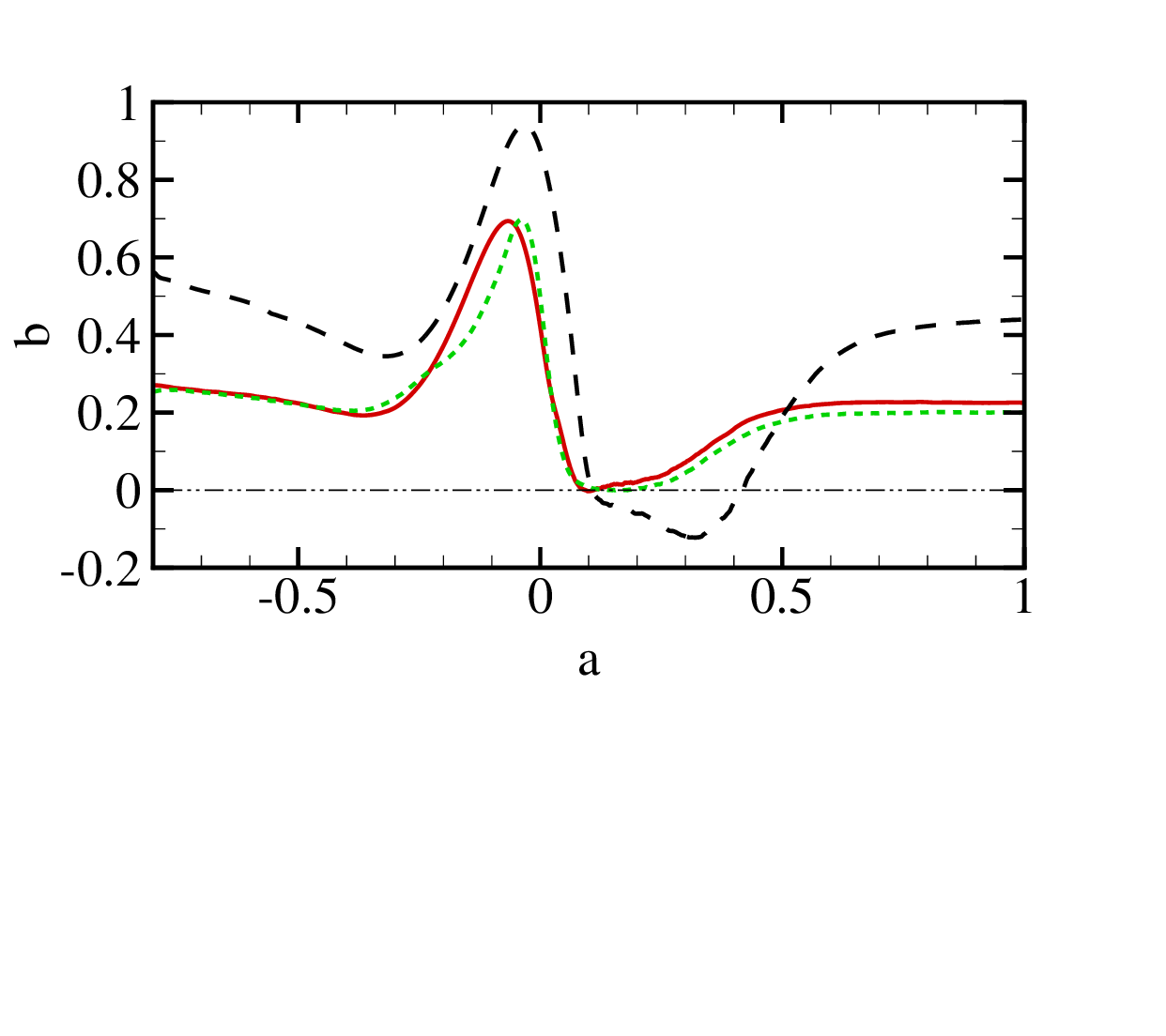}
}
\end{subfigure}
\begin{subfigure}{0.495\textwidth}
\subcaption{~~~~~~~~~~~~~~~~~~~~~~~~~~~~~~~~~~~~~~~~~~~~~~~~~~~~~~~~~~~~~~~~~}
\vspace{-1pt}
{\psfrag{a}[][]{{$x/L$}}
\psfrag{b}[][]{{$\left<u_1 \right>/U_{\infty}$}}
\includegraphics[width=\textwidth,trim={1.5 6.6cm 1.6cm 1.3cm},clip]{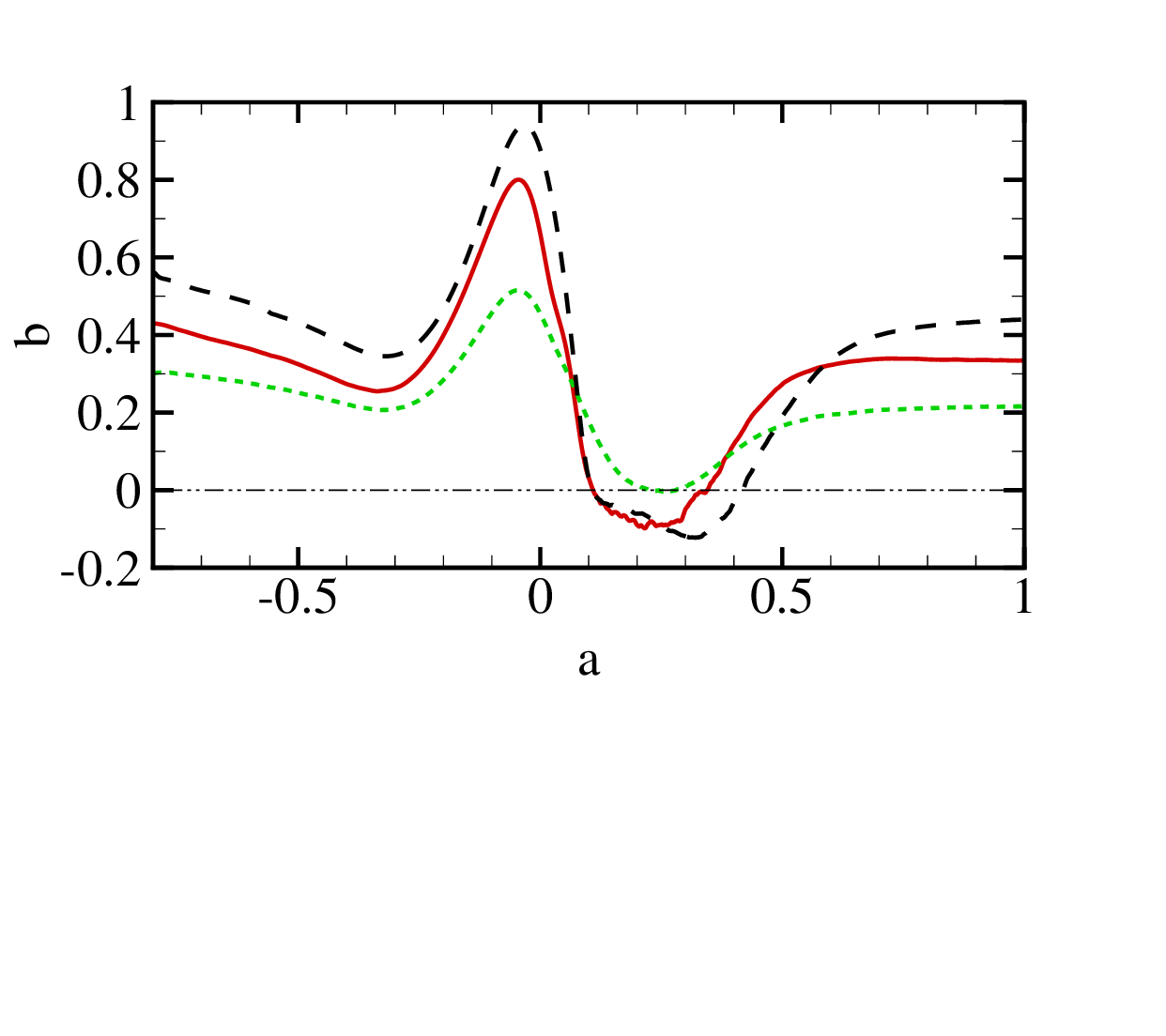}
}
\end{subfigure}
\begin{subfigure}{0.495\textwidth}
\subcaption{~~~~~~~~~~~~~~~~~~~~~~~~~~~~~~~~~~~~~~~~~~~~~~~~~~~~~~~~~~~~~~~~~}
\vspace{-1pt}
{\psfrag{a}[][]{{$x/L$}}
\psfrag{b}[][]{{$\left<u_1 \right>/U_{\infty}$}}
\includegraphics[width=\textwidth,trim={1.5 6.6cm 1.6cm 1.3cm},clip]{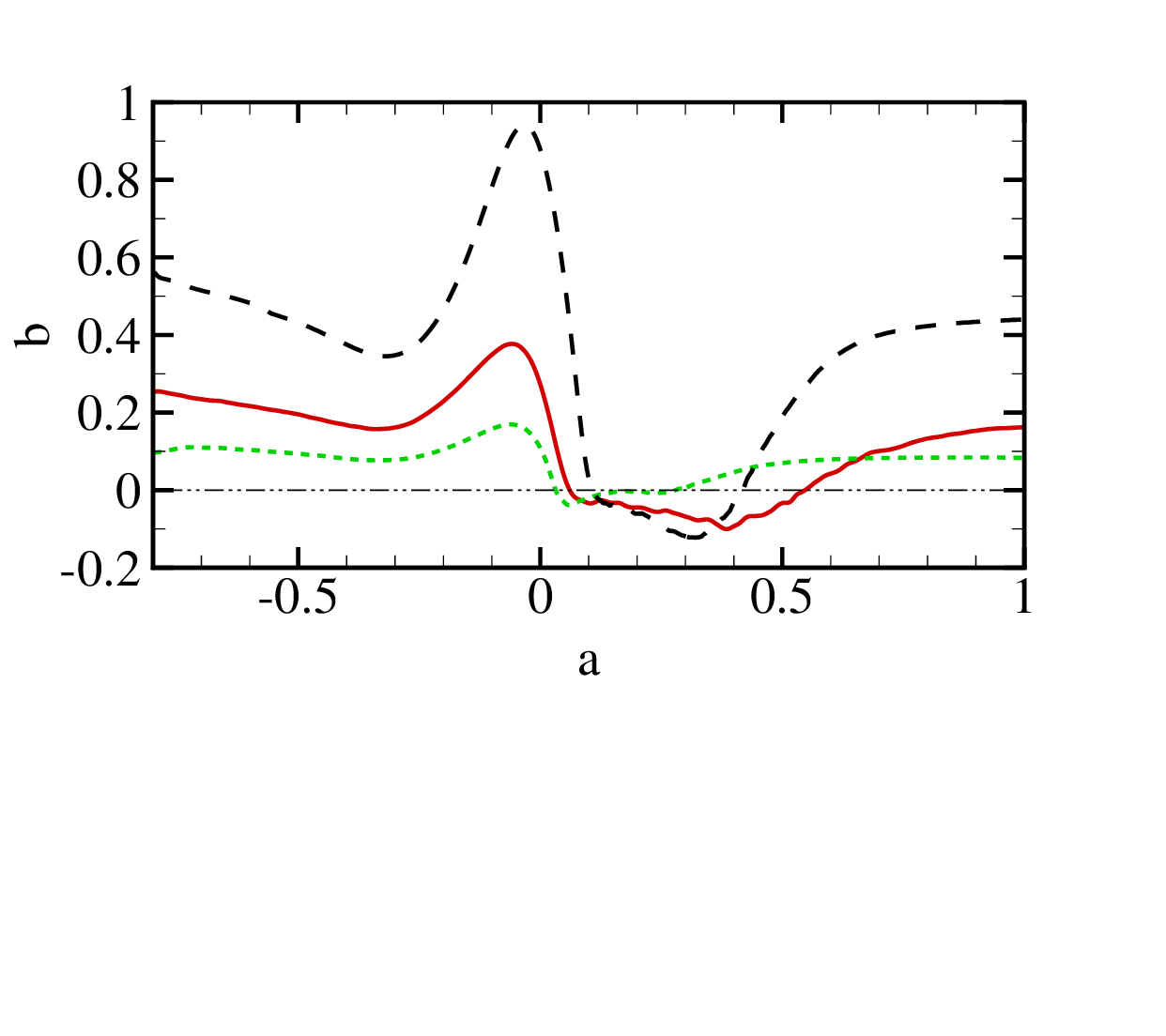}
}
\end{subfigure}
\vskip -0.4cm
\caption{{The profiles of mean streamwise velocity at $x_2/L=2.38\times 10^{-4}$ from the baseline-mesh simulations using (a) the AMD model, (b) the Vreman model ($c=0.025$), and (c) the MSM with different types of mesh, alongside a reference profile from the DNS \citep{uzun2022high} at the same wall-normal location. Lines indicate {\redsolid}, WMLES with the isotropic mesh; {\greendashed}, WMLES with the anisotropic mesh; {\blackdashed}, DNS~\citep{uzun2022high}; {\blackdashdotdot}, $\left<u_1 \right>/U_{\infty}=0$.}}
\label{wallnormal_cut_base_aniso}
\end{figure}

\begin{figure}
\centering
\begin{subfigure}{0.325\textwidth}
\subcaption{~~~~~~~~~~~~~~~~~~~~~~~~~~~~~~~~~~~~~~~~~~~~~~}
\vspace{-1pt}
{\psfrag{b}[][]{{$x_2/L$}}
\psfrag{a}[][]{{$\left<u_1 \right>/U_{\infty}$}}
\includegraphics[width=\textwidth,trim={2 1.6cm 8cm 0.3cm},clip]{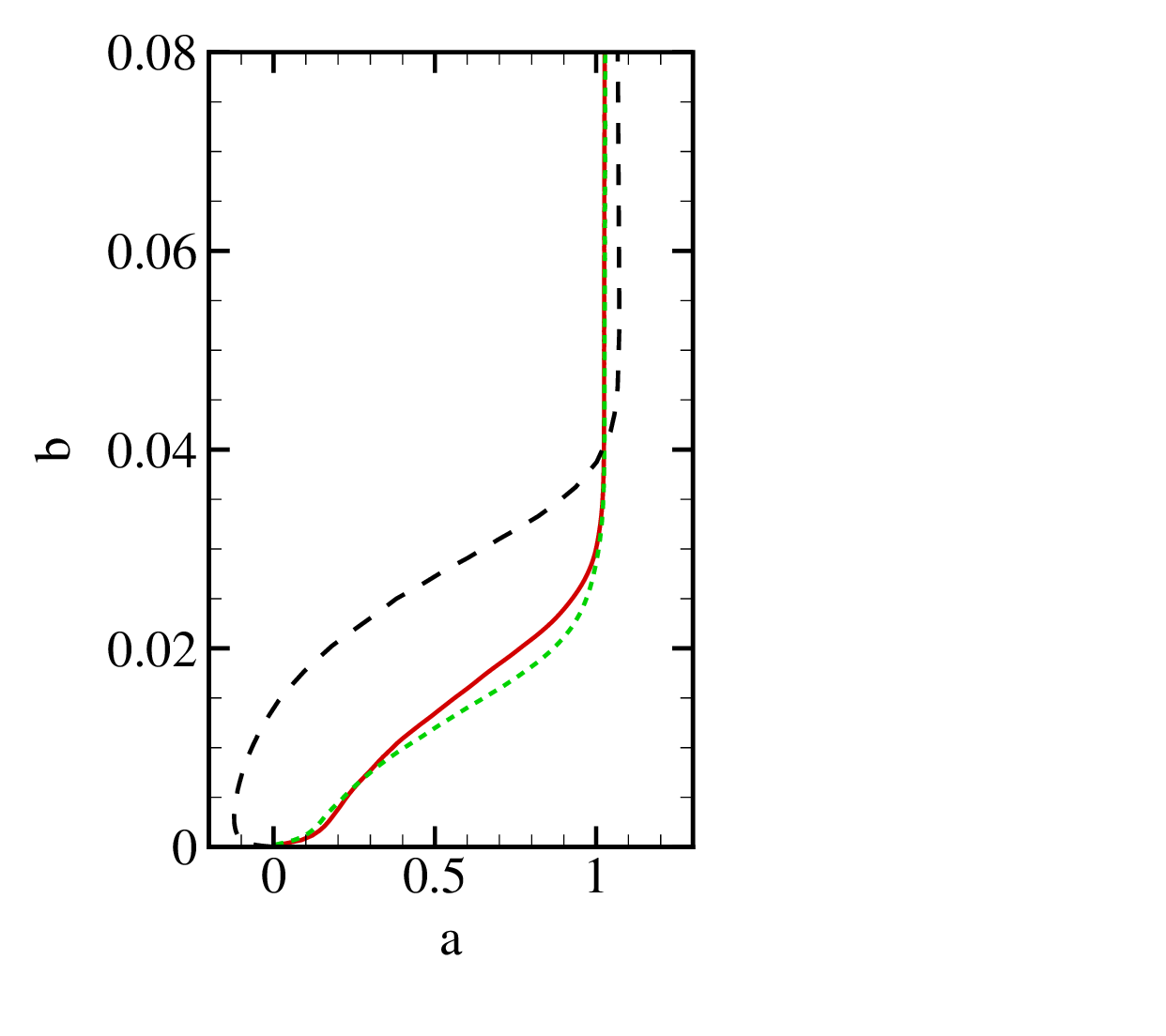}
}
\end{subfigure}
\begin{subfigure}{0.325\textwidth}
\subcaption{~~~~~~~~~~~~~~~~~~~~~~~~~~~~~~~~~~~~~~~~~~~~~~}
\vspace{-1pt}
{\psfrag{b}[][]{{$x_2/L$}}
\psfrag{a}[][]{{$\left<u_1 \right>/U_{\infty}$}}
\includegraphics[width=\textwidth,trim={2 1.6cm 8cm 0.3cm},clip]{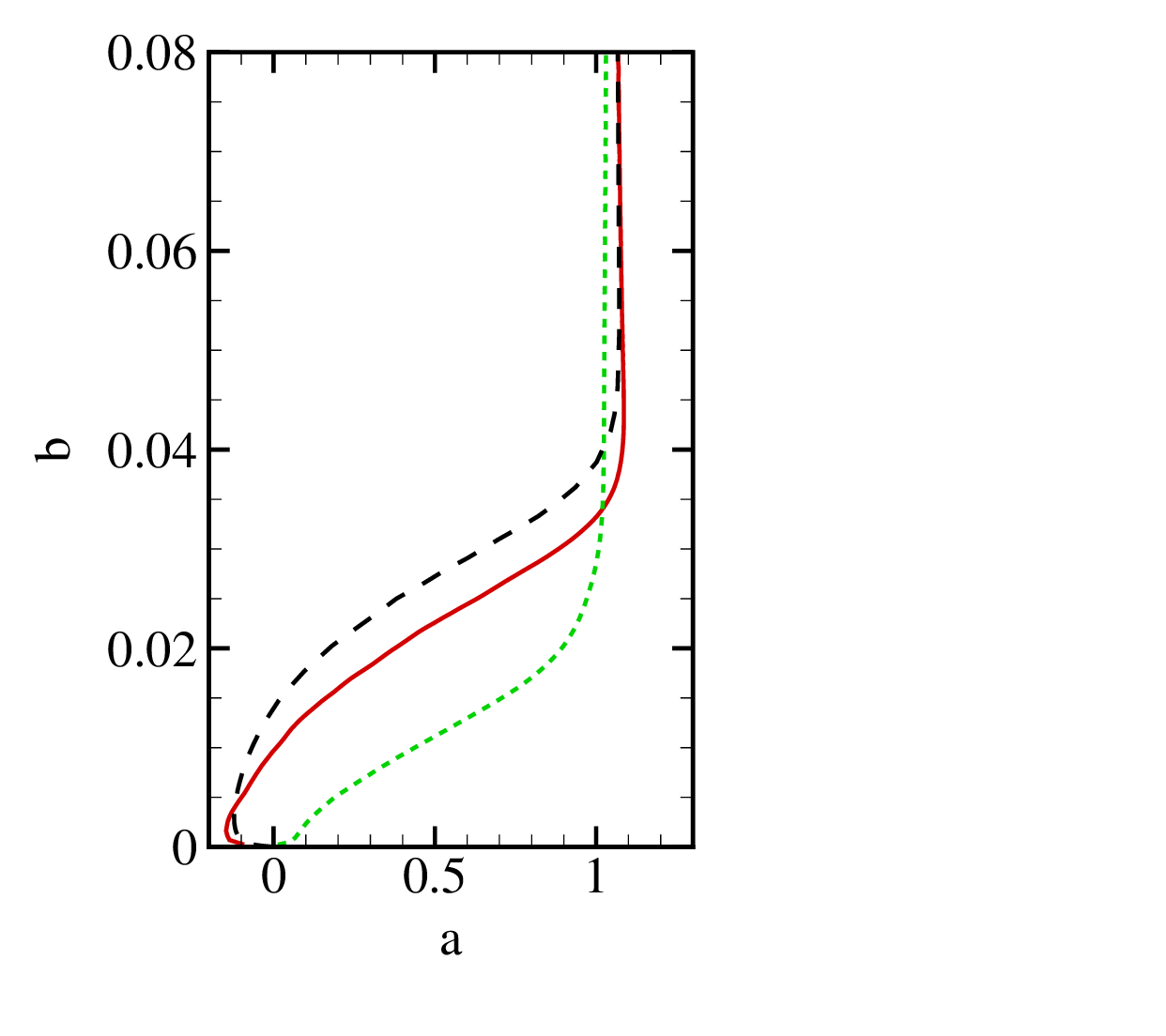}
}
\end{subfigure}
\begin{subfigure}{0.325\textwidth}
\subcaption{~~~~~~~~~~~~~~~~~~~~~~~~~~~~~~~~~~~~~~~~~~~~~~}
\vspace{-1pt}
{\psfrag{b}[][]{{$x_2/L$}}
\psfrag{a}[][]{{$\left<u_1 \right>/U_{\infty}$}}
\includegraphics[width=\textwidth,trim={2 1.6cm 8cm 0.3cm},clip]{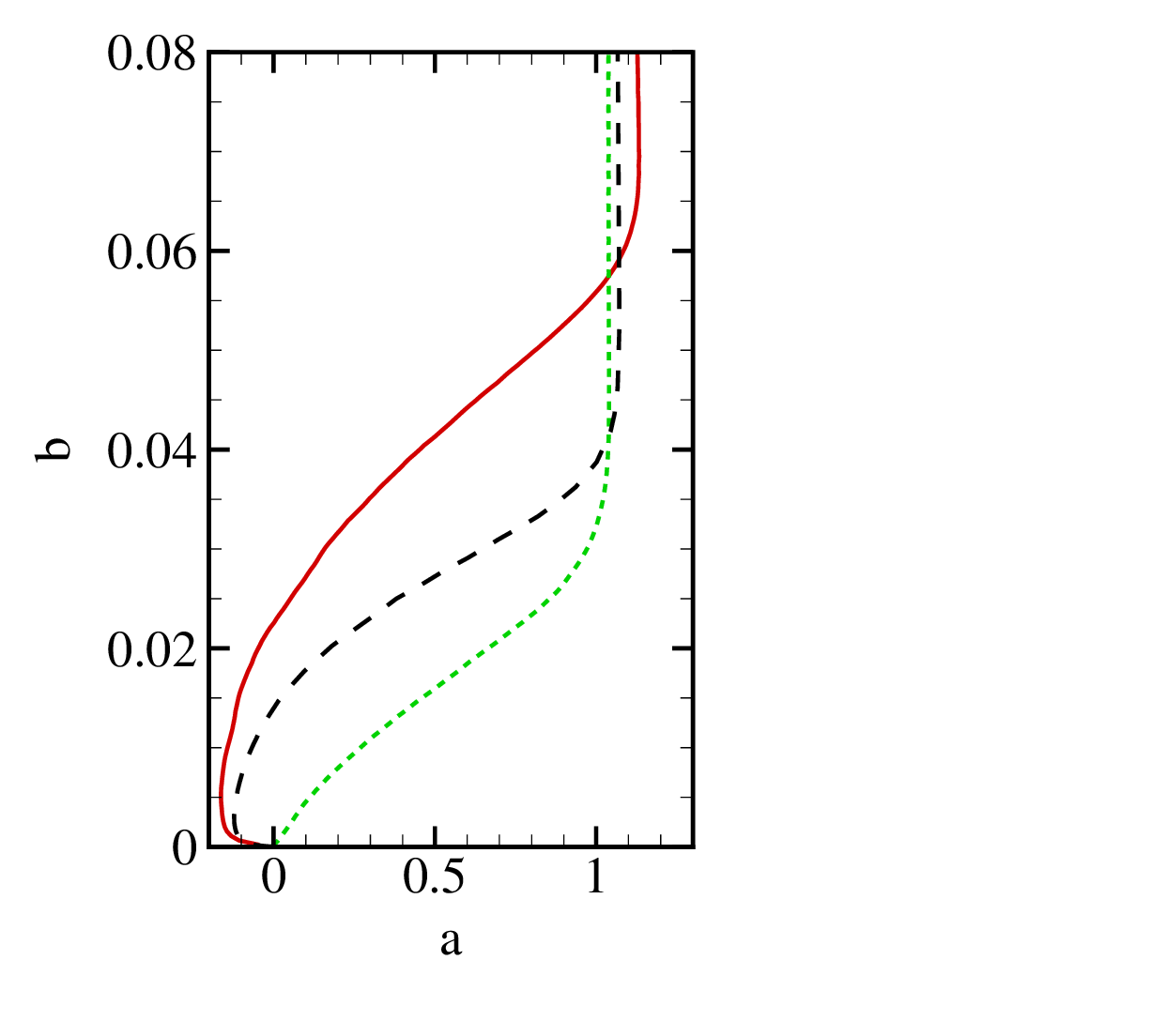}
}
\end{subfigure}
\vskip -0.3cm
\caption{{The profiles of mean streamwise velocity at $x/L=0.2$ from the baseline-mesh simulations using (a) the AMD model, (b) the Vreman model ($c=0.025$), and (c) the MSM with different types of mesh. Lines indicate {\redsolid}, WMLES with the isotropic mesh; {\greendashed}, WMLES with the anisotropic mesh; {\blackdashed}, DNS~\citep{uzun2022high}.}}
\label{streamwise_cut8_base_aniso}
\end{figure}

{Figure~\ref{wallnormal_cut_base_aniso} illustrates the comparison of the mean streamwise velocity at a wall-normal location, $x_2/L=2.38\times 10^{-4}$, corresponding to the center of the first off-wall cell in the baseline isotropic mesh.} The baseline-mesh simulations using the AMD model reveal only minor variations in response to changes in mesh type. {However, for the simulations employing the Vreman model ($c=0.025$) and the MSM, the velocity prediction changes dramatically across the entire domain.} When an anisotropic mesh is used, the separation bubble size substantially decreases. Figure~\ref{streamwise_cut8_base_aniso} further explores this by displaying the comparison of mean streamwise velocity profiles at $x/L=0.2$, including the DNS results for reference. {At this location, pronounced variations in velocity prediction are observed in simulations utilizing both the Vreman model and the MSM, with the previously separated TBL becoming attached or nearly attached when the anisotropic mesh is employed.} In contrast, the simulations conducted with the AMD model continue to exhibit minimal changes in the velocity profile, regardless of the mesh type used. Similar patterns are evident in simulations using coarser meshes; however, for the sake of brevity, those results are not presented here.

From the discussions above, it is evident that the impact of the mesh anisotropy on the simulation results is strongly contingent on the specific SGS model utilized. {It should be noted that, as in many LES frameworks, our numerical method implicitly applies low-pass spatial filtering to the incompressible Navier-Stokes equations. A conventional approach to estimating the filter width is to use the geometric mean of the grid size, derived from the volume of the mesh cell. This method is utilized in the SGS models under investigation, with the exception of the AMD model. This approximation is generally suitable for isotropic mesh cells. However, in the case of anisotropic mesh cells, this approach to estimating the filter width may introduce errors due to the SGS model's sensitivity to filter width variations. As a result, when employing these SGS models, the impact of mesh type becomes more pronounced and significant. In contrast, the AMD model, which is an optimized minimum-dissipation model specifically designed for anisotropic meshes, mitigates the potential error from filter width approximation, thus reducing the influence of mesh type.} This highlights the importance of incorporating the effect of mesh-cell properties when designing a comprehensive SGS model that can accurately capture turbulent flow physics with the use of an anisotropic mesh.

\section{Conclusions}\label{sec:4}

A parametric analysis to assess the sensitivities of WMLES for separated turbulent flow has been conducted. The study focuses on the flow over a two-dimensional Gaussian-shaped bump and specifically delves into the effects of SGS models, mesh resolution, wall boundary conditions and mesh anisotropy. The core findings of the investigation are summarized below:

1. The simulations display a clear dependence on SGS model. The influence of SGS model is profound, not just for the near-wall flow field within the attached-flow region, but crucially on the accurate capture of the separation bubble. 

{2. The simulations approach grid convergence with a fine mesh that features WRLES-like resolutions in the streamwise and spanwise directions. Beyond this limit, however, the simulation results demonstrate intricate sensitivities to mesh resolution.} The consistency of results convergence with mesh refinement is significantly influenced by the particular SGS model employed. For instance, non-monotonic convergence in capturing the separation bubble is discernible with certain SGS models like the AMD model and the Vreman model ($c=0.07$), while this behavior is less observable within the attached-flow region. {Furthermore, simulations using the MSM demonstrate more consistent capture of the separation bubble across different mesh resolutions, suggesting that accounting for the anisotropy of SGS stress may mitigate the non-monotonic convergence issue.}

3. The flow simulation's dependence on the SGS model remains significant, even when different {idealized} wall boundary conditions are employed. {While these wall boundary conditions can noticeably affect the simulations of separated turbulent flow, the extent of their influence varies with the employed SGS model and is mainly limited to the skin friction on the wall.}

4. Anisotropic mesh cells can adversely affect the prediction of separated turbulent flow. Moreover, the impact of the mesh anisotropy is contingent on the SGS model employed. Specifically, for the AMD model, which is optimized to mitigate the effects of mesh anisotropy, its influence on flow predictions is relatively weak. 

5. The simulations utilizing isotropic meshes indicate that the Vreman model with $c=0.025$ offers a comparatively robust performance. This model not only accurately predicts the attached TBL and the separation bubble when employing a reasonably resolved mesh, but also demonstrates a monotonic convergence of its simulation predictions toward the DNS results as the mesh is refined. Nonetheless, users should be cautious with the model due to its pronounced sensitivity to the mesh anisotropy.

{However, it is important to note that the simulations in this study utilize boundary conditions designed as idealized wall models, which aim to ensure that the applied wall-shear stress matches the mean values from DNS. Incorrect wall-shear stress could lead to additional errors, making it crucial to employ an appropriate wall model that guarantees accurate wall-shear stress in practical WMLES. Additionally, it is essential to recognize the coupling between the wall model and the SGS model in WMLES, given that the wall model relies on data from the near-wall flow field to predict wall quantities as boundary conditions for the simulation.}

Overall, the analysis has deepened our understanding of the sensitivities associated with separated-flow WMLES. Yet, there remain many open questions that need to be answered through further exploration. The parametric study conducted here emphasizes the impacts of a select set of variables. This leaves room for similar studies that might delve into other factors such as mesh topology, explicit filtering, and Reynolds number. Our current simulations also suggest the importance of a closer examination of the DSM's near-wall behavior in WMLES, including its sensitivities to the applied filtering and the implementation at the off-wall cells.

Given the insights drawn from the present results, there is a pressing need to enhance the robustness of the existing SGS models in WMLES, which encompasses refining the model capability to predict flow separation, quickening and ensuring the monotonicity of convergence and enhancing prediction accuracy with anisotropic meshes. This progress demands a deep understanding of the SGS model behavior at mesh resolutions not originally designed for resolving near-wall turbulence. Furthermore, when developing future SGS models, it is imperative to consider these applications and needs. Recent work in this field, notably the non-Boussinesq SGS model proposed by \citet{agrawal2022non} and the dynamic nonlinear SGS model introduced by \citet{uzun2022dynamic}, exhibit robust capability for separated-flow LES and present promising trajectories for enhanced SGS motion modeling. {In addition, to sidestep the adverse coupling effects between the wall model and the SGS model and enhance the robustness of flow predictions, one avenue could be the development of a unified model for both SGS and wall modeling, as demonstrated by models such as the building-block flow model \citep{arranz2023wall, ling2022wall}.}


\section*{Acknowledgments}\label{sec:5}
This work was supported by National Science Foundation (NSF) grant No.~2152705. Computer time was provided by the Discover project at Pittsburgh Supercomputing Center through allocation PHY230012 from the Advanced Cyberinfrastructure Coordination Ecosystem: Services \& Support (ACCESS) program, which is supported by NSF grants No.~2138259, No.~2138\\286, No.~2138307, No.~2137603, and No.~2138296. The authors sincerely thank Dr.~Ali Uzun and Dr.~Mujeeb Malik for generously sharing their DNS data. We also extend special gratitude to Dr.~Meng Wang for his invaluable help.

\appendix

\section{Influence of TBL inflow condition on the simulations} \label{append}

\begin{figure}
\centering
\begin{subfigure}{0.325\textwidth}
\subcaption{~~~~~~~~~~~~~~~~~~~~~~~~~~~~~~~~~~~~~~~~~~~~~~}
\vspace{-1pt}
{\psfrag{b}[][]{{$x_2/L$}}
\psfrag{a}[][]{{$\left<u_1 \right>/U_{\infty}$}}
\includegraphics[width=\textwidth,trim={2.2 1.8cm 8.2cm 0.4cm},clip]{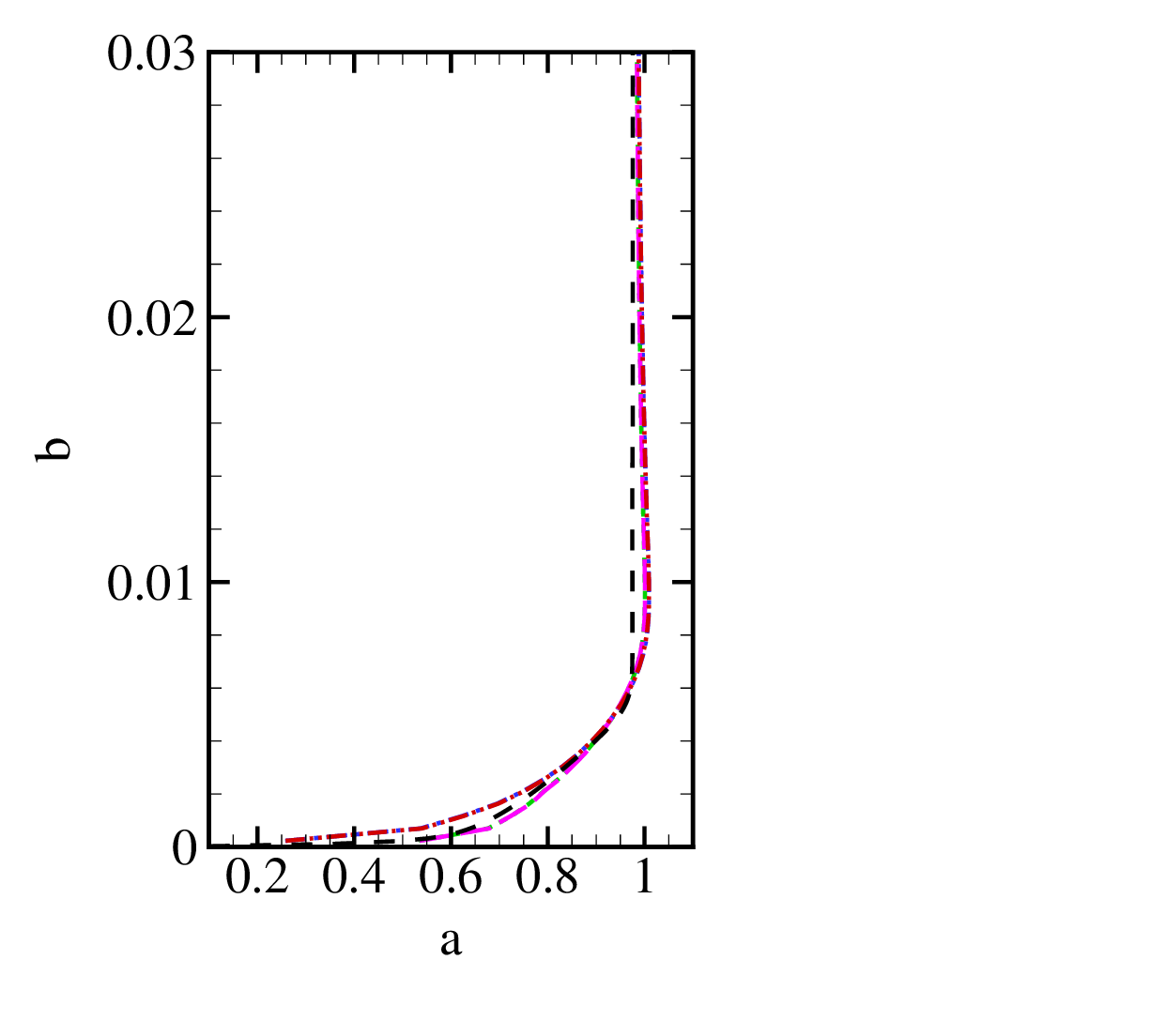}
}
\end{subfigure}
\begin{subfigure}{0.325\textwidth}
\subcaption{~~~~~~~~~~~~~~~~~~~~~~~~~~~~~~~~~~~~~~~~~~~~~~}
\vspace{-1pt}
{\psfrag{b}[][]{{$x_2/L$}}
\psfrag{a}[][]{{$\left<u_1 \right>/U_{\infty}$}}
\includegraphics[width=\textwidth,trim={2.2 1.8cm 8.2cm 0.4cm},clip]{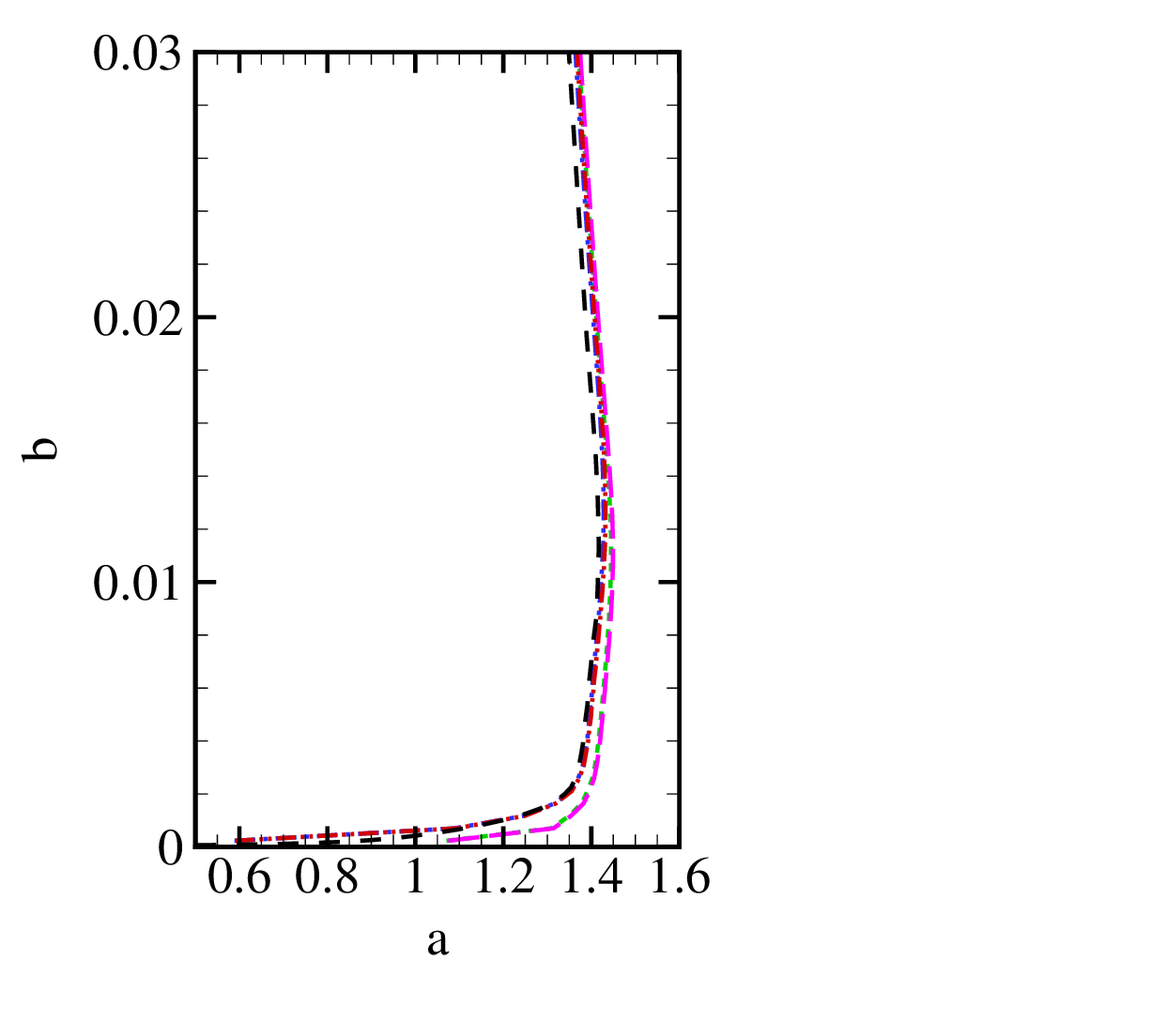}
}
\end{subfigure}
\begin{subfigure}{0.325\textwidth}
\subcaption{~~~~~~~~~~~~~~~~~~~~~~~~~~~~~~~~~~~~~~~~~~~~~~}
\vspace{-1pt}
{\psfrag{b}[][]{{$x_2/L$}}
\psfrag{a}[][]{{$\left<u_1 \right>/U_{\infty}$}}
\includegraphics[width=\textwidth,trim={2.2 1.8cm 8.2cm 0.4cm},clip]{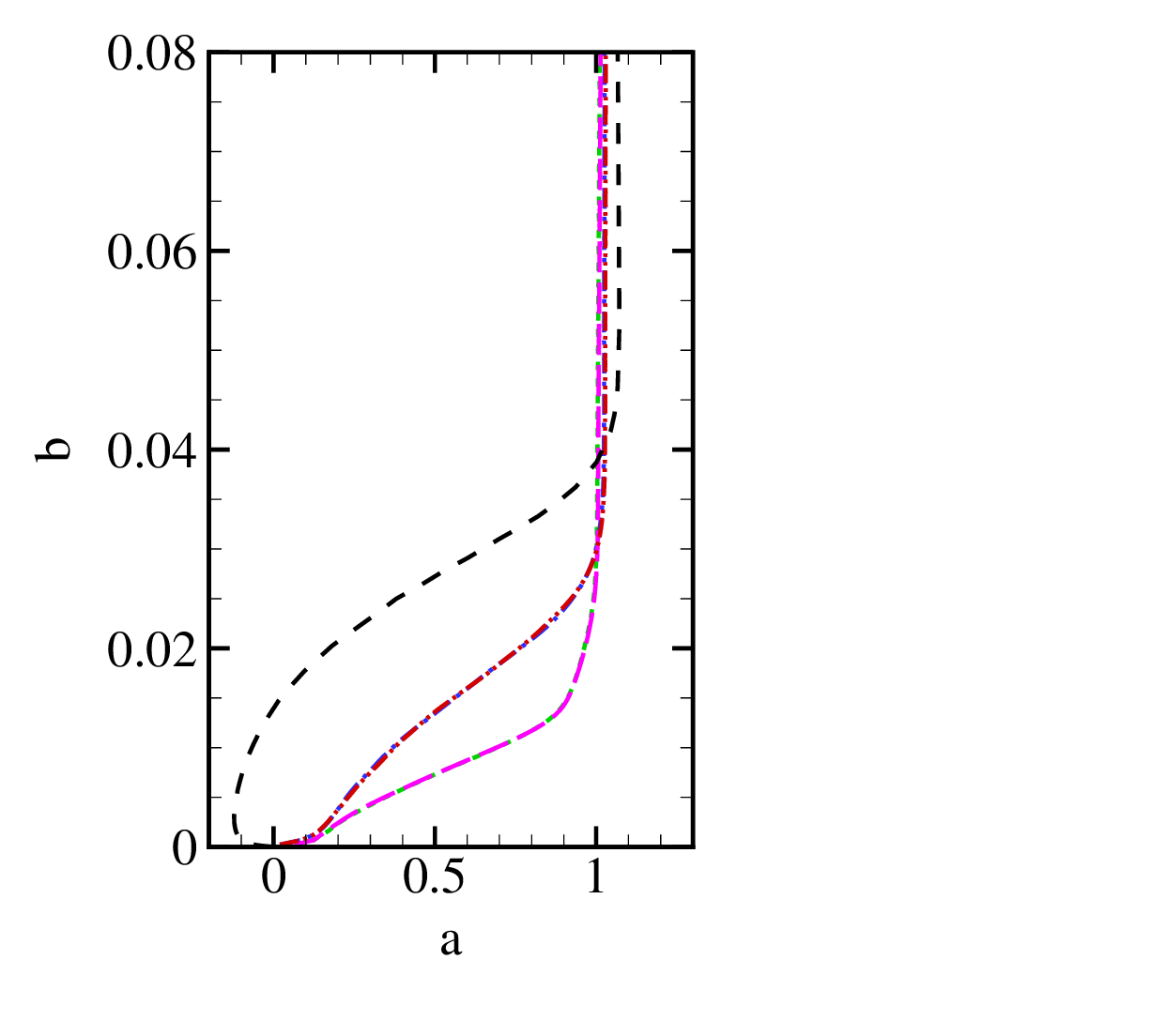}
}
\end{subfigure}
\vskip -0.3cm
\caption{Comparison of the mean streamwise velocity profiles from the Group 1 baseline-mesh simulations using the inflow generated with different SGS models at (a) $x/L=-0.709$, (b) $x/L=-0.025$ and (c) $x/L=0.2$. Lines indicate {\greendashed}, WMLES with the DSM and the inflow generated by the DSM; {\bluedashdotted}, WMLES with the AMD model and the inflow generated by the DSM; {\magentadashed}, WMLES with the DSM and the inflow generated by the AMD model; {\reddashdotdot}, WMLES with the AMD model and the inflow generated by the AMD model; {\blackdashed}, DNS~\citep{uzun2022high}.}
\label{streamwise_cuts158_base_AMDinflow}
\end{figure}

To check the dependence of the SGS model behavior on the TBL inflow, we collect different TBL inflow data and use it to run the simulations of flow over the bump. More specifically, the new TBL inflow is generated from the simulation using the AMD model.  {The new inflow features a boundary layer thickness of $\delta_{in}/L=0.0062$, a momentum thickness Reynolds number of $Re_{\theta} \approx 1126$, and a friction Reynolds number of $Re_\tau \approx 630$, all of which are larger than those associated with the original inflow.} 
Figure~\ref{streamwise_cuts158_base_AMDinflow} provides a comparison of the mean streamwise velocity profiles from the bump-flow simulations with different TBL inflows, extracted at three stations along the $x$ direction. For conciseness, only the results from the baseline-mesh simulations utilizing the velocity Neumann boundary condition with the AMD model and the DSM are shown. The consistency between the results from the simulations with different TBL inflows is strong at all stations. These comparative findings suggest that the influence of the TBL inflow condition on the performance of the SGS models is negligible. 

\begin{thebibliography}{52}
\expandafter\ifx\csname natexlab\endcsname\relax\def\natexlab#1{#1}\fi
\providecommand{\url}[1]{\texttt{#1}}
\providecommand{\href}[2]{#2}
\providecommand{\path}[1]{#1}
\providecommand{\DOIprefix}{doi:}
\providecommand{\ArXivprefix}{arXiv:}
\providecommand{\URLprefix}{URL: }
\providecommand{\Pubmedprefix}{pmid:}
\providecommand{\doi}[1]{\href{http://dx.doi.org/#1}{\path{#1}}}
\providecommand{\Pubmed}[1]{\href{pmid:#1}{\path{#1}}}
\providecommand{\bibinfo}[2]{#2}
\ifx\xfnm\relax \def\xfnm[#1]{\unskip,\space#1}\fi
\bibitem[{Rumsey et~al.(2022)Rumsey, Slotnick, and Woeber}]{rumsey2022over}
\bibinfo{author}{C.~L. Rumsey}, \bibinfo{author}{J.~P. Slotnick}, \bibinfo{author}{C.~Woeber},
\newblock \bibinfo{title}{Hlpw-4/gmgw-3: Overview and workshop summary},
\newblock in: \bibinfo{booktitle}{AIAA Aviation 2022 Forum}, \bibinfo{year}{2022}, p. \bibinfo{pages}{3295}.
\bibitem[{Larsson et~al.(2016)Larsson, Kawai, Bodart, and Bermejo-Moreno}]{larsson2016large}
\bibinfo{author}{J.~Larsson}, \bibinfo{author}{S.~Kawai}, \bibinfo{author}{J.~Bodart}, \bibinfo{author}{I.~Bermejo-Moreno},
\newblock \bibinfo{title}{Large eddy simulation with modeled wall-stress: recent progress and future directions},
\newblock \bibinfo{journal}{Mech. Eng. Rev.} \bibinfo{volume}{3} (\bibinfo{year}{2016}) \bibinfo{pages}{15--00418}.
\bibitem[{Bose and Park(2018)}]{bose2018wall}
\bibinfo{author}{S.~T. Bose}, \bibinfo{author}{G.~I. Park},
\newblock \bibinfo{title}{Wall-modeled large-eddy simulation for complex turbulent flows},
\newblock \bibinfo{journal}{Annu. Rev. Fluid Mech.} \bibinfo{volume}{50} (\bibinfo{year}{2018}) \bibinfo{pages}{535--561}.
\bibitem[{Wang and Moin(2002)}]{wang2002dynamic}
\bibinfo{author}{M.~Wang}, \bibinfo{author}{P.~Moin},
\newblock \bibinfo{title}{Dynamic wall modeling for large-eddy simulation of complex turbulent flows},
\newblock \bibinfo{journal}{Phys. Fluids} \bibinfo{volume}{14} (\bibinfo{year}{2002}) \bibinfo{pages}{2043--2051}.
\bibitem[{Kawai and Larsson(2013)}]{kawai2013dynamic}
\bibinfo{author}{S.~Kawai}, \bibinfo{author}{J.~Larsson},
\newblock \bibinfo{title}{Dynamic non-equilibrium wall-modeling for large eddy simulation at high \uppercase{R}eynolds numbers},
\newblock \bibinfo{journal}{Phys. Fluids} \bibinfo{volume}{25} (\bibinfo{year}{2013}).
\bibitem[{Bose and Moin(2014)}]{bose2014dynamic}
\bibinfo{author}{S.~T. Bose}, \bibinfo{author}{P.~Moin},
\newblock \bibinfo{title}{A dynamic slip boundary condition for wall-modeled large-eddy simulation},
\newblock \bibinfo{journal}{Phys. Fluids} \bibinfo{volume}{26} (\bibinfo{year}{2014}).
\bibitem[{Park and Moin(2014)}]{park2014improved}
\bibinfo{author}{G.~I. Park}, \bibinfo{author}{P.~Moin},
\newblock \bibinfo{title}{An improved dynamic non-equilibrium wall-model for large eddy simulation},
\newblock \bibinfo{journal}{Phys. Fluids} \bibinfo{volume}{26} (\bibinfo{year}{2014}).
\bibitem[{Bae et~al.(2019)Bae, Lozano-Dur{\'a}n, Bose, and Moin}]{bae2019dynamic}
\bibinfo{author}{H.~J. Bae}, \bibinfo{author}{A.~Lozano-Dur{\'a}n}, \bibinfo{author}{S.~T. Bose}, \bibinfo{author}{P.~Moin},
\newblock \bibinfo{title}{Dynamic slip wall model for large-eddy simulation},
\newblock \bibinfo{journal}{J. Fluid Mech.} \bibinfo{volume}{859} (\bibinfo{year}{2019}) \bibinfo{pages}{400--432}.
\bibitem[{Bodart et~al.(2013)Bodart, Larsson, and Moin}]{bodart2013large}
\bibinfo{author}{J.~Bodart}, \bibinfo{author}{J.~Larsson}, \bibinfo{author}{P.~Moin},
\newblock \bibinfo{title}{Large eddy simulation of high-lift devices},
\newblock in: \bibinfo{booktitle}{21st AIAA Computational Fluid Dynamics Conference}, \bibinfo{year}{2013}, p. \bibinfo{pages}{2724}.
\bibitem[{Park(2017)}]{park2017wall}
\bibinfo{author}{G.~I. Park},
\newblock \bibinfo{title}{Wall-modeled large-eddy simulation of a high \uppercase{R}eynolds number separating and reattaching flow},
\newblock \bibinfo{journal}{AIAA J.} \bibinfo{volume}{55} (\bibinfo{year}{2017}) \bibinfo{pages}{3709--3721}.
\bibitem[{Whitmore et~al.(2021)Whitmore, Griffin, Bose, and Moin}]{whitmore2021large}
\bibinfo{author}{M.~Whitmore}, \bibinfo{author}{K.~Griffin}, \bibinfo{author}{S.~Bose}, \bibinfo{author}{P.~Moin},
\newblock \bibinfo{title}{Large-eddy simulation of a \uppercase{G}aussian bump with slip-wall boundary conditions},
\newblock in: \bibinfo{booktitle}{Center for Turbulence Research Annual Research Briefs}, \bibinfo{year}{2021}, pp. \bibinfo{pages}{45--58}.
\bibitem[{Iyer and Malik(2022)}]{iyer2022wall}
\bibinfo{author}{P.~S. Iyer}, \bibinfo{author}{M.~R. Malik},
\newblock \bibinfo{title}{Wall-modeled \uppercase{LES} of turbulent flow over a two dimensional \uppercase{G}aussian bump},
\newblock in: \bibinfo{booktitle}{ICCFD11 Paper}, volume \bibinfo{volume}{204}, \bibinfo{year}{2022}.
\bibitem[{Rezaeiravesh et~al.(2019)Rezaeiravesh, Mukha, and Liefvendahl}]{rezaeiravesh2019systematic}
\bibinfo{author}{S.~Rezaeiravesh}, \bibinfo{author}{T.~Mukha}, \bibinfo{author}{M.~Liefvendahl},
\newblock \bibinfo{title}{Systematic study of accuracy of wall-modeled large eddy simulation using uncertainty quantification techniques},
\newblock \bibinfo{journal}{Comput. Fluids} \bibinfo{volume}{185} (\bibinfo{year}{2019}) \bibinfo{pages}{34--58}.
\bibitem[{Lozano-Dur{\'a}n and Bae(2019)}]{lozano2019error}
\bibinfo{author}{A.~Lozano-Dur{\'a}n}, \bibinfo{author}{H.~J. Bae},
\newblock \bibinfo{title}{Error scaling of large-eddy simulation in the outer region of wall-bounded turbulence},
\newblock \bibinfo{journal}{J. Comput. Phys.} \bibinfo{volume}{392} (\bibinfo{year}{2019}) \bibinfo{pages}{532--555}.
\bibitem[{Whitmore et~al.(2020)Whitmore, Lozano-Dur{\'a}n, and Moin}]{whitmore2020requirements}
\bibinfo{author}{M.~Whitmore}, \bibinfo{author}{A.~Lozano-Dur{\'a}n}, \bibinfo{author}{P.~Moin},
\newblock \bibinfo{title}{Requirements and sensitivity analysis of \uppercase{RANS}-free wall-modeled \uppercase{LES}},
\newblock in: \bibinfo{booktitle}{Center for Turbulence Research Annual Research Briefs}, \bibinfo{year}{2020}, pp. \bibinfo{pages}{97--108}.
\bibitem[{Slotnick(2019)}]{slotnick2019integrated}
\bibinfo{author}{J.~P. Slotnick},
\newblock \bibinfo{title}{Integrated \uppercase{CFD} validation experiments for prediction of turbulent separated flows for subsonic transport aircraft},
\newblock in: \bibinfo{booktitle}{Separated Flow: Prediction, Measurement and Assessment for Air and Sea Vehicles, STO Publication STO-MP-AVT-307-06}, \bibinfo{year}{2019}.
\bibitem[{Williams et~al.(2020)Williams, Samuell, Sarwas, Robbins, and Ferrante}]{williams2020experimental}
\bibinfo{author}{O.~Williams}, \bibinfo{author}{M.~Samuell}, \bibinfo{author}{E.~S. Sarwas}, \bibinfo{author}{M.~Robbins}, \bibinfo{author}{A.~Ferrante},
\newblock \bibinfo{title}{Experimental study of a \uppercase{CFD} validation test case for turbulent separated flows},
\newblock in: \bibinfo{booktitle}{AIAA Scitech 2020 Forum}, \bibinfo{year}{2020}, p. \bibinfo{pages}{0092}.
\bibitem[{Gray et~al.(2021)Gray, Gluzman, Thomas, Corke, Lakebrink, and Mejia}]{gray2021new}
\bibinfo{author}{P.~D. Gray}, \bibinfo{author}{I.~Gluzman}, \bibinfo{author}{F.~Thomas}, \bibinfo{author}{T.~Corke}, \bibinfo{author}{M.~Lakebrink}, \bibinfo{author}{K.~Mejia},
\newblock \bibinfo{title}{A new validation experiment for smooth-body separation},
\newblock in: \bibinfo{booktitle}{AIAA Aviation 2021 Forum}, \bibinfo{year}{2021}, p. \bibinfo{pages}{2810}.
\bibitem[{Gray et~al.(2022{\natexlab{a}})Gray, Gluzman, Thomas, and Corke}]{gray2022experimental}
\bibinfo{author}{P.~D. Gray}, \bibinfo{author}{I.~Gluzman}, \bibinfo{author}{F.~O. Thomas}, \bibinfo{author}{T.~C. Corke},
\newblock \bibinfo{title}{Experimental characterization of smooth body flow separation over wall-mounted \uppercase{G}aussian bump},
\newblock in: \bibinfo{booktitle}{AIAA Scitech 2022 Forum}, \bibinfo{year}{2022}{\natexlab{a}}, p. \bibinfo{pages}{1209}.
\bibitem[{Gray et~al.(2022{\natexlab{b}})Gray, Gluzman, Thomas, Corke, Lakebrink, and Mejia}]{gray2022benchmark}
\bibinfo{author}{P.~D. Gray}, \bibinfo{author}{I.~Gluzman}, \bibinfo{author}{F.~O. Thomas}, \bibinfo{author}{T.~C. Corke}, \bibinfo{author}{M.~T. Lakebrink}, \bibinfo{author}{K.~Mejia},
\newblock \bibinfo{title}{Benchmark characterization of separated flow over smooth \uppercase{G}aussian bump},
\newblock in: \bibinfo{booktitle}{AIAA Aviation 2022 Forum}, \bibinfo{year}{2022}{\natexlab{b}}, p. \bibinfo{pages}{3342}.
\bibitem[{Gluzman et~al.(2022)Gluzman, Gray, Mejia, Corke, and Thomas}]{gluzman2022simplified}
\bibinfo{author}{I.~Gluzman}, \bibinfo{author}{P.~Gray}, \bibinfo{author}{K.~Mejia}, \bibinfo{author}{T.~C. Corke}, \bibinfo{author}{F.~O. Thomas},
\newblock \bibinfo{title}{A simplified photogrammetry procedure in oil-film interferometry for accurate skin-friction measurement over arbitrary geometries},
\newblock \bibinfo{journal}{Exp. Fluids} \bibinfo{volume}{63} (\bibinfo{year}{2022}) \bibinfo{pages}{118}.
\bibitem[{Balin and Jansen(2021)}]{balin2021direct}
\bibinfo{author}{R.~Balin}, \bibinfo{author}{K.~E. Jansen},
\newblock \bibinfo{title}{Direct numerical simulation of a turbulent boundary layer over a bump with strong pressure gradients},
\newblock \bibinfo{journal}{J. Fluid Mech.} \bibinfo{volume}{918} (\bibinfo{year}{2021}) \bibinfo{pages}{A14}.
\bibitem[{Uzun and Malik(2021)}]{uzun2021simulation}
\bibinfo{author}{A.~Uzun}, \bibinfo{author}{M.~R. Malik},
\newblock \bibinfo{title}{Simulation of a turbulent flow subjected to favorable and adverse pressure gradients},
\newblock \bibinfo{journal}{Theor. Comp. Fluid Dyn.} \bibinfo{volume}{35} (\bibinfo{year}{2021}) \bibinfo{pages}{293--329}.
\bibitem[{Rizzetta and Garmann(2023)}]{rizzetta2023wall}
\bibinfo{author}{D.~P. Rizzetta}, \bibinfo{author}{D.~J. Garmann},
\newblock \bibinfo{title}{Wall-resolved large-eddy simulation of flow over a three-dimensional \uppercase{g}aussian bump},
\newblock in: \bibinfo{booktitle}{AIAA Scitech 2023 Forum}, \bibinfo{year}{2023}, p. \bibinfo{pages}{0286}.
\bibitem[{Wright et~al.(2021)Wright, Balin, Jansen, and Evans}]{wright2021unstructured}
\bibinfo{author}{J.~R. Wright}, \bibinfo{author}{R.~Balin}, \bibinfo{author}{K.~E. Jansen}, \bibinfo{author}{J.~A. Evans},
\newblock \bibinfo{title}{Unstructured \uppercase{LES\_DNS} of a turbulent boundary layer over a \uppercase{G}aussian bump},
\newblock in: \bibinfo{booktitle}{AIAA Scitech 2021 Forum}, \bibinfo{year}{2021}, p. \bibinfo{pages}{1746}.
\bibitem[{Uzun and Malik(2022)}]{uzun2022high}
\bibinfo{author}{A.~Uzun}, \bibinfo{author}{M.~R. Malik},
\newblock \bibinfo{title}{High-fidelity simulation of turbulent flow past \uppercase{G}aussian bump},
\newblock \bibinfo{journal}{AIAA J.} \bibinfo{volume}{60} (\bibinfo{year}{2022}) \bibinfo{pages}{2130--2149}.
\bibitem[{Agrawal et~al.(2022)Agrawal, Whitmore, Griffin, Bose, and Moin}]{agrawal2022non}
\bibinfo{author}{R.~Agrawal}, \bibinfo{author}{M.~P. Whitmore}, \bibinfo{author}{K.~P. Griffin}, \bibinfo{author}{S.~T. Bose}, \bibinfo{author}{P.~Moin},
\newblock \bibinfo{title}{Non-\uppercase{B}oussinesq subgrid-scale model with dynamic tensorial coefficients},
\newblock \bibinfo{journal}{Phys. Rev. Fluids} \bibinfo{volume}{7} (\bibinfo{year}{2022}) \bibinfo{pages}{074602}.
\bibitem[{Zhou et~al.(2023)Zhou, Whitmore, Griffin, and Bae}]{zhou2023large}
\bibinfo{author}{D.~Zhou}, \bibinfo{author}{M.~P. Whitmore}, \bibinfo{author}{K.~P. Griffin}, \bibinfo{author}{H.~J. Bae},
\newblock \bibinfo{title}{Large-eddy simulation of flow over \uppercase{B}oeing \uppercase{G}aussian bump using multi-agent reinforcement learning wall model},
\newblock in: \bibinfo{booktitle}{AIAA Aviation 2023 Forum}, \bibinfo{year}{2023}, p. \bibinfo{pages}{3985}.
\bibitem[{Arranz et~al.(2023)Arranz, Ling, and Lozano-Duran}]{arranz2023wall}
\bibinfo{author}{G.~Arranz}, \bibinfo{author}{Y.~Ling}, \bibinfo{author}{A.~Lozano-Duran},
\newblock \bibinfo{title}{Wall-modeled \uppercase{LES} based on building-block flows: Application to the \uppercase{G}aussian bump},
\newblock in: \bibinfo{booktitle}{AIAA Aviation 2023 Forum}, \bibinfo{year}{2023}, p. \bibinfo{pages}{3984}.
\bibitem[{Balin et~al.(2020)Balin, Jansen, and Spalart}]{balin2020wall}
\bibinfo{author}{R.~Balin}, \bibinfo{author}{K.~E. Jansen}, \bibinfo{author}{P.~R. Spalart},
\newblock \bibinfo{title}{Wall-modeled \uppercase{LES} of flow over a \uppercase{G}aussian bump with strong pressure gradients and separation},
\newblock in: \bibinfo{booktitle}{AIAA Aviation 2020 Forum}, \bibinfo{year}{2020}, p. \bibinfo{pages}{3012}.
\bibitem[{You et~al.(2008)You, Ham, and Moin}]{you2008discrete}
\bibinfo{author}{D.~You}, \bibinfo{author}{F.~Ham}, \bibinfo{author}{P.~Moin},
\newblock \bibinfo{title}{Discrete conservation principles in large-eddy simulation with application to separation control over an airfoil},
\newblock \bibinfo{journal}{Phys. Fluids} \bibinfo{volume}{20} (\bibinfo{year}{2008}) \bibinfo{pages}{101515}.
\bibitem[{Van~der Vorst(1992)}]{van1992bi}
\bibinfo{author}{H.~A. Van~der Vorst},
\newblock \bibinfo{title}{\uppercase{B}\lowercase{i}-\uppercase{CGSTAB}: \uppercase{A} fast and smoothly converging variant of \uppercase{B}\lowercase{i}-\uppercase{CG} for the solution of nonsymmetric linear systems},
\newblock \bibinfo{journal}{SIAM J. Sci. Stat. Comp.} \bibinfo{volume}{13} (\bibinfo{year}{1992}) \bibinfo{pages}{631--644}.
\bibitem[{Vreman(2004)}]{vreman2004eddy}
\bibinfo{author}{A.~W. Vreman},
\newblock \bibinfo{title}{An eddy-viscosity subgrid-scale model for turbulent shear flow: Algebraic theory and applications},
\newblock \bibinfo{journal}{Phys. Fluids} \bibinfo{volume}{16} (\bibinfo{year}{2004}) \bibinfo{pages}{3670--3681}.
\bibitem[{Germano et~al.(1991)Germano, Piomelli, Moin, and Cabot}]{germano1991dynamic}
\bibinfo{author}{M.~Germano}, \bibinfo{author}{U.~Piomelli}, \bibinfo{author}{P.~Moin}, \bibinfo{author}{W.~H. Cabot},
\newblock \bibinfo{title}{A dynamic subgrid-scale eddy viscosity model},
\newblock \bibinfo{journal}{Phys. Fluids A-Fluid} \bibinfo{volume}{3} (\bibinfo{year}{1991}) \bibinfo{pages}{1760--1765}.
\bibitem[{Lilly(1992)}]{lilly1992proposed}
\bibinfo{author}{D.~K. Lilly},
\newblock \bibinfo{title}{A proposed modification of the \uppercase{G}ermano subgrid-scale closure method},
\newblock \bibinfo{journal}{Phys. Fluids A-Fluid} \bibinfo{volume}{4} (\bibinfo{year}{1992}) \bibinfo{pages}{633--635}.
\bibitem[{Rozema et~al.(2015)Rozema, Bae, Moin, and Verstappen}]{rozema2015minimum}
\bibinfo{author}{W.~Rozema}, \bibinfo{author}{H.~J. Bae}, \bibinfo{author}{P.~Moin}, \bibinfo{author}{R.~Verstappen},
\newblock \bibinfo{title}{Minimum-dissipation models for large-eddy simulation},
\newblock \bibinfo{journal}{Phys. Fluids} \bibinfo{volume}{27} (\bibinfo{year}{2015}).
\bibitem[{Bardina(1983)}]{bardina1983improved}
\bibinfo{author}{J.~Bardina}, \bibinfo{title}{Improved turbulence models based on large eddy simulation of homogeneous, incompressible turbulent flows}, \bibinfo{publisher}{Stanford University}, \bibinfo{year}{1983}.
\bibitem[{Sarghini et~al.(1999)Sarghini, Piomelli, and Balaras}]{sarghini1999scale}
\bibinfo{author}{F.~Sarghini}, \bibinfo{author}{U.~Piomelli}, \bibinfo{author}{E.~Balaras},
\newblock \bibinfo{title}{Scale-similar models for large-eddy simulations},
\newblock \bibinfo{journal}{Phys. Fluids} \bibinfo{volume}{11} (\bibinfo{year}{1999}) \bibinfo{pages}{1596--1607}.
\bibitem[{Meneveau and Katz(2000)}]{meneveau2000scale}
\bibinfo{author}{C.~Meneveau}, \bibinfo{author}{J.~Katz},
\newblock \bibinfo{title}{Scale-invariance and turbulence models for large-eddy simulation},
\newblock \bibinfo{journal}{Annu. Rev. Fluid Mech.} \bibinfo{volume}{32} (\bibinfo{year}{2000}) \bibinfo{pages}{1--32}.
\bibitem[{Smagorinsky(1963)}]{smagorinsky1963general}
\bibinfo{author}{J.~Smagorinsky},
\newblock \bibinfo{title}{General circulation experiments with the primitive equations: I. the basic experiment},
\newblock \bibinfo{journal}{Mon. Weather Rev.} \bibinfo{volume}{91} (\bibinfo{year}{1963}) \bibinfo{pages}{99--164}.
\bibitem[{Yang and Wang(2013)}]{yang2013boundary}
\bibinfo{author}{Q.~Yang}, \bibinfo{author}{M.~Wang},
\newblock \bibinfo{title}{Boundary-layer noise induced by arrays of roughness elements},
\newblock \bibinfo{journal}{J. Fluid Mech.} \bibinfo{volume}{727} (\bibinfo{year}{2013}) \bibinfo{pages}{282--317}.
\bibitem[{Zhou et~al.(2020)Zhou, Wang, and Wang}]{zhou2020large}
\bibinfo{author}{D.~Zhou}, \bibinfo{author}{K.~Wang}, \bibinfo{author}{M.~Wang},
\newblock \bibinfo{title}{Large-eddy simulation of an axisymmetric boundary layer on a body of revolution},
\newblock in: \bibinfo{booktitle}{AIAA Aviation 2020 Forum}, \bibinfo{year}{2020}, p. \bibinfo{pages}{2989}.
\bibitem[{Zhou et~al.(2022)Zhou, Wang, and Wang}]{zhou2022computational}
\bibinfo{author}{D.~Zhou}, \bibinfo{author}{K.~Wang}, \bibinfo{author}{M.~Wang},
\newblock \bibinfo{title}{Computational analysis of noise generation by a rotor at the rear of an axisymmetric body of revolution},
\newblock in: \bibinfo{booktitle}{28th AIAA/CEAS Aeroacoustics 2022 Conference}, \bibinfo{year}{2022}, p. \bibinfo{pages}{3090}.
\bibitem[{Bae and Lozano-Dur{\'a}n(2021)}]{bae2021effect}
\bibinfo{author}{H.~J. Bae}, \bibinfo{author}{A.~Lozano-Dur{\'a}n},
\newblock \bibinfo{title}{Effect of wall boundary conditions on a wall-modeled large-eddy simulation in a finite-difference framework},
\newblock \bibinfo{journal}{Fluids} \bibinfo{volume}{6} (\bibinfo{year}{2021}) \bibinfo{pages}{112}.
\bibitem[{Zhou et~al.(2022)Zhou, Whitmore, Griffin, and Bae}]{zhou2022RLWM}
\bibinfo{author}{D.~Zhou}, \bibinfo{author}{M.~P. Whitmore}, \bibinfo{author}{K.~P. Griffin}, \bibinfo{author}{H.~J. Bae},
\newblock \bibinfo{title}{Multi-agent reinforcement learning for wall modeling in \uppercase{LES} of flow over periodic hills},
\newblock in: \bibinfo{booktitle}{Proceedings of the CTR Summer Program}, \bibinfo{year}{2022}, pp. \bibinfo{pages}{25--34}.
\bibitem[{Lund et~al.(1998)Lund, Wu, and Squires}]{lund1998generation}
\bibinfo{author}{T.~S. Lund}, \bibinfo{author}{X.~Wu}, \bibinfo{author}{K.~D. Squires},
\newblock \bibinfo{title}{Generation of turbulent inflow data for spatially-developing boundary layer simulations},
\newblock \bibinfo{journal}{J. Comput. Phys.} \bibinfo{volume}{140} (\bibinfo{year}{1998}) \bibinfo{pages}{233--258}.
\bibitem[{Inagaki et~al.(2023)Inagaki, Kobayashi et~al.}]{inagaki2023analysis}
\bibinfo{author}{K.~Inagaki}, \bibinfo{author}{H.~Kobayashi}, et~al.,
\newblock \bibinfo{title}{Analysis of anisotropic subgrid-scale stress for coarse large-eddy simulation},
\newblock \bibinfo{journal}{Phys. Rev. Fluids} \bibinfo{volume}{8} (\bibinfo{year}{2023}) \bibinfo{pages}{104603}.
\bibitem[{Geurts and Fr{\"o}hlich(2002)}]{geurts2002framework}
\bibinfo{author}{B.~J. Geurts}, \bibinfo{author}{J.~Fr{\"o}hlich},
\newblock \bibinfo{title}{A framework for predicting accuracy limitations in large-eddy simulation},
\newblock \bibinfo{journal}{Phys. Fluids} \bibinfo{volume}{14} (\bibinfo{year}{2002}) \bibinfo{pages}{L41--L44}.
\bibitem[{Bae(2018)}]{bae2018investigation}
\bibinfo{author}{H.~J. Bae}, \bibinfo{title}{Investigation of dynamic subgrid-scale and wall models for turbulent boundary layers}, \bibinfo{publisher}{Stanford University}, \bibinfo{year}{2018}.
\bibitem[{Iyer and Malik(2021)}]{Iyer2021wall}
\bibinfo{author}{P.~S. Iyer}, \bibinfo{author}{M.~R. Malik},
\newblock \bibinfo{title}{Wall-modeled \uppercase{LES} of flow over a gaussian bump},
\newblock in: \bibinfo{booktitle}{AIAA Scitech 2021 Forum}, \bibinfo{year}{2021}, p. \bibinfo{pages}{1438}.
\bibitem[{Uzun and Malik(2022)}]{uzun2022dynamic}
\bibinfo{author}{A.~Uzun}, \bibinfo{author}{M.~R. Malik}, \bibinfo{title}{A dynamic nonlinear subgrid-scale model for large-eddy simulation of complex turbulent flows}, \bibinfo{type}{{NASA Report}} \bibinfo{number}{TM–20220013891}, National Aeronautics and Space Administration, \bibinfo{year}{January~2022}.
\bibitem[{Ling et~al.(2022)Ling, Arranz, Williams, Goc, Griffin, and Lozano-Dur{\'a}n}]{ling2022wall}
\bibinfo{author}{Y.~Ling}, \bibinfo{author}{G.~Arranz}, \bibinfo{author}{E.~Williams}, \bibinfo{author}{K.~Goc}, \bibinfo{author}{K.~Griffin}, \bibinfo{author}{A.~Lozano-Dur{\'a}n},
\newblock \bibinfo{title}{Wall-modeled large-eddy simulation based on building-block flows},
\newblock in: \bibinfo{booktitle}{Proceedings of the CTR Summer Program}, \bibinfo{year}{2022}, pp. \bibinfo{pages}{5--14}.

\end{thebibliography}



\end{document}